\documentclass[usenames,dvipsnames,11pt]{article}

\usepackage{graphicx}
\usepackage{amsmath}
\usepackage{amssymb}
\usepackage{latexsym}
\usepackage{mathrsfs}
\usepackage{amsthm}
\usepackage{setspace}
\usepackage{epsfig}
\usepackage{subfigure}
\usepackage{authblk}
\usepackage{amsfonts}
\usepackage{url}
\usepackage{float}
\usepackage{multibib}

\newcites{appendix}{Reference}

\usepackage[authoryear,round]{natbib}	
\usepackage{mathtools}														 
\mathtoolsset{showonlyrefs=true}									 

\usepackage[dvips]{color}

\usepackage[margin=1in]{geometry}
\newtheorem{proposition}{Proposition}[section]

\newtheorem{definition}{Definition}[section]
\newtheorem{assumption}{Assumption}[section]
\numberwithin{equation}{section}
\newtheorem{theorem}{Theorem}[section]
\newtheorem{remark}{Remark}[section]
\newtheorem{vg}{Example}[section]

\newcommand{\be}{\begin{equation}}
\newcommand{\ee}{\end{equation}}
\newcommand{\nee}{\nonumber\end{equation}}
\newcommand{\eel}[1]{\label{#1}\end{equation}}
\newcommand{\brmk}[1]{\begin{remark}\label{#1}\begin{em} }
\newcommand{\ermk}{ $\quad\triangleleft$\end{em}\end{remark}}
\newcommand{\bvg}[1]{\begin{vg}\label{#1}\begin{em} }
\newcommand{\evg}{ $\quad\triangleleft$\end{em}\end{vg}}

\usepackage{lipsum}

\newcommand\blfootnote[1]{%
  \begingroup
  \renewcommand\thefootnote{}\footnote{#1}%
  \addtocounter{footnote}{0}%
  \endgroup
}

\usepackage{natbib}
\begin{document}
\bibliographystyle{plainnat}
\begin{titlepage}

\title{{\bf{\Huge Positive Eigenfunctions of Markovian Pricing Operators: Hansen-Scheinkman Factorization, Ross Recovery and Long-Term Pricing}}\blfootnote{
The authors thank Peter Carr for bringing Ross¡¯ Recovery Theorem to their attention and for numerous stimulating discussions, and Torben Andersen, Jaroslav Borovicka, Peter Carr, Lars Peter Hansen, Stephen Ross, Jose Scheinkman, Viktor Todorov, the two anonymous referees, and the anonymous Associate Editor for useful comments that helped improve this paper. This research was supported in part by the grant CMMI 1536503 from the National Science Foundation.}}

\author{Likuan Qin\footnote{likuanqin2012@u.northwestern.edu}
}
\author{Vadim Linetsky\footnote{linetsky@iems.northwestern.edu}
}
\affil{\emph{Department of Industrial Engineering and Management Sciences}\\
\emph{McCormick School of Engineering and Applied Sciences}\\
\emph{Northwestern University}}
\date{}

\end{titlepage}

\maketitle

\begin{abstract}
This paper develops a spectral theory of Markovian asset pricing models where the underlying economic uncertainty follows a continuous-time Markov process $X$ with a general state space (Borel right process (BRP)) and the stochastic discount factor (SDF) is a positive semimartingale multiplicative functional of $X$.
A key result is the uniqueness theorem for a positive eigenfunction of the pricing operator such that $X$ is recurrent under a new probability measure associated with this eigenfunction (recurrent eigenfunction).
As economic applications, we prove uniqueness of the \citet{hansen_2009} factorization of the Markovian SDF corresponding to the recurrent eigenfunction, extend the Recovery Theorem of \citet{ross_2011} from discrete time, finite state irreducible Markov chains to
recurrent BRPs, and obtain the long-maturity asymptotics of the pricing operator.
When an asset pricing model is specified by given risk-neutral probabilities together with a short rate function of the Markovian state, we give sufficient conditions for existence of a recurrent eigenfunction and
provide explicit examples in a number of important financial models, including affine and quadratic diffusion models and an affine model with jumps. These examples show that the recurrence assumption, in addition to fixing uniqueness, rules out unstable economic dynamics, such as the short rate asymptotically going to infinity or to a zero lower bound trap without possibility of escaping.
\end{abstract}

\section{Introduction}

In frictionless markets free of arbitrage, the pricing relationship assigning prices to uncertain future payoffs is a linear operator. In particular, if all uncertainty is generated by a time-homogeneous Markov process $X$ and the stochastic discount factor (also known as the pricing kernel) is a positive multiplicative functional of $X$, the pricing operators indexed by time between the present and the future payoff date form an operator semigroup when payoffs viewed as functions of the future Markov state  are assumed to lay in an appropriate function space. Early contributions on Markovian pricing semigroups include \citet{garman_1985} and \citet{duffie_1991}. \citet{hansen_2009} give a comprehensive study of Markovian pricing semigroups in financial economics. \citet{linetsky_2004} and \citet{linetsky_2008} survey a range of applications in financial engineering.

If the Markov process used to model the underlying economic uncertainty belongs to the class of symmetric Markov processes (cf. \citet{chen_2011} and \citet{fukushima_2010})  and one limits oneself to square-integrable payoffs so that the pricing operators are symmetric in the corresponding $L^2$ space, one can harness the power of the Spectral Theorem for self-adjoint operators in Hilbert spaces to construct a spectral resolution of the pricing operator. If the spectrum is purely discrete, one obtains convenient eigenfunction expansions that expand square-integrable payoffs in the basis of {\em eigen-payoffs} that are eigenfunctions of the pricing operator. The prices then also expand in the eigenfunction basis.
The concept of {\em eigen-securities}, contingent claims with eigen-payoffs, is explicitly introduced in \citet{davydov_2003}. Early applications of  eigenfunction expansions in finance appear in \citet{beaglehole_1992}.  Applications of the spectral method to pricing a wide variety of securities in financial engineering can be found in \citet{lewis_1998}, \citet{lewis_2000option}, \citet{lipton_2001mathematical},  \citet{albanese_2001black}, \citet{lipton_2002universal}, \citet{davydov_2003}, \citet{albanese_2004unifying}, \citet{gorovoi_2004},  \citet{linetsky_2004}, \citet{linetsky_2004assian}, \citet{linetsky_2004bessel}, \citet{linetsky_2006pricing}, \citet{boyarchenko_2007eigenfunction}, \citet{gorovoy_2007intensity}, \citet{rafael_2010}, \citet{mendoza_2011pricing},  \citet{fouque_2011spectral}, \citet{lorig_2011time}, \citet{rafael_2013},  \citet{li_2013optimal}, \citet{mendoza_2014multivariate}, \citet{lingfei_2012}.

In this paper we depart from this literature in the following ways. First, we do not impose any structural assumptions on the Markov process, other than assuming that it is a Borel right process, the most general Markov process that can serve as the Markovian stochastic driver of an arbitrage-free economy. In particular, we do not make any symmetry assumptions. Furthermore, we do not restrict the space of payoffs other than Borel measurability. On the other hand, in this paper we focus only on strictly positive eigenfunctions and, in particular, feature and investigate eigen-securities with strictly positive eigen-payoffs.

Our focus on positive eigenfunctions is due to two important recent developments in financial economics. First, \citet{hansen_2009} introduce the following remarkable factorization of the Markovian stochastic discount factor (SDF). If $\pi(x)$ is a positive eigenfunction of the pricing operator
\be
{\mathscr P}_tf(x):={\mathbb E}_x^{\mathbb P}[S_t f(X_t)]
\eel{pp}
mapping time-$t$ payoffs to time-zero prices with the eigenvalue $e^{-\lambda t}$ for some real $\lambda$ ($-\lambda$ is the eigenvalue of the properly defined infinitesimal generator of the pricing semigroup), i.e.,
\be
{\mathscr P}_t \pi(x)=e^{-\lambda t}\pi(x),
\eel{eigen}
then the SDF or pricing kernel (PK) $S_t$ admits a factorization:
\be
S_t = e^{-\lambda t}\frac{\pi(X_0)}{\pi(X_t)}M_t^\pi, \quad \text{where}\quad
M_t^\pi  = e^{\lambda t}\frac{\pi(X_t)}{\pi(X_0)}S_t
\eel{HS}
 is a positive martingale with $M_0^\pi=1$
(our $\lambda=-\rho$ in Section 6 of  \citet{hansen_2009}). \citet{hansen_2009} use $M^\pi$ to introduce a new probability measure, which we call the {\em eigen-measure} and denote by ${\mathbb Q}^\pi$ to signify that it is associated with the eigenfunction $\pi$. Each ${\mathbb Q}^\pi$  is characterized by the property that the eigen-security $e^{\lambda t}\frac{\pi(X_t)}{\pi(X_0)}$ associated with the eigenfunction $\pi$ serves as the numeraire asset under that measure (see \citet{geman_1995changes} for changes of measure corresponding to the changes of numeraire). Every positive eigenfunction leads to such a factorization, so in general we have a set of eigen-measures $({\mathbb Q}^\pi)_\pi$ indexed by all positive eigenfunctions.

  \citet{hansen_2009} show that imposing certain stability (ergodicity) assumptions on the dynamics of $X$ under ${\mathbb Q}^\pi$ singles out a unique $\pi$ (and ${\mathbb Q}^\pi$), if it exists. They further give sufficient conditions for existence of such an eigenfunction. Moreover, under their ergodicity assumptions they identify the corresponding
factorization of the SDF with the long-term factorization featured in \citet{alvarez_2005using} in discrete time that decomposes the SDF into discounting at the rate of return on the zero-coupon bond of asymptotically long maturity (the long bond) and a further risk adjustment accomplished by an additional martingale component. The long-term factorization is of central interest in financial economics, as it furnishes an explicit decomposition of risk premia in the economy into the risk premia earned from holding the long bond and additional risk premia. In particular, this long-term risk decomposition is of central importance in macro-finance, the discipline at the intersection of financial economics and macroeconomics.
\citet{hansen_2012}, \citet{hansen_2012pricing}, \citet{hansen_2012recursive}, \citet{hansen_2013}, \citet{borovicka_2013exam}, \citet{borovicka_2014mis} and \citet{linetsky_2014long} provide theoretical developments, and \citet{bakshi_2012} provide some further empirical evidence complementing the original empirical results of \citet{alvarez_2005using}.
The mathematics underlying these developments is the Perron-Frobenius type theory governing positive eigenfunctions for certain classes of positive linear operators in function spaces.


Another closely related recent development that inspired this paper is
the Recovery Theorem of \citet{ross_2011}.
Ross poses a theoretically interesting and practically important question: under what assumptions can one uniquely recover the market participants' beliefs about  physical probabilities  from Arrow-Debreu state prices implied by  observed market prices of options? Such an identification would be of great interest to finance researchers and market participants, as
it would open  avenues for extracting market's assessment of physical probabilities that could be incorporated in investment decisions and supply scenarios for risk management.
Ross' recovery theorem provides the following answer to this question. If all uncertainty in an arbitrage-free, frictionless economy follows a finite state, discrete time irreducible Markov chain and the pricing kernel satisfies a structural assumption of {\em transition independence}, then Ross shows that there exists a unique recovery of physical probabilities from given Arrow-Debreu state prices.
Ross' proof  crucially relies on the celebrated Perron-Frobenius theorem establishing existence and uniqueness of a positive eigenvector of an irreducible non-negative matrix.
\citet{ross_2014talk} also extends his recovery result to the case where the state space is continuous and the pricing kernel is both bounded from above by a constant and bounded away from zero by applying the Krein-Rutman theorem.

\citet{carr_2012} observe that  Ross' recovery result can be extended to 1D diffusions on  bounded intervals with regular boundaries at both ends by observing that the infinitesimal generator of such a diffusion is a regular Sturm-Liouville operator that has a unique positive eigenfuction. They then rely on the regular Sturm-Liouville theory to show uniqueness of Ross recovery. They also provide further insights into the recovery result in the diffusion setting. \citet{dubynskiy_2013} further explore 1D diffusion models with reflecting boundary conditions. \citet{walden_2013} studies recovery for 1D diffusions on  the whole real line, shows that Ross' recovery is possible if both boundaries are non-attracting, and studies the recovery in the classical equilibrium setting with von Neumann-Morgenstern preferences. \citet{audrino_2014empirical} develop a non-parametric estimation procedure for recovery from option prices in the framework of Markov chains with finite state space and conduct an empirical analysis of recovery from S\&P 500 options data.

\citet{hansen_2013} and \citet{borovicka_2014mis} point out that Ross' assumption of transition independence of the Markovian pricing kernel amounts to the specialization of the factorization in  \citet{hansen_2009} to the case where the martingale component is degenerate, specifically $M^\pi=1$ in Eq.\eqref{HS}.
They discuss economic limitations of this assumption. In particular they demonstrate that transition independence does not generally hold in structural models with recursive preferences and/or  non-stationary consumption. They further point out that, if the pricing kernel is not transition independent, then what is recovered via the Perron-Frobenius theory is not the physical probability measure, but rather the eigen-measure ${\mathbb Q}^\pi$, which, under further ergodicity assumptions is identified with the long forward measure featured in \citet{alvarez_2005using} and \citet{hansen_2009}.
   \citet{martin_2013}, working in discrete time, ergodic finite-state Markov chain environments, also discuss identification of Ross' recovered probability measure with the long forward measure. This is further developed by \citet{linetsky_2014long} in general semimartingale environments.

The present paper develops a spectral theory of Markovian pricing operators and, as applications of this theory, extends and complements \citet{hansen_2009} results on factorizations of Markovian pricing kernels and, as a consequence, extends \citet{ross_2011} recovery  to continuous-time Markov processes with general state spaces.
The contributions and structure of this paper are as follows.
Section \ref{framework} presents our setting of Markovian pricing operators associated with a Borel right process (BRP), a Markov process on a state space with a Borel sigma algebra with right-continuous paths and having the strong Markov property. Working in the framework of BRPs allows us to apply the results of \citet{cinlar_1980} on stochastic calculus of semimartingales defined over a right process and ensure that all results hold for all initial states in the state space (and, indeed, for all initial distributions). The BRP framework is general enough to encompass all Markov processes that arise in continuous-time finance, including continuous-time Markov chains, diffusions in the whole Euclidean space, as well as in domains with boundaries and prescribed boundary behavior, and pure jump and jump-diffusion processes in the whole Euclidean space or in domains with boundaries. At the same time, due to the results of \citet{cinlar_1980} on stochastic calculus for Markov processes, it is essentially the most general class of Markov processes that make sense as stochastic drivers of arbitrage-free continuous-time asset pricing models. Thus, our choice of BRP as the stochastic driver in a Markovian economy is fundamental.

The key result of the paper is the uniqueness Theorem \ref{unique_eigen} in Section \ref{unique_r_eigen} for {\em recurrent positive eigenfunctions} of Markovian pricing operators. We prove that there exists at most one positive eigenfunction $\pi_R$ such that the BRP is recurrent under the corresponding eigen-measure ${\mathbb Q}^{\pi_R}$ (we call such an eigenfunction and the corresponding eigen-measure recurrent). {\em  This theorem is an extension of the uniqueness part of the Perron-Frobenius theorem to Feynman-Kac-type operators associated with positive multiplicative functionals of BRPs.}
This result yields uniqueness of a recurrent Hansen-Scheinkman factorization of the Markovian SDF, as well as uniqueness of the recurrent eigen-measure ${\mathbb Q}^{\pi_R}$ (Theorem \ref{unique_HS}). As a consequence, it immediately yields the uniqueness part of Ross' recovery theorem under assumptions that the Markovian driver is a recurrent Borel right process and the SDF is transition independent via identification of the physical measure ${\mathbb P}$ with the recurrent eigen-measure ${\mathbb Q}^{\pi_R}$ under these assumptions (Theorem \ref{unique_theorem_1}). Under additional ergodicity assumptions, it further yields a long-maturity asymptotics of the Markovian pricing operator:
\be
{\mathscr P}_tf(x)=c_f e^{-\lambda_R t}\pi_R(x)+O(e^{-(\lambda_R+\alpha)t}),
\eel{LTA}
where the constant $c_f$ depends on the payoff $f$ (Theorem \ref{long_pricing}), thus further complementing the results of \citet{hansen_2009}.

In Section \ref{riskless_rate} we further show that if we assume that the instantaneous riskless interest rate (short rate) exists, then it is uniquely identified from the knowledge of any positive eigenfunction and the corresponding eigenvalue of the pricing operator. This further leads to the identification of the corresponding risk-neutral probabilities under which all asset prices are martingales when taken relative to the riskless asset (money market account) earning interest at the instantaneous riskless interest rate.

Owing to the importance of the short rate modeling approach in financial engineering, in Section \ref{exist} we start with a {\em given} risk-neutral measure ${\mathbb Q}$ governing the risk-neutral dynamics of the Markov process $X$ and a given short rate function $r(x)$ and study existence of a recurrent eigenfunction of the pricing operator defined by the given pair $(X,r)$.
In particular, we provide three distinct sets of sufficient conditions for existence of a recurrent eigenfunction when $X$ is specified under the risk-neutral measure together with a given short rate function: for Hunt processes with duals and Hilbert-Schmidt semigroups, for one-dimensional diffusions on (finite or infinite) intervals, and for diffusions in ${\mathbb R}^d$.
The first result is based on Jentzsch's theorem, which is a counterpart of Perron-Frobenius theorem for integral operators in $L^2$ spaces and follows the recent work of \citet{zhang_2013} on quasi-stationarity and quasi-ergodicity.  The second result is based on the application of the singular Sturm-Liouville theory (in particular, Sturm's oscillations of solutions) to 1D diffusions (cf. \citet{linetsky_2004}, \citet{linetsky_2008}). The third result is based on the theory of positive harmonic functions and diffusions in \citet{pinsky_1995}.

In Section \ref{examples5} we give examples of parametric models with short rates, where the recurrent eigenfunction can be determined in closed form, yielding an explicit recurrent Hansen-Scheinkman factorization and, under the additional assumption of transition independence, Ross recovery. These include a variety of 1D diffusion models, multi-dimensional affine and quadratic diffusion models, as well as a CIR model with jumps.
Our examples show that the recurrence assumption, in addition to fixing uniqueness, rules out unstable economic dynamics, such as the risk-free interest rate asymptotically going to infinity or to a zero lower bound trap with no possibility of escaping.
The e-companion contain proofs, additional technical materials, and further examples.


\section{Markovian Pricing Kernels: Eigen-Securities, Hansen-Scheinkman Factorization and Ross Recovery}
\label{framework}


The stochastic driver of all economic uncertainty in our model is a conservative {\em Borel right process} (BRP) $X=(\Omega,{\mathscr F},({\mathscr F}_{t})_{t\geq 0},(X_t)_{t\geq 0},({\mathbb P}_x)_{x\in E})$.
A BRP is a continuous-time, time-homogeneous Markov process taking values in a Borel subset $E$ of some metric space (so that $E$ is equipped with a Borel sigma-algebra ${\mathscr E}$; the reader can think of $E$ as a Borel subset of the Euclidean space ${\mathbb R}^d$), having right-continuous paths and possessing the strong Markov property (i.e., the Markov property extended to stopping times). The probability measure  ${\mathbb P}_x$ governs the behavior of the process $(X_t)_{t\geq 0}$ when started from $x\in E$ at time zero.
If the process starts from a probability distribution $\mu$, the corresponding measure is denoted ${\mathbb P}_\mu$. A statement concerning $\omega\in \Omega$ is said to hold ${\mathbb P}$-almost surely if it is true ${\mathbb P}_x$-almost surely for all $x\in E$. The information filtration  $({\mathscr F}_{t})_{t\geq 0}$  in our model is the filtration generated by $X$ completed with ${\mathbb P}_\mu$-null sets for all initial distributions $\mu$ of $X_0$. It is right continuous and, thus, satisfies the usual hypothesis of stochastic calculus.
Appendix \ref{prelim_borel_right} in the e-companion gives precise definitions. $X$ is assumed to be conservative, meaning that ${\mathbb P}_x(X_t\in E)=1$ for each initial $x\in E$ and all $t\geq 0$ (the process does not exit the state space $E$ in finite time, i.e. there is no killing or explosion).

Our choice of Borel right processes as the class of Markov processes we work with is due to the work of  \citet{cinlar_1980} (see also Chapter VI of \citet{sharpe_1988}) who develop
stochastic calculus for semimartingales defined over a right process. When dealing with a Markov process, we have a family of probability measures $({\mathbb P}_x)_{x\in E}$ indexed by the initial state $x\in E$.
 \citet{cinlar_1980} show that stochastic calculus of semimartingales defined over a right process can be set up so that all key properties hold simultaneously for all starting points $x\in E$ and, in fact, for all initial distributions $\mu$ of $X_0$.
In particular, an $({\mathscr F}_{t})_{t\geq 0}$-adapted process $S$ is an ${\mathbb P}_x$-semimartingale (local martingale, martingale) {\em simultaneously for all} $x\in E$ and, in fact, for all ${\mathbb P}_\mu$, where $\mu$ is the initial distribution (of $X_0$).
With some abuse of notation, in this section we simply write ${\mathbb P}$ where, in fact, we are dealing with the family of measures $({\mathbb P}_x)_{x\in E}$ indexed by the initial state $x$. Correspondingly, we simply say that a process is a ${\mathbb P}$-semimartingale (local martingale, martingale), meaning that it is a ${\mathbb P}_x$-semimartingale (local martingale, martingale) for each $x\in E$.
The advantage of working in this generality of Borel right processes is that we can treat processes with discrete state spaces (Markov chains), diffusions in the whole Euclidean space or in a domain with a boundary and some boundary behavior, as well as pure jump and jump-diffusion processes in the whole Euclidean space or in a domain with a boundary, all in a unified fashion.

We assume a frictionless, arbitrage-free economy with a positive $((\mathscr{F}_t)_{t\geq 0},{\mathbb P})$-semimartingale pricing kernel (PK) $(S_t)_{t\geq 0}$ (see \citet{duffie_2001}, \citet{hansen_2012risk}, \citet{hansen_2009pricing}, \citet{hansen_2009} and \citet{rogers_1998origins}
 for surveys of PKs).
\begin{assumption}\label{finite_exp}
In this paper the PK $(S_t)_{t\geq 0}$ is assumed to be a strictly positive semimartingale multiplicative functional of the BRP $X$, i.e. $S_{t+s}(\omega)=S_t(\omega)S_s(\theta_t(\omega))$, where  $\theta_s: \Omega\rightarrow \Omega$ is the shift operator, $X_s(\theta_t(\omega))=X_{t+s}(\omega)$, $S$ is  normalized so that $S_0=1$, the process of its left limits is also assumed to be strictly positive, $S_{-}>0$, and ${\mathbb E}^{\mathbb P}_x[S_t]<\infty$ for all $t>0$ and all $x\in E$.
\end{assumption}
The shift operator is defined and discussed in Appendix \ref{prelim_borel_right} in the e-companion. The multiplicative property of the pricing kernel ensures time consistency of pricing in time-homogeneous Markovian environments. Namely,
under Assumption \ref{finite_exp} the time-$s$ price of a payoff $f(X_t)$ at time $t\geq s\geq 0$ is
\be
{\mathbb E}^{\mathbb P}\left[(S_t/ S_s) f(X_{t})  \left|{\mathscr F}_s\right.\right]={\mathbb E}_{X_s}^{\mathbb P}\left[S_{t-s}f(X_{t-s})\right]={\mathscr P}_{t-s}f(X_s),
\eel{pricing}
where we used the Markov property and time homogeneity of $X$ and the multiplicative property of $S$ and introduced a family of {\em pricing operators} $({\mathscr P}_t)_{t\geq 0}$ given by Eq.\eqref{pp},
where ${\mathbb E}_{x}^{\mathbb P}$ denotes the expectation with respect to  ${\mathbb P}_x$.
The pricing operator ${\mathscr P}_t$ maps the payoff function $f$ at time $t$ into its price (present value) function at time zero as the function of the initial state $X_0=x$ and enjoy the semigroup property, ${\mathscr P}_t {\mathscr P}_t={\mathscr P}_{t+s}$. The collection of pricing operators $({\mathscr P}_t)_{t\geq 0}$ is then referred to as the pricing semigroup. The time-$0$ price of an Arrow-Debreu security that pays one unit of account at time $t\geq 0$ if the state $X_t$ is in the Borel set $B\in {\mathscr E}$ and nothing otherwise is
${\mathscr P}_t(x,B)=({\mathscr P}_{t}1_B)(x).$ We call this measure on the state-space the {\em Arrow-Debreu (AD) state-price  measure}. The pricing operators are expressed in terms of the AD state price measures by:
${\mathscr P}_t f(x)=\int_E f(y){\mathscr P}_t(x,dy).$
Thus, the AD measures indexed by time $t$ are the kernels of the pricing operators.

Under Assumption \ref{finite_exp}, the prices of zero-coupon bonds of all maturities are finite,
$P(x,t):={\mathscr P}_t(x,E)={\mathbb E}_{x}^{\mathbb P}[S_t]<\infty$ for all $t>0$ and  $x\in E$.


Suppose the pricing operators ${\mathscr P}_t$ possess a positive eigenfunction $\pi$, i.e. $\pi(x)$ is a strictly positive, finite  Borel function, $0<\pi(x)<\infty$ for all $x\in E$, such that the equation \eqref{eigen}
is satisfied
for all $t\geq 0$ and $x \in E$ and some real $\lambda$.
The key observation of \citet{hansen_2009} is that then the PK
admits a multiplicative factorization \eqref{HS}.
We also call \eqref{HS} the {\em eigen-factorization} of the PK. \citet{hansen_2009} use the martingale $M^\pi$ to define a new probability measure
${\mathbb Q}^\pi|_{{\mathscr F}_t}=M_t^\pi {\mathbb P}|_{{\mathscr F}_t}$
locally equivalent to ${\mathbb P}$ on each ${\mathscr F}_t$.
We call ${\mathbb Q}^\pi$ an {\em eigen-measure}. We note that in the set-up of BRPs $M^\pi$ is a ${\mathbb P}_x$-martingale for each $x\in E$, and ${\mathbb Q}_x^\pi$ is locally equivalent to ${\mathbb P}_x$ for each $x\in E$, and, in fact, for each initial distribution $\mu$. We thus sometimes omit explicit dependence on the initial state $x$. The transition operator $Q_t^\pi$ of $X$ under ${\mathbb Q}^\pi$ reads in terms of the pricing operator:
\be
Q^\pi_t f(x)={\mathbb E}^{{\mathbb Q}^\pi}_x[f(X_t)]={\mathbb E}_x^{\mathbb P}[M_t^\pi f(X_t)]=e^{\lambda t}\frac{1}{\pi(x)}{\mathscr P}_t(\pi f)(x).
\eel{Ppi}


Now consider a security that delivers the payoff $e^{\lambda T}\pi(X_T)/\pi(X_0)$ at the fixed time $T>0$. Since $\pi$ is the eigenfunction of the pricing operator, the time-$t$ price of this security is $e^{\lambda t}\pi(X_t)/\pi(X_0)$ for any $t\leq T$.
Following \citet{davydov_2003}, we call such securities {\em eigen-securities}. Here we normalize their payoffs so that their price at time zero is equal to one unit of account. Denote by $E^{T,\pi}=(E_t^{T,\pi})_{t\in [0,T]}$ the price (value) process of the eigensecurity associated with the eigenfunction $\pi$ and paying off at time $T>0$. It is immediate that if $E^{T_1,\pi}$ and $E^{T_2,\pi}$ are two eigen-securities associated with the same eigenfunction but paying off at different times $T_1$ and $T_2$, then their price processes coincide on $T_1\wedge T_2$. We can thus consider an infinitely-lived eigen-security with the price process $E^\pi_t=e^{\lambda t}\pi(X_t)/\pi(X_0),$ where we now omit dependence on maturity in our notation.
It can be understood as the eigen-security paying off in the far distant future. For each fixed $T>0$, we also consider a trading strategy that invests at time zero in a $T$-maturity eigensecurity. At time $T$, it rolls over the proceeds into a new investment in the eigen-security with maturity at time $2T$. At time $2T$, it rolls over the proceeds into an eigensecurity with maturity at $3T$, etc.  It is easy to see that the value process of each of these trading strategies in eigen-securities coincides with the value process of infinitely-lived eigen-security $E^\pi_t=e^{\lambda t}\pi(X_t)/\pi(X_0)$ and is independent of $T$.
Eigensecurities were introduced by \citet{davydov_2003}. In that paper, given a pricing operator that was assumed to be a self-adjoint operator in an appropriately defined $L^2$ space of payoffs, the authors considered eigensecurities with not necessarily non-negative payoffs and used them as the basis for the eigenfunction expansion of other securities with $L^2$  payoffs via the Spectral Theorem. In the present paper we focus on eigen-securities with strictly positive payoffs, do not assume any $L^2$ structure, and use positive  eigensecurities to define eigen-measures.

The eigen-factorization \eqref{HS} can be re-written in the form
$S_t = (1/E_t^\pi)M_t^\pi,$
where the first factor discounts at the rate of return earned on holding the eigen-security of asymptotically long maturity, while the second factor is a martingale encoding further risk premia. Under ${\mathbb Q}^\pi$ the pricing operator reads:
\be
{\mathscr P}_{t}f(x):=e^{-\lambda t}\pi(x){\mathbb E}_{x}^{{\mathbb Q}^\pi}\left[ \frac{f(X_t)}{\pi(X_t)}\right]={\mathbb E}_{x}^{{\mathbb Q}^\pi}\left[ \frac{f(X_t)}{E_t^\pi}\right].
\eel{pqpi}
Thus, the eigen-security serves as the {\em numeraire asset} under the corresponding eigen-measure ${\mathbb Q}^\pi$.
We now consider a special sub-class of Markovian PKs.
\begin{definition}{\bf (Transition Independent Pricing Kernel)}\label{def_ti}
A PK is  said to be {\em transition independent} if there is a strictly positive, finite Borel function $\pi$ and a real constant $\lambda$ such that the PK takes the form:
\be
S_t=e^{-\lambda t}\frac{\pi(X_0)}{\pi(X_t)}.
\eel{pk}
\end{definition}
From the previous discussion it is immediate that $\pi$ is a positive eigenfunction of the pricing operator ${\mathscr P}_t$ with the eigenvalue $e^{-\lambda t}$. Furthermore, it is immediate that in an economy with a transition-independent pricing kernel the eigen-security $E_t^\pi$ serves as the numeraire asset under ${\mathbb P}$, i.e. $S_t=1/E_t^\pi$, $M_t^\pi=1$ and  ${\mathbb P}={\mathbb Q}^\pi$. We then also immediately have the following result.
\begin{proposition}
\label{eigen_growh_optimal}
In an economy with the transition-independent PK of the form \eqref{pk} the eigen-security $E^\pi$ associated with the same eigenfunction $\pi$ is growth optimal, i.e. it has the highest expected log return.
\end{proposition}
{\bf Proof.}
Consider the value process $V_t$ of an asset or of a self-financing portfolio normalized so that $V_0=1$ and such that $S_t V_t=V_t/E_t^\pi$ is a $\mathbb{P}$-martingale. By the martingale property,
$\mathbb{E}[V_t/E_t^\pi]=1.$
By Jensen's inequality,
$\mathbb{E}[\log(V_t/E_t^\pi)]\leq \log \mathbb{E}[V_t/E_t^\pi]=0,$
which immediately implies
$\mathbb{E}[\log(V_t)]\leq\mathbb{E}[\log(E_t^\pi)]$ for all $t$, i.e., the eigen-security has the highest expected log return. $\Box$\\

The model with a representative agent with the consumption process $C_t=C(X_t)$ taken to be a function of the Markov state and with the representative agent's von Neumann-Morgenstern utility function $U$ and constant discount rate $\lambda$ gives a canonical example of the transition independent pricing kernel \eqref{pk} $S_t=e^{-\lambda t}U^\prime(C(X_t))/U^\prime(C(X_0))$
with $\pi(x)=1/U^\prime(C(x))$.

The SAINTS model of \citet{constantinides_1992theory} is apparently the first instance in the literature of constructing a continuous-time Markovian asset pricing model by directly specifying the pricing kernel in the transition-independent form \eqref{pk}. \citet{constantinides_1992theory} takes $X$ to be a Markov process in ${\mathbb R}^{n+1}$ with $n$ coordinates specified to be Ornstein-Uhlenbeck diffusions and one coordinate specified to be a 1D Brownian motion. He then explicitly calculates the bond prices and derives the process for the short rate $r_t=r(X_t)$, which turns out to be quadratic in the $n$ OU factors.

\citet{rogers_1997} gives a far-reaching generalization of this approach to explicitly constructing asset pricing models by directly specifying the pricing kernel as a positive supermartingale (in fact, a potential) in the transition-independent form for some function $\pi$ under an auxiliary probability measure that can be identified with the eigen-measure ${\mathbb Q}^\pi$ (the approach of \citet{flesaker_1996positive} is closely related; see also \citet{jin_2001equilibrium} for connections with the Heath-Jarrow-Morton approach and supporting equilibrium models). The work of \citet{rogers_1997} is an important precursor to the work on eigen-factorization of Markovian pricing kernels and recovery that is the focus of the present paper. Rogers assumes the PK in the transition independent form under {\em some} probability measure, without identifying it with the physical measure. Thus, his assumption can be seen as the pre-cursor of the Hansen-Scheinkman eigen-factorization, with Ross' recovery as the special case when the probability measure under which the PK has the transition independent form identified with the physical measure.

 \citet{hansen_2009} start with a general positive semimartingale multiplicative functional pricing kernel and consider the factorization \eqref{HS} when the pricing kernel possesses a positive eigenfunction. Their pricing kernels are not, in general, supermartingales, and do not, in general, admit the factorization into the product of the discount factor $e^{-\int_0^t r(X_s)ds}$ with some non-negative short rate function $r(x)$ and a positive martingale, as in \citet{rogers_1997}. Their framework encompasses models with the short rate allowed to become negative, as well as models where no short rate exists (such models include both the situation where the riskless asset (the savings account) with the predictable price process of finite variation exists, but is not absolutely continuous \footnote{This situation arises  in models with agents having finite marginal utility from
consumption at the origin; see \citet{karatzas_1991} and \citet{schweizer_2000}.}, as well as models where no riskless asset with the predictable price process of finite variation exists), while Rogers' framework specifically focuses on models with the non-negative short rate.

In general, the PK may possess multiple positive eigenfunctions. Suppose $\pi_1$ and $\pi_2$ are two distinct positive eigenfunctions with the respective eigenvalues $\lambda_1$ and $\lambda_2$. Then the corresponding martingales $M_t^{\pi_i}$, $i=1,2$ can be used to define eigen-measures ${\mathbb Q}^{\pi_1}$ and ${\mathbb Q}^{\pi_2}$ locally equivalent to ${\mathbb P}$, and locally equivalent to each other:
\be
\left.{\mathbb Q}^{\pi_2}_x\right|_{{\mathscr F}_t}=e^{(\lambda_2-\lambda_1)t}\frac{\pi_2(X_t)\pi_1(x)}{\pi_1(X_t)\pi_2(x)}\left.{\mathbb Q}^{\pi_1}_x\right|_{{\mathscr F}_t}
\eel{P2P1}
for each $x\in E$.
The result of \citet{hansen_2009} (Proposition 7.2) is that, while there may be multiple positive eigenfunctions, there is at most one positive eigenfunction $\pi$ such that $X$ has certain stochastic stability (ergodicity) properties under the corresponding eigen-measure ${\mathbb Q}^\pi$ (we discuss this in detail in the next section). The factorization of the pricing kernel corresponding to this eigenfunction leading to the ergodic dynamics of $X$ under ${\mathbb Q}^\pi$  is extensively applied in \citet{hansen_2012}, \citet{hansen_2012pricing},  \citet{borovicka_2013exam} and \citet{hansen_2013}.


We now turn to the Recovery Theorem of \citet{ross_2011}.
In contrast to \citet{rogers_1997} and \citet{hansen_2009}, \citet{ross_2011} assumes that the PK has the transition independent form \eqref{pk} {\em directly} under the physical probability measure ${\mathbb P}$.
According to the discussion earlier in this section, Ross' assumption implies that $M_t^\pi=1$ in Hansen and Scheinkman's factorization \eqref{HS}, ${\mathbb P}={\mathbb Q}^\pi$ for some positive eigenfunction $\pi$ of the pricing kernel, and, hence, that the corresponding eigen-security $E^\pi$ is growth-optimal (by Proposition \ref{eigen_growh_optimal}).

Under the structural assumption of transition independence, when $X$ is a discrete time, finite state irreducible Markov chain, Ross shows that if the state prices are known, then there exists a unique physical probability measure compatible with these state prices and such that the PK is in the form \eqref{pk}, and it can be explicitly recovered from this knowledge of state prices. Ross' proof of existence and uniqueness relies on the Perron-Frobenius theorem for irreducible non-negative matrices.

In more detail, Ross' recovery problem is to recover the physical transition probabilities of $X$ from the given state prices.
Assuming the AD state-price measures ${\mathscr P}_t(x,B)$ are given, under Ross' assumption that the pricing kernel is in the transition independent form \eqref{pk}, so that $\pi$ is a positive eigenfunction of the pricing operators, as long as the positive eigenfunction is unique (up to an overall constant multiplicative factor), Ross' recovery succeeds, and
 the physical transition operators of $X$ are recovered via equating the physical transition operator $P_t$ of $X$ under the physical measure with the transition operator $Q^\pi_t$ of $X$ under the eigen-measure ${\mathbb Q}^\pi$ given by \eqref{Ppi}.
In Ross' setting of Markov chains with finite state spaces, irreducibility of the chain is a crucial assumption that fixes uniqueness of the positive eigenvector via the Perron-Frobenius theorem for irreducible non-negative matrices.
While there is a  notion of irreducibility for BRPs (see Appendix \ref{relation_recur} in the e-companion), it is insufficient to fix uniqueness for general state spaces.
We thus need to give new sufficient conditions that fix uniqueness for general Markov processes. Fortunately, we are able to prove uniqueness of a positive eigenfunction such that $X$ is recurrent under the associated eigen-measure ${\mathbb Q}^\pi$. We call this eigenfunction, if it exists, {\em recurrent}. Recall that an irreducible finite-state Markov chain is recurrent. Thus, in fact, Ross' recovery theorem for Markov chains already implicitly assumes recurrence. As we show in the next Section, recurrence serves as the sufficient condition for uniqueness in continuous-time Markovian models with general state spaces.

\section{Uniqueness of a Recurrent Eigenfunction and its Economic Implications}
\label{unique_r_eigen}

Here we work with the definition of recurrence of a BRP in \citet{getoor_1980} that follows \citet{azema_1966} and \citet{azema_1969}.
Let $X$ be a BRP with the state space $E$ with the Borel sigma-algebra ${\mathscr E}$.
For a Borel set $B\in {\mathscr E}$,
we define the {\em occupation time} of $B$ by
$\eta_B:=\int_0^\infty 1_B(X_s)ds$ and
define the {\em Green's} or {\em potential measure} of $X$ (in order to interchange the expectation and integration with respect to time, we use  the fact  that a Markov process is progressively measurable (cf. Lemma A.1.13 in \citet{chen_2011}) and Tonelli's theorem):
$$
R(x,B):={\mathbb E}_x[\eta_B]=\int_0^\infty P_t(x,B)dt,
$$
where $P_t(x,B)$ is the transition probability (i.e. probability for $X_t$ to be in $B$ at time $t$ if started at $x$ at time zero).
The potential measure is interpreted as the expected time the process $X$ spends in the set $B$ during its lifetime when started from $x\in E$.
It can also be defined on the larger sigma-algebra ${\mathscr E}^*$ of universally measurable subsets of $E$ (i.e. subsets measurable with respect to all complete probability measures on $E$).
We now give the definition of a recurrent BRP following Proposition 2.4 in \citet{getoor_1980} (cf. \citet{blumenthal_1968} p.89 or \citet{sharpe_1988} p.60).
\begin{definition}
\label{recurrent_getoor} ({\bf Recurrence of a Borel right process})
Assume that $E$ has at least two points. A Borel right process $X$ is said to be {\em recurrent}, if for each $B\in {\mathscr E}^*$
$R(x,B)=0$ or $R(x,B)=\infty$
for all $x\in E$.
\end{definition}
On average during its lifetime a recurrent BRP is expected to spend either a zero amount of time or an infinite amount of time  in  every universally measurable subset of the state space $E$.
It spends an infinite amount of time on average in all ``large enough" sets. Appendix \ref{relation_recur} in the e-companion gives useful sufficient conditions to verify recurrence in the sense of Definition \ref{recurrent_getoor}, as well as discusses relationships with other definitions of recurrence for Markov processes and, in particular, for diffusions.
The following theorem is the key result of this paper.
\begin{theorem}{\bf (Uniqueness of a Recurrent Eigenfunction)}\label{unique_eigen}
Let $X$ be a Borel right process, $S$ a positive semimartingale multiplicative functional of $X$ satisfying Assumption 3.1, and $({\mathscr P}_t)_{t\geq 0}$ a family of operators acting on Borel functions by Eq.\eqref{pp}.
Then there exists at most one  positive finite Borel function $\pi_R(x)$ (up to a multiplicative constant) such that $\pi_R$ is a positive eigenfunction of each ${\mathscr P}_t$ (i.e. Eq.\eqref{eigen} holds for some real $\lambda$ for all $t>0$ and $x\in E$) and
$X$ is recurrent in the sense of Definition \ref{recurrent_getoor} under ${\mathbb Q}^{\pi_R}$.
\end{theorem}
{\bf Proof.}
We prove by contradiction. Suppose the pricing operators have two recurrent eigenfunctions $\pi_i$ with eigenvalues $\lambda_i$, $i=1,2$.
We first assume that $\lambda_1\not=\lambda_2$. Without loss of generality, we assume that $\lambda_2<\lambda_1$.
${\mathbb Q}_x^{\pi_i}$ are equivalent to each other on each ${\mathscr F}_t$ and are related by Eq.\eqref{P2P1}. We denote by $Q^{\pi_i}_t$ the transition operators of $X$ under ${\mathbb Q}^{\pi_i}$ and by $R^{\pi_i}(x,\cdot)$ their corresponding potential measures.
Since we assumed that $X$ is recurrent under both ${\mathbb Q}_x^{\pi_t}$, $R^{\pi_i}(x,B)=0$ or $R^{\pi_i}(x,B)=\infty$ for each Borel set $B$ and all $x\in E$.

Consider Borel sets $B_{n}= \{x\in E: \pi_1(x) \geq 1/n\quad\text{and}\quad \pi_2(x)\leq n\},$ $n=1,2,\ldots$. Since $\pi_i$ is a strictly positive finite Borel function, $0<\pi_i(x)<\infty$ for all $x\in E$ and, hence, $B_n\nearrow E$.
Since $X$ is assumed to be conservative under ${\mathbb Q}$, it is also conservative under ${\mathbb Q}^{\pi_i}$, and $R^{\pi_i}(x,E)=\infty$ for all $x\in E$. Since for each $n$, $R^{\pi_i}(x,B_n)=0$ or $R^{\pi_i}(x,B_n)=\infty$ for all $x\in E$, there exists $N$ such that $R^{\pi_i}(x,B_n)=\infty$ for all $x\in E$ and $n\geq N$.
On the other hand, for the set $B_{N}$ we can write:
$$
R^{\pi_2}(x,B_{N})=\int_0^\infty {\mathbb E}_x^{{\mathbb Q}^{\pi_2}}[{\bf 1}_{B_{N}}(X_t)]dt
=\int_0^\infty e^{(\lambda_2-\lambda_1)t} {\mathbb E}_x^{{\mathbb Q}^{\pi_1}}\left[\frac{\pi_2(X_t)\pi_1(x)}{\pi_1(X_t)\pi_2(x)}{\bf 1}_{B_{N}}(X_t)\right]dt
$$
$$
\leq N^2\frac{\pi_1(x)}{\pi_2(x)}\int_0^t e^{(\lambda_2-\lambda_1)t}
Q^{\pi_1}_t (x,B_{N})dt
\leq N^2\frac{\pi_1(x)}{\pi_2(x)}\int_0^t e^{(\lambda_2-\lambda_1)t} dt<\infty,
$$
where we used Eq.\eqref{P2P1} and the fact that $\pi_1(x)\geq 1/N$ and $\pi_2(x)\leq N$ on $B_{N}$.
Thus, we have a contradiction and we cannot have two positive eigenfunctions $\pi_1$ and $\pi_2$ with eigenvalues $\lambda_2<\lambda_1$ such that $X$ is recurrent under both probabilities ${\mathbb P}^1$ and ${\mathbb P}^2$.

We next assume that $\lambda_1=\lambda_2=:\lambda$. Denote $f(x):=\pi_1(x)/\pi_2(x)$.  From Eq.\eqref{P2P1} we get
\be
{\mathbb E}_x^{{\mathbb Q}^{\pi_2}}[f(X_t)]=f(x).
\eel{invariant_function}
Thus, $f(x)$ is an {\em invariant function} of the transition semigroup $(Q^{\pi_2}_t)_{t\geq 0}$ of $X$ under ${\mathbb Q}^{\pi_2}$.
By Proposition 2.4 of \citet{getoor_1980} (also \citet{blumenthal_1968} p.89 or \citet{sharpe_1988} p.60), each excessive function of a recurrent process is constant on $E$. The invariant function is excessive. Thus, the invariant function $f(x)$
is constant.
Hence, $\pi_2(x)$ is a constant multiple of $\pi_1(x)$ and ${\mathbb Q}^{\pi_1}$ and ${\mathbb Q}^{\pi_2}$ coincide. $\Box$\\

Theorem 3.1 plays the role of the uniqueness part of the Perron-Frobenius theorem for Feynman-Kac-type operators \eqref{pp}, with the irreducibility of the Markov chain replaced with the assumption that $X$ is recurrent in the sense of Definition \ref{recurrent_getoor} under  ${\mathbb Q}^\pi$.
We call $\pi_R$ defined in Theorem \ref{unique_eigen} a {\em recurrent eigenfunction} and the corresponding ${\mathbb Q}^{\pi_R}$ a {\em recurrent eigen-measure}.


We now come back to the Hansen-Scheinkman eigen-factorization \eqref{HS}. We call a Hansen-Scheinkman eigen-factorization of the pricing kernel {\em recurrent} if a BRP $X$ is recurrent in the sense of Definition \ref{recurrent_getoor} under ${\mathbb Q}^\pi$.
By Theorem \ref{unique_eigen} we have the following result.
\begin{theorem} ({\bf Uniqueness of a Recurrent Hansen-Scheinkman Eigen-Factorization})\label{unique_HS}
If all uncertainty in the economy is generated by a BRP $X$ and if the pricing kernel is a positive semimartingale multiplicative functional of $X$ satisfying Assumption 3.1, then the PK admits at most one recurrent Hansen-Scheinkman factorization \eqref{HS}.
\end{theorem}
This result is close to Proposition 7.2 in \citet{hansen_2009}, but is distinct from it, as our stability assumptions are different.
Proposition 7.2 in \citet{hansen_2009} establishes that there exists at most one positive eigenfunction such that
under the eigen-measure ${\mathbb Q}^\pi$ $X$ has a stationary probability distribution $\hat{\varsigma}$ (cf. Assumption 7.2 of \citet{hansen_2009}),
the discretely sampled skeleton process $X_{\Delta j}$ is $\hat{\varsigma}$-irreducible for some $\Delta>0$  (cf. Assumption 7.3 of \citet{hansen_2009}), and $X$ is Harris recurrent (cf. Assumption 7.4 of \citet{hansen_2009}).
Here our assumption is recurrence of the BRP $X$ in the sense of Definition \ref{recurrent_getoor} under ${\mathbb Q}^\pi$. Our proof is also different from the proof of Proposition 7.2 in Appendix B of \citet{hansen_2009}. It relies only on recurrence.
It establishes that, under our assumptions, the ratio of any two positive eigenfunctions corresponding to the same eigenvalue is constant {\em everywhere on} $E$, which immediately follows from the fact that excessive functions of a recurrent right process are constant. 
While the stability assumptions made in \citet{hansen_2009} are natural in their context of analyzing long-term risk (see also \citet{linetsky_2014long}), recurrence in the sense of Definition \ref{recurrent_getoor} is already sufficient to fix uniqueness of the eigen-factorization.

We can now give the counterpart of \citet{ross_2011} Theorem 1 (Ross' Recovery Theorem) in the setting of BRPs.
\begin{theorem}\label{unique_theorem_1}({\bf Ross' Recovery Theorem for Recurrent BRPs})
If all uncertainty in the economy is generated by a recurrent Borel right process $X$, if there is no arbitrage, and if the Arrow-Debreu state prices ${\mathscr P}_t(x,dy)$ are generated by a transition independent pricing kernel of the form in Definition \ref{def_ti}, then there exists a unique solution to the problem of recovering the physical probabilities of X and the pricing kernel $S$ in the transition independent form \eqref{pk} with the positive eigenfunction $\pi$ and eigenvalue $\lambda$ from the knowledge of the Arrow-Debreu state prices.
\end{theorem}
{\bf Proof.}
It is assumed that $X$ is a recurrent BRP, and the PK is in the form \eqref{pk} for some $\pi$ and $\lambda$ to be determined. At the same time,
by Theorem \ref{unique_HS} the PK admits at most one  Hansen-Scheinkman factorization \eqref{HS} such that $X$ is recurrent under ${\mathbb Q}^\pi$. Thus, the transition independent PK in the form \eqref{pk} is identified with its recurrent Hansen-Scheinkman factorization \eqref{HS} with $M_t^\pi=1$ and, thus, the physical measure is identified with the recurrent eigen-measure under these assumptions, $\mathbb{P}=\mathbb{Q}^\pi$, and $\pi$ and $\lambda$ are identified with the unique recurrent eigenfunction $\pi_R$ and the corresponding eigenvalue $\lambda_R$. The latter exist by assumptions of recurrence of $X$ and transition independence of $S$ explicitly made in the theorem. The pricing kernel  then has the form \eqref{pk} with $\pi_R$ and $\lambda_R$. The transition probabilities of $X$ under ${\mathbb P}$ can then be uniquely recovered by equalizing the transition operator of $X$ under $\mathbb{P}$ with the transition operator $Q^{\pi_R}$ given by  \eqref{Ppi} under $\mathbb{Q}^{\pi_R}$, the latter already expressed in terms of the Arrow-Debreu state-price measures ${\mathscr P}_t(x,dy)$.   $\Box$\\
\\
Theorem \ref{unique_theorem_1} establishes uniqueness of Ross recovery under the assumption that $X$ is recurrent under the physical measure ${\mathbb P}$ in the sense of Definition \ref{recurrent_getoor}, in addition to Ross' assumption of transition independence of the PK. The recurrence assumption on the BRP $X$ is a sufficient replacement of the irreducibility assumption on the finite state Markov chain in the context of BRPs \footnote{We note that the result of \citet{carr_2012} on Ross' recovery for one-dimensional diffusions on bounded intervals with reflected boundary conditions is consistent with Theorem \ref{unique_theorem_1}, as such a diffusion is recurrent. The result of \citet{walden_2013} is also consistent with Theorem \ref{unique_theorem_1}, as 1D diffusions under his smoothness assumptions, along with his assumption of non-attracting boundaries, are also recurrent.}.

The above discussion reveals economic meaning of the transition independence assumption, necessarily fixing growth optimality of the  eigen-security $E^\pi$ under the physical measure. Under the recurrence assumption, the eigensecurity is associated with the recurrent eigenfunction. Furthermore, under stronger ergodicity assumptions \citet{linetsky_2014long} prove that the recurrent eigen-security can be identified with the long bond (pure discount bond of asymptotically long maturity), thus fixing the growth optimality of the long bond. This result extends a closely related result of \citet{hansen_2009} (see also \citet{martin_2013} and
\citet{borovicka_2014mis} for related results).

We stress that recurrence is a sufficient, but not necessary, condition to fix uniqueness. Suppose we relax the recurrence assumption in Theorem \ref{unique_theorem_1} and the state prices ${\mathscr P}_t(x,dy)$ are given. Ross' recovery problem is to separately identify the transition probabilities $P_t(x,dy)$ of $X$ under the physical measure and the eigenfunction $\pi(x)$  and eigenvalue $\lambda$ defining the pricing kernel in the transition independent form. {\em In general}, there may be multiple solutions $(Q_t^\pi(x,dy),\pi,\lambda)$ to this problem if the pricing operator defined by the state prices ${\mathscr P}_t(x,dy)$ admits multiple positive eigenfunctions  (here $Q_t^\pi(x,dy)$ is the transition function of $X$ under the eigen-measure corresponding to $\pi$). {\em A priori} there is no way to fix one of the solutions and, thus, to choose one eigen-measure among ${\mathbb Q}^\pi$ to identify with {\em the} physical measure ${\mathbb P}$. Recurrence of a BRP $X$  is a sufficient condition that ensures uniqueness.  For example, it is possible that a particular specification of state prices is such that the corresponding pricing operator already possesses a unique positive eigenfunction. It may or may not be recurrent (an example of a unique positive eigenfunction that also turns out to be recurrent is provided by 1D diffusions on a bounded interval with reflection at both ends). In that special case no additional conditions are needed to fix uniqueness.
Thus, our Theorem \ref{unique_theorem_1} is not the most general form of Ross' recovery, but rather a formulation under conditions sufficient to ensure uniqueness of recovery in the general setting of BRPs.

We next turn to pricing securities with long maturities.
\begin{theorem}
\label{long_pricing} {\bf (Long-term Pricing)}
Suppose the pricing kernel admits a recurrent eigenfunction $\pi_R$ and let $\mathbb{Q}^{\pi_R}$ denote the associated recurrent eigen-measure. Suppose further there exists a probability measure $\varsigma$ on $E$, a Borel function $\beta(x)$ and some positive constants $\alpha$ and $t_0$, such that the following exponential ergodicity estimate holds for all bounded Borel functions $f(x)$:
\be
\left|\mathbb{E}_x^{\mathbb{Q}^{\pi_R}}\left[\frac{f(X_t)}{\pi_R(X_t)}\right]-c_f\right|\leq \beta(x)\|f\|_{\infty} e^{-\alpha t}
\eel{expo_ergo}
for all $t\geq t_0$ and each $x\in E$, where
\be
c_f:=\varsigma\left(\frac{f}{\pi_R}\right)=\int_E \frac{f(y)}{\pi_R(y)}\varsigma(dy).
\eel{cf}
Then the following long maturity pricing estimate holds for all $t\geq t_0$
\be
\left|\mathscr{P}_t f(x)-c_f e^{-\lambda_R t}\pi_R(x)\right| \leq \beta(x)\pi_R(x)\|f\|_{\infty} e^{-(\lambda_R+\alpha) t}
\eel{LTpricing}
for any bounded Borel payoff $f$ and each $x\in E$.
\end{theorem}
{\bf Proof.}
Since the recurrent eigenfunction and, hence, the recurrent eigen-measure are assumed to exist in formulation of the theorem, we
can write the pricing operator in the form Eq.\eqref{pqpi}. The long-term pricing estimate \eqref{LTpricing} then immediately follows from the exponential ergodicity estimate \eqref{expo_ergo}.  $\Box$\\
\\
The long-term pricing relationship reveals that when the recurrent eigenfunction exists, under the exponential ergodicity assumption current prices of payoffs occurring at long maturities depend on the current state   approximately as the eigenfunction $\pi_R(x)$ and decay in time approximately exponentially at the rate  $\lambda_R$. Only the overall constant factor $c_f$ depends on the payoff $f$ via \eqref{cf}.
The approximation error for securities with long but finite maturities decays exponentially at the rate $\alpha$ in the exponential ergodicity estimate as maturity increases.
In particular, the annualized yield on the zero-coupon bond with unit payoff $f=1$ and long maturity is approximated by the eigenvalue $\lambda_R$, while the bond's gross return process defined by $P(X_t,T-t)/P(X_0,T)$ (tracking gross return from time $0$ to time $t$ on the zero-coupon bond with maturity $T>>t$ in the distant future) is approximated by the value process of the recurrent eigensecurity $E_t^{\pi_R}=e^{\lambda t}\pi_R(X_t)/\pi_R(X_0)$.
In the context of discrete-time, finite-state Markov chains, this is explained in detail in \citet{martin_2013}. A close result in continuous Markovian environments appears in Proposition 7.1 in  \citet{hansen_2009}.
This result also leads to the identification of the long-term forward measure ${\mathbb L}$ with the recurrent eigen-measure ${\mathbb Q}^\pi$, as shown by \citet{linetsky_2014long}.

In the context of pricing square-integrable payoffs in 1D diffusion term structure models with the pricing operator possessing a spectral expansion, the long-term pricing asymptotics of Theorem 3.4 is the first term in the spectral expansion corresponding to the principal eigenfunction (cf. \citet{davydov_2003}, \citet{gorovoi_2004}, \citet{linetsky_2006pricing} and Section 5.2). In that case the exponent $\alpha$ is equal to the spectral gap between the lowest eigenvalue and the next eigenvalue. With empirically realistic rates of mean reversion in such term structure models, the long term asymptotics may already closely approximate prices of payoffs at twenty to thirty year maturities.
In contrast to these works on eigenfunction expansions for diffusions, Theorem \ref{long_pricing} formulates the long-term pricing asymptotics for a much wider class of Markovian models and payoffs, not assuming any underlying $L^2$ structure.

{\bf Remark 3.1.} In this paper our focus is on the factorization of the pricing semigroup. However, Theorem 3.1 is a general result establishing uniqueness of recurrent eigenfunctions of Feynman-Kac type semigroups associated with positive multiplicative semimartingale functionals of BRPs. As such, the result applies to all of the semigroups studied in \citet{hansen_2009}, including the semigroups modeling economic growth.

\section{Riskless Rate and Risk-Neutral Probabilities}
\label{riskless_rate}

We now turn our attention to PKs that admit a risk-neutral factorization with a short rate.
\begin{assumption}\label{RN_factor}{\bf (Risk-Neutral Factorization)}
In addition to Assumption \ref{finite_exp}, in this Section we assume that the PK admits a factorization in the form
\be
S_t=e^{-\int_0^t r(X_s)ds}M_t,
\eel{pkfactorization}
where $r(x)$ is a Borel function such that $\int_0^t |r(X_s)|ds<\infty$ a.s. for each finite $t>0$ (not assumed to be non-negative in order to accommodate affine models with OU-type factors) and $M_t$ is a positive martingale with $M_0=1$.
\end{assumption}
The function $r(x)$ defines the short rate process $r_t=r(X_t)$, and the martingale $M$ can be used to change over to the risk-neutral measure $\left.{\mathbb Q}\right|_{{\mathscr F}_t}=M_t \left.{\mathbb P}\right|_{{\mathscr F}_t},$ under which the pricing operator reads
\be
{\mathscr P}_tf(x) = {\mathbb E}_x^{\mathbb Q}[e^{-\int_0^t r(X_s)ds} f(X_t)].
\eel{Prn}
Note that existence of a short rate is an  additional assumption and is not automatic. Assumption \ref{finite_exp}, together with the assumption that $S$ is a special semimartingale, imply that there exists a predictable additive functional of finite variation $A$ and a positive multiplicative local martingale functional $M$ such that $S_t=e^{-A_t}M_t$. The factorization \eqref{pkfactorization} requires the local martingale $M$ to be a true martingale and the positive additive functional $A$ to be absolutely continuous. In the rest of this paper we make Assumption \ref{RN_factor} in addition to Assumption \ref{finite_exp}.  Uniqueness of the risk-neutral factorization for positive semimartingale PKs is proved by \citet{doberlein_2001savings}.

The next result shows how to explicitly recover the short rate from a PK given in the form \eqref{HS}. This result is a counterpart of the result in \citet{rogers_1997}, but does not assume that the short rate is non-negative and the PK is a supermartingale potential.
First recall that a pair of Borel functions $h$ and $f$ on $E$ with $\int_0^t |f(X_s)|ds<\infty$ a.s. for all $t>0$  is said to belong to the domain of an {\em extended generator} of $X$ if the process $h(X_t)-\int_0^t f(X_s)ds$ is a local martingale, and one writes $f(x)={\cal G}h(x)$ (cf. \citet{palmowski_2002} or \citet{ethier_2005} for details).
\begin{theorem}{\bf (Short Rate and Risk-Neutral Measure)}\label{recover_short}
Suppose the PK satisfies Assumptions \ref{finite_exp} and \ref{RN_factor} and admits a representation in the form \eqref{HS}. Denote $h(x)=1/\pi(x)$.
Then the short rate function $r(x)$ is recovered via
\be
r(x)=\lambda-{\cal G}^\pi h(x)/h(x),
\eel{shortrate}
where ${\cal G}^\pi$ is the extended generator of the transition semigroup $(Q^\pi_t)_{t\geq 0}$ of $X$ under  ${\mathbb Q}^\pi$ given by \eqref{Ppi},
 and the transition semigroup of $X$ under ${\mathbb Q}$ reads
\be
Q_t f(x) = {\mathbb E}_x^{\mathbb Q}[f(X_t)]=\pi(x){\mathbb E}_x^{\mathbb Q^\pi}\left[e^{\int_0^t r(X_s)ds-\lambda t}\frac{f(X_t)}{\pi(X_t)}\right]
\eel{Q}
with the extended generator ${\cal G}^Q f={\cal G}^\pi f+\Gamma^\pi(h,f)/h,$
where $\Gamma^\pi(h,f)$ is the {\em carr\'{e} du champ} (squared field) operator of $X$ under ${\mathbb Q}^\pi$ (see \citet{palmowski_2002} for a detailed account and references):
$\Gamma^\pi(h,f)(x)={\cal G}^\pi (fh)(x)-f(x){\cal G}^\pi h(x) - h(x){\cal G}^\pi f(x).$
\end{theorem}
The proof is given in Appendix \ref{proof_short} in the e-companion. When the short rate exists, Theorem \ref{recover_short} allows us to explicitly recover it from any given positive eigenfunction and the corresponding eigenvalue of the PK and, thus, also explicitly recover the risk neutral probabilities from the knowledge of the eigenvalue and eigenfunction.

In financial economics one typically starts with the pricing kernel derived in a structural general equilibrium model and then extracts the implied short rate.
A reduced-form approach more commonly followed in the financial engineering literature, as well as in the financial markets practice, is to start directly with a class of Markovian risk-neutral laws ${\mathbb Q}$ and a class of explicitly specified short rate functions $r(x)$ and calibrate the pricing operators under ${\mathbb Q}$ to market-observed security prices.
To link these two approaches, in the next section we investigate the question of existence of a recurrent positive eigenfunction in a  given short rate model.
In this approach we start with a BRP $X$ with the {\em given} risk-neutral probability law $({\mathbb Q}_x)_{x\in E}$ and a {\em given} short rate function $r(x)$ on $E$. The pricing operators are then defined by \eqref{Prn}. The question we are faced with is whether the pricing operators
possess a positive eigenfunction $\pi(x)$ satisfying \eqref{eigen}
for some real $\lambda$ and all $t>0$ and $x\in E$ and
such that under the locally equivalent probability measure $\left.{\mathbb Q}^\pi\right|_{{\mathscr F}_t}=\tilde{M}_t^\pi \left.{\mathbb Q}\right|_{{\mathscr F}_t}$
defined by the positive ${\mathbb Q}$-martingale
\be
\tilde{M}_t^\pi=e^{-\int_0^t r(X_s)ds+\lambda t} \pi(X_t)/\pi(X_0)
\eel{M}
the process $X$ is recurrent.

We remark that, while $M_t^\pi$ is a ${\mathbb P}$-martingale  changing measure from ${\mathbb P}$ to ${\mathbb Q}^\pi$, $\tilde{M}_t^\pi$ is a ${\mathbb Q}$-martingale changing measure from ${\mathbb Q}$ to ${\mathbb Q}^\pi$. It is then immediate that $M_t^\pi/\tilde{M}_t^\pi=e^{\int_0^t r(X_s)ds} S_t=M_t$ is the ${\mathbb P}$-martingale in the risk-neutral factorization \eqref{pkfactorization} changing measure from ${\mathbb P}$ to ${\mathbb Q}$.

We further remark that the risk-neutral stochastic discount factor under the risk-neutral measure ${\mathbb Q}$ itself possesses an eigen-factorization
$$e^{-\int_0^t r(X_s)ds}=\tilde{M}_t^\pi e^{-\lambda t}\pi(X_0)/\pi(X_t)$$
that shares the same eigenfunction $\pi$ with the eigen-factorization of the stochastic discount factor $S$ \eqref{HS} under ${\mathbb P}$, but with the ${\mathbb Q}$-martingale factor $\tilde{M}^\pi$ in place of the ${\mathbb P}$-martingale $M^\pi$ appearing in the eigen-factorization of $S_t$ under the physical measure.

\section{Existence of a Recurrent Eigenfunction in Short Rate Models}
\label{exist}



Building on the work of \citet{nummelin_1984} and  \citet{meyn_2005}, \citet{hansen_2009}   develop sufficient conditions for existence of a positive eigenfunction for the semigroup of Markovian pricing operators such that $X$ satisfies their ergodicity assumptions under ${\mathbb Q}^\pi$ (their Section 9 and Appendix D). Their sufficient conditions are formulated at the level of general operator semigroups. While working in less generality, here we give some explicit and easy to verify sufficient conditions for existence of a recurrent positive eigenfunction for several classes of short rate models important in applications.

\subsection{The $L^2(E,m)$ Approach}
\label{exist_1_1}


In this section we assume that, under the {\em given} risk-neutral measure, $X$ is a conservative {\em Hunt process} on a locally compact separable metric space $E$. This entails making additional assumptions that the Borel right process $X$ on $E$ has sample paths with left limits and is {\em quasi-left continuous} (no jumps at predictable stopping times, and fixed times in particular).
In this section we further assume that the given short rate function $r(x)$ is non-negative.
Let $X^r$ denote $X$ killed  at the rate $r$ (i.e. the process is killed (sent to an isolated cemetery state) at the first time the positive continuous additive functional $\int_0^t  r(X_s)ds$ exceeds an independent unit-mean exponential random variable).  It is a {\em Borel standard process} (see Definition A.1.23 and Theorem A.1.24 in \citet{chen_2011}) since it shares the sample path with the Hunt process $X$ prior to the killing time. The pricing semigroup $({\mathscr P}_t)_{t\geq 0}$ is then identified with the transition semigroup of the Borel standard process $X^r$.

In this section we assume that there is a positive sigma-finite reference measure $m$ with full support on $E$ such that
$X^r$ has a {\em dual} with respect to $m$. That is, there is a strong Markov process $\hat{X}^r$ on $E$ with semigroup $(\hat{\mathscr P}_t)_{t\geq 0}$
such that for any $t>0$ and non-negative functions
$f$ and $g$:
\be
\int_E f(x){\mathscr P}_tg(x)m(dx)=\int_E g(x)\hat{\mathscr P}_tf(x)m(dx).
\ee
We further make the following assumptions.
\begin{assumption}
\label{L2assumption}
(i) There exists a family of continuous and strictly positive functions $p(t,\cdot,\cdot)$ on $E\times E$ such that for any $(t,x)\in(0,\infty)\times E$ and any non-negative function $f$ on $E$,
\be
{\mathscr P}_tf(x)=\int_E p(t,x,y)f(y)m(dy),\qquad \hat{\mathscr P}_tf(x)=\int_E p(t,y,x)f(y)m(dy).
\ee
(ii) The density satisfies:
\be
\label{exist_2_assumption1}
\int_E\int_E p^2(t,x,y) m(dx)m(dy)<\infty,\quad \forall t>0.
\ee
(iii) There exists some $T>0$ such that
\be
\label{exist_2_assumption2}
\sup_{x\in E} \int_E p^2(t,x,y) m(dy)<\infty,\quad \sup_{x\in E}\int_E p^2(t,y,x) m(dy)<\infty,\quad \forall t\geq T.\\
\ee
\end{assumption}

Under these assumptions, we have the following results.
\begin{theorem}
\label{exist_theorem_1}
Suppose Assumption \ref{L2assumption} is satisfied.
(i) The process $X$ is $m$-irreducible and satisfies the absolute continuity assumption \ref{ac_condition} with respect to the reference measure $m$ (see Appendix \ref{relation_recur} in the e-companion).\\
(ii) The pricing operator ${\mathscr P}_t$ and the dual operator $\hat{{\mathscr P}}_t$ possess unique positive, continuous, bounded eigenfunctions ${\pi}(x)$ and $\hat{\pi}(x)$ belonging to $L^2(E,m)$:
\be
\label{exist_2_eigenfunction}
\int_E p(t,x,y) \pi(y)m(dy)=e^{-\lambda t}\pi(x),\quad
\int_E p(t,y,x) \hat{\pi}(y)m(dy)=e^{-\lambda t}\hat{\pi}(x)
\ee
with some $\lambda\geq 0$ for each $t>0$ and every $x\in E$.\\
(iii)
Let $C:=\int_E \pi(x)\hat{\pi}(x)m(dx)$.
There exist constants $c,\alpha>0$ and $T^\prime>0$ such that for $t\geq T^\prime$ we have the estimate for the density
\be
|C e^{\lambda t}p(t,x,y)- \pi(x)\hat{\pi}(y)|\leq ce^{-\alpha t}, \quad x,y\in E.
\eel{limiting_dist}
(iv) The process $X$ is recurrent in the sense of Definition \ref{recurrent_getoor} and in the sense of Definition \ref{recurrent_tweedie} under ${\mathbb Q}^\pi$ defined by the martingale \eqref{M}.
Moreover, $X$ is {\em positive recurrent} under ${\mathbb Q}^\pi$ with the stationary distribution
$\varsigma(dx)=C^{-1}\pi(x) \hat{\pi}(x)m(dx).$\\
(v) If in addition $m$ is a finite measure, i.e. $m(E)<\infty$, then for any payoff $f\in L^2(E,m)$ we have the following long maturity estimate for all $t\geq T'$:
\be
\left|\mathscr{P}_tf(x)-c_fe^{-\lambda t}\pi(x)\right|\leq K\|f\|_{L^2(E,m)}e^{-(\lambda+\alpha)t}
\eel{long_pricing_hunt}
with $c_f=\int_E (f(x)/\pi(x))\varsigma(dx)=C^{-1}\int_E f(x)\hat{\pi}(x)m(dx)$,
and $K$ is a constant independent of $f$, $x$ and $t$.
\end{theorem}
The proof is given in Appendix \ref{proof_zhang} in the e-companion and is based on \citet{zhang_2013} which, in turn, is based on Jentzsch's theorem, a counterpart of the Perron-Frobenius theorem for integral operators in $L^2$ spaces.

Theorem \ref{exist_theorem_1} immediately yields existence of a recurrent Hansen-Scheinkman factorization for Hunt processes under Assumption \ref{L2assumption}.
Part v in
Theorem \ref{exist_theorem_1} is a consequence of Theorem \ref{long_pricing} with $\beta(x)=1/\pi(x)$. In Theorem \ref{long_pricing} we directly assume existence of a recurrent eigenfunction, and then prove long term pricing under the additional assumption of exponential ergodicity, while in this section we give sufficient conditions on the process and the pricing kernel such that the recurrent eigenfunction is guaranteed to exist and exponential ergodicity holds under the recurrent eigen-measure.


In the special case when ${\mathscr P}_t=\hat{\mathscr P}_t$, i.e. the pricing operators are symmetric with respect to the measure $m$, $({\mathscr P}_t)_{t\geq 0}$ can be interpreted as the transition semigroup of a {\em symmetric Markov process} $X^r$ killed at the rate $r$ (cf. \citet{chen_2011} and \citet{fukushima_2010}). In particular, essentially all one-dimensional diffusions are symmetric Markov processes with the speed measure $m$ acting as the symmetry measure. We come back to this in Section \ref{exist_2}.

In the symmetric case, Assumption \ref{L2assumption} (ii) implies that for each $t>0$ the pricing operator ${\mathscr P}_t$ is a  symmetric Hilbert-Schmidt operator in $L^2(E,m)$. It further implies that the pricing semigroup is {\em trace class} (cf. \citet{davies_2007} Section 7.2) and, hence, for each $t>0$  the pricing operator ${\mathscr P}_t$ has a purely discrete spectrum $\{e^{-\lambda_n t},n=1,2,\ldots\}$ with $0\leq \lambda_1\leq\lambda_2\leq \ldots$ repeated according to the eigenvalue multiplicity  with the finite trace
${\rm tr}{\mathscr P}_t=\int_E p(t,x,x)m(dx)=\sum_{n=1}^\infty e^{-\lambda_n t}<\infty.$
Using the symmetry of the density, $p(t,x,y)=p(t,y,x)$, and the Chapman-Kolmogorov equation, Assumption \ref{L2assumption} (iii) reduces to the assumption that there exists a constant $T>0$ such that
\be
\sup_{x\in E}p(t,x,x)<\infty\quad \text{for all}\quad t\geq T.
\eel{l2assumption_equ}

\subsection{One-Dimensional Diffusions}
\label{exist_2}

In this section we consider the case where $X$ is a conservative 1D diffusions on an interval $I$ with
with left and right end-points $l$ and $r$ that can be either finite or infinite, $-\infty\leq l < r \leq \infty$. If an endpoint is finite, we assume that it is either inaccessible (either a {\em natural} or an {\em entrance} boundary) or a {\em regular boundary} specified as {\em instantaneously reflecting} (see Chapter II of \citet{borodin_2002} for Feller's classification of boundaries and other details about 1D diffusions). If a boundary is inaccessible, then it is not included in the state space ($I$ is open at an inaccessible boundary). If a boundary is instantaneously reflecting, it is included in the state space ($I$ is closed at a reflecting boundary), since the process can reach the boundary from the interior.
In particular, we exclude from consideration exit and regular killing boundaries since $X$ is assumed to be conservative, and here we also exclude absorbing boundaries since it is {\em a priori} clear that an absorbing boundary remains absorbing under any locally equivalent measure transformation, thus ensuring that $X$ is not recurrent under any locally equivalent measure.

Every conservative 1D diffusion has two basic characteristics: the {\em speed measure} $m$ and the {\em scale function} $S$ \footnote{In the words of Feller, a one-dimensional diffusion process $X$ travels
according to a road map indicated by its scale $S$ and with speed indicated by its
speed measure $m$. See \citet{borodin_2002} and \citet{karlin_1981} for further details.
}.
The speed measure $m$ is a measure on the Borel sigma-algebra of $I$ such that $0<m((a,b))<\infty$ for any $l<a<b<r$. For every $t>0$ and $x\in I$ the transition measure of $X$ is absolutely continuous with respect to $m$, i.e. $P_t(x,A)=\int_I p(t,x,y)m(dy)$. The density $p(t,x,y)$ may be taken to be positive and jointly continuous in $x,y,t$ and symmetric in $x,y$, i.e. $p(t,x,y)=p(t,y,x)$ (this was first proved by \citet{mckean_1956}). Due to this symmetry, a 1D diffusion is a symmetric Markov process. Moreover, $X$ is $m$-irreducible with respect to the speed measure $m$, and satisfies the absolute continuity Assumption \ref{ac_condition} (see Appendix \ref{relation_recur} in the e-companion for definitions of irreducibility) due to existence of a positive continuous density $p(t,x,y)$ with respect to the speed measure $m$.
Thus, the results of Section \ref{exist_1_1} can be applied to 1D diffusions.

However, for 1D diffusions we are able to formulate more general and easier to verify sufficient conditions under some additional assumptions based on the Sturm-Liouville theory. To this end, we consider here the special case in which the speed measure is absolutely continuous with respect to the Lebesgue measure on $I$, i.e. $m(dx)=m(x)dx$, and the scale function is $S(x)=\int^x s(y)dy$, where the speed and scale densities $m(x)$ and $s(x)$ are continuous and positive. Moreover, we also assume that $s(x)$ is continuously differentiable. In that case the infinitesimal generator of the transition semigroup of the 1D diffusion acting on $C_b(I)$ (continuous bounded functions on $I$) can be written in the form
\be
{\cal G}f(x)=\frac{1}{2}\sigma^2(x)f''(x)+\mu(x)f'(x),
\ee
where $\sigma(x)$ and $\mu(x)$ are volatility and drift functions related to the speed and scale densities by:
\be
m(x)=\frac{2}{\sigma^2(x)s(x)},\quad s(x)=e^{-\int^x \frac{2\mu(y)dy}{\sigma^2(y)}}.
\ee
The domain of the generator of the transition semigroup on $C_b(I)$ is $D({\cal G})=\{f,{\cal G}f\in C_b(I), \text{b.c.}\}$, where the boundary conditions (b.c.) can be found in \citet{borodin_2002}.

We also assume in this section that there is a non-negative short rate $r(x)\geq 0$ on $I$.
The infinitesimal generator of the pricing semigroup $({\mathscr P}_t)_{t\geq 0}$ on $C_b(I)$ can be written in the following  {\em formally} self-adjoint form when acting on $C^2_c(l,r)$ functions (twice-differentiable functions with compact supports in $(l,r)$):
\be
{\cal A}f(x)={\cal G}f(x)-r(x)f(x)=\frac{1}{m(x)}\left(\frac{f^\prime(x)}{s(x)}\right)^\prime-r(x)f(x).
\eel{SLop}
Furthermore, the pricing semigroup in $C_b(I)$ restricted to $C_b(I)\cap L^2(I,m)$ extends uniquely to a strongly continuous semigroup of self-adjoint contractions on $L^2(I,m)$. Its infinitesimal generator is an unbounded self-adjoint, non-positive operator on $L^2(I,m)$ with domain given in \citet{mckean_1956}, p.526 and \citet{langer_1990}, p.15, or \citet{linetsky_2008}, p.232.  With some abuse of notation we use the same notation for the pricing semigroup and its generator when considered in different function spaces $C_b(I)$ and $L^2(I,m)$. We observe that the generator of the pricing semigroup can be interpreted as the {\em Sturm-Liouville (SL) operator}.
The theory of SL operators can be brought to bear to establish a spectral classification of 1D diffusions (with killing) and, hence, pricing semigroups. This classification is given in \citet{linetsky_2004} and \citet{linetsky_2008}, Sections 3.4-3.6 based  on {\em Sturm's theory of oscillations of solutions of the SL ordinary differential equation} (for general background on the SL theory see \citet{amerin_2005} and references therein):
\be
-{\cal A}f(x)=\lambda f(x),\quad x\in (l,r),
\eel{SLeq}
where ${\cal A}$ is the second-order differential operator \eqref{SLop}.

\begin{theorem}
\label{1d_sufficient}
Under the assumptions on $X$ and $r$ in this section, if the set of eigenvalues of the SL operator $-{\cal A}$ in $L^2(I,m)$ is non-empty, then:\\
(i) the lowest eigenvalue $\lambda_0$ (principal eigenvalue) is non-negative and the corresponding eigenfunction (principal eigenfunction) $\pi_0(x)$ is strictly positive on $I$. Moreover, $\pi_0(x)$ is also an eigenfunction of the pricing operator ${\mathscr P}_t$ with the eigenvalue $e^{-\lambda_0 t}\leq 1$.\\
(ii) Under ${\mathbb Q}^{\pi_0}$ $X$ is a {\em positively recurrent} 1D diffusion on $I$ with the generator ${\cal G}^{\pi_0}$ acting on $C^2_c(l,r)$ by
\be
{\cal G}^{\pi_0} f(x)=\frac{1}{2}\sigma^2(x)f^{\prime \prime}(x)+\mu^{\pi_0}(x)f^\prime(x),\quad
\mu^{\pi_0}(x)=\mu(x)+\sigma^2(x)\frac{\pi_0^\prime(x)}{\pi_0(x)},
\eel{1dgpi0}
the stationary distribution $\varsigma(dx)=\pi_0^2(x)m(x)dx$ (where $\pi_0$ is normalized so that $\int_I \pi_0^2(x)m(x)dx=1$) and scale density $s(x)/\pi_0^2(x)$.\\
(iii) If in addition there is a spectral gap $\alpha>0$ between $\lambda_0$ and the bottom of the spectrum of $-{\cal A}$ in $L^2(I,m)$ above $\lambda_0$, and the density of the pricing kernel $\mathscr{P}_t$ satisfies \eqref{exist_2_assumption1} for some $t=T>0$ (and, hence, for all $t\geq T$), then for any payoff function $f\in L^2(I,m)$ the long-maturity pricing estimate holds for $t\geq 2T$
\be
\left|\mathscr{P}_t f(x)-c_f e^{-\lambda_0 t}\pi_0(x)\right|\leq Kp(2T,x,x)\|f\|_{L^2(I,m)}e^{-(\lambda+\alpha)t},
\ee
where $c_f=\int_I f(y)\pi_0(y)m(dy)$, and $K$ is a constant independent of $f$, $x$ and $t$.
\end{theorem}
The proof is given Appendix \ref{proof_1D} in the e-companion. Theorem \ref{1d_sufficient} reduces the question of existence of a recurrent positive eigenfunction in the 1D diffusion setting to the question of existence of an $L^2(I,m)$-eigenfunction of the corresponding SL equation.
Appendix \ref{proof_1D} in the e-companion gives sufficient conditions in terms of the asymptotic properties of $\sigma(x)$, $\mu(x)$ and $r(x)$ near the end-points of the interval $I$. Part iii gives sufficient conditions for the long term pricing formula for 1D diffusions that are significantly less stringent than the assumptions in Section \ref{exist_1_1} (in particular here we require that the pricing semigroup is only {\em eventually}  Hilbert-Schmidt for $t\geq T$ for some $T>0$, as opposed to {\em immediately} Hilbert-Schmidt (for all $t>0$), as in Section \ref{exist_1_1}).

\subsection{Multi-Dimensional Diffusions in ${\mathbb R}^d$}
\label{exist_3}

In this section we assume that the (risk-neutral) process $X$ is a diffusion in  $E={\mathbb R}^d$ in the sense that
$X$ is constructed as a unique solution of the Stroock-Varadhan martingale problem for a second-order differential operator on ${\mathbb R}^d$ under conditions in Theorem 10.4 on page 32 of \citet{pinsky_1995}.
Namely, let $a_{ij}(x)=a_{ji}(x)$, $i,j=1,\ldots,d$, and $b_i(x)$, $i=1,...,d$, be measurable locally bounded functions on ${\mathbb R}^d$, and assume that $a_{ij}(x)$ are continuous and the matrix $(a_{ij}(x))$ is locally elliptic, i.e. $\sum_{i,j=1}^d a_{ij}(x)v_i v_j>0$ for all $x\in {\mathbb R}^d$ and all $v\in {\mathbb R}^d-\{0\}$. Let ${\cal G}$ be the differential operator of the form:
\be
{\cal G}=\frac{1}{2}\displaystyle{\sum_{i,j=1}^d}a_{ij}\frac{\partial^2}{\partial x_i\partial x_j}+\displaystyle{\sum_{i=1}^d}b_{i}\frac{\partial}{\partial x_i}.
\ee
Then, by Theorem 10.4 on page 32 of \citet{pinsky_1995}, there exists at most one solution to the Stroock-Varadhan martingale problem for ${\cal G}$ on ${\mathbb R}^d$. The existence is ensured by an additional {\em non-explosion} condition (10.4) on page 33 of \citet{pinsky_1995}. Under this condition, the unique solution $({\mathbb Q}_x)_{x\in {\mathbb R}^d}$ to the martingale problem is such that the process $X$ with continuous paths in ${\mathbb R}^d$  is conservative and possesses the strong Markov property. Furthermore, ${\mathbb Q}_x(X_t\in B)$ possesses a density $p(t,x,y)$ with respect to the Lebesgue measure and, for any finite stopping time $\tau$,
${\mathbb Q}_x(X_{t+\tau}\in B|{\mathscr F}_\tau)=\int_B p(t,X_\tau,y)dy$ for all $t>0$ and Borel sets $B$ in ${\mathbb R}^d$ (\citet{pinsky_1995}, p.36 Theorem 10.6). In this section, we assume in addition that $a_{ij}$ and $b_i$ are locally H\"{o}lder continuous on $\mathbb{R}^d$.

Examples of diffusion processes are provided by solutions of stochastic differential equations of the form
\be
dX_t=b(X_t)dt + \sigma(X_t)dB_t
\eel{SDE}
with some measurable volatility matrix $\sigma(x)$ such that $\sigma(x)\sigma^\top(x)=a(x)$. In particular, if $b$ and $\sigma$ satisfy the sufficient conditions for the existence and uniqueness of a non-exploding weak solution in ${\mathbb R}^d$, these yield examples of diffusions we work with in this section.

We further assume there is a short rate $r_t=r(X_t)$, where $r(x)$ is a given
short rate function.

\begin{theorem}
\label{exist_3_main}
Assume that, in addition to the standing assumptions about $X$ in this section, $a$, $b$ and $r$ are all locally H\"{o}lder continuous on $\mathbb{R}^d$, and $r$ is such that there exists an exhausting domain sequence $(D_n)_{n\geq 1}$ in ${\mathbb R}^d$ such that $D_n\subset\subset D_{n+1}$ with $D_n \nearrow {\mathbb R}^d$ and
\be
r_n\rightarrow\infty\quad, \text{where}\quad  r_n:=\inf \{r(x):x\in {\mathbb R}^d-D_n\}.
\eel{rlimit}
Then the pricing operator ${\mathscr P}_t$ has a strictly positive eigenfunction $\pi$ having continuous and H\"{o}lder continuous
second derivatives and such that under ${\mathbb Q}^\pi=({\mathbb Q}^\pi_x)_{x\in \mathbb{R}^d}$ the diffusion process $X$ is recurrent both in the sense of Definition \ref{recurrent_getoor} and in the sense of Definition \ref{recurrent_tweedie} with $\psi={\rm Leb}$, the Lebesgue measure on $\mathbb{R}^d$.
Furthermore, $({\mathbb Q}_x^\pi)_{x\in \mathbb{R}^d}$ solves the martingale problem for the operator
\be
{\cal G}^\pi =\frac{1}{2}\displaystyle{\sum_{i,j=1}^d}a_{ij}\frac{\partial^2}{\partial x_i\partial x_j}+\displaystyle{\sum_{i=1}^d}b^\pi_{i}\frac{\partial}{\partial x_i}\quad \text{with}\quad b^\pi_{i}(x)=b_i(x)+\sum_{j=1}^d a_{ij}(x)\frac{\nabla_j\pi(x)}{\pi(x)}.
\eel{bP}
\end{theorem}
The proof is based on the theory of second-order elliptic operators and associated diffusion processes presented in \citet{pinsky_1995}. It is given in Appendix \ref{proof_mD} in the e-companion.
In particular, if the diffusion $X$ is a unique non-exploding solution of the SDE \eqref{SDE}, we can also obtain the drift \eqref{bP} by directly applying Girsanov's theorem. The process
$B^{\pi,i}_t =B_t^i-\int_0^t \lambda_s^i ds,$
where
\be
\lambda_t^i = \sum_{j=1}^d \sigma_{ji}(x)\frac{\nabla_j\pi(x)}{\pi(x)},
\eel{market_price}
is a standard Brownian motion under ${\mathbb Q}^\pi$ (here $B$ is a Brownian motion under ${\mathbb Q}$). The process $\lambda_t^i$ plays the role of the market price of risk in this diffusion model.

Theorem \ref{exist_3_main} can be generalized by replacing ${\mathbb R}^d$ with an open domain $D\subseteq {\mathbb R}^d$ and requiring that $X$ is a diffusion on $D$ constructed by solving the martingale problem on $D$ and such that $X$ does not reach the boundary of $D$ when started from  any $x\in D$ (the non-explosion condition in ${\mathbb R}^d$ is replaced with the requirement that $X$ does not reach the boundary of $D$, i.e. it does not exit the open domain $D$).
All conditions on the coefficients $a,b,r$ are formulated by replacing ${\mathbb R}^d$ with the open domain $D$.

We note that the sufficient condition \eqref{rlimit} is satisfied in quadratic term structure models (QTSM) under a non-degeneracy condition, where the short rate is quadratic in the state variable and  $r(x)\rightarrow \infty$ as $\|x\|\rightarrow \infty$, but is not satisfied in affine term structure models (ATSM).  Fortunately, we are able to prove existence in ATSM under appropriate assumptions directly by using their special properties (see Appendix \ref{affine_appendix} in the e-companion). We also remark that the sufficient conditions for existence of a recurrent positive eigenfunction we are able to give for 1D diffusions in Section \ref{exist_2} are much sharper than the sufficient conditions in ${\mathbb R}^d$ in this section. In particular, for 1D diffusions we do not need to assume that the short rate tends to infinity at the boundary.

\section{Examples of Short Rate Models}
\label{examples5}



\subsection{Recurrent Eigenfunctions in One-Dimensional Diffusion Models}

In this section we treat some popular 1D short rate diffusion models. Here $X$ is a 1D diffusion with the specified risk-neutral dynamics, and the short rate function $r(x)$ is specified. For 1D diffusions we are able to give a detailed treatment of positive eigenfunctions. We start with the Sturm-Liouville ODE \eqref{SLeq} associated with the generator ${\cal A}$ (5.5) of the pricing semigroup ${\mathscr P}$. Each positive eigenfunction $\pi(x)$ of the SL equation \eqref{SLeq} gives rise to a positive {\em local} martingale $\tilde{M}^\pi$ in the form \eqref{M}.
If the eigenfunction of the generator is also an eigenfunction of the semigroup, then $\tilde{M}^\pi$ is a positive martingale, and we can define a new probability measure. Among all these probability measures, there is at most one such that $X$ is recurrent under it. We will also see that there are parametric families of additional positive eigenfunctions such that $X$ is transient under the associated probability measures, and such models exhibit unstable economic behavior, where the riskless rate either asymptotically runs off to infinity (asymptotic hyperinflation) or to zero (zero lower bound trap).

\subsubsection{CIR Model}
\label{CIRin5}

Consider a CIR SDE under the risk-neutral measure ${\mathbb Q}$
\be dX_t=(a+b X_t)dt+\sigma\sqrt{X_t}dB_t^{\mathbb Q}
\eel{SDE1_meanCIR}
with $a>0$, $b\in {\mathbb R}$, $\sigma>0$. The short rate is $r_t=X_t$. A detailed discussion of the CIR model is given in Appendix \ref{appendix_cir} in the e-companion, where Assumption \ref{L2assumption} is explicitly verified. Thus, the CIR model possesses a recurrent eigenfunction.

It is instructive to give a more detailed treatment of positive eigenfunctions in the CIR model.
For simplicity in what follows we assume that the Feller condition holds, i.e. $2a\geq\sigma^2$, so that the process stays strictly positive (does not hit zero).
We start with the CIR Sturm-Liouville ODE (here $b=-\kappa$)
\be
\frac{1}{2}\sigma^2 x\pi^{\prime\prime}+(a- \kappa x)\pi^\prime - x\pi=-\lambda \pi
\eel{ODE_meanCIR}
with $\sigma>0$, $a>0$, $\kappa=-b\in {\mathbb R}$, and $\lambda\in {\mathbb R}$.
It can be reduced to the confluent hypergeometric equation, and its solutions can be expressed in terms of Kummer and Tricomi confluent hypergeometric functions.
We first characterize all solutions (not necessarily positive). Denote $\gamma:=\sqrt{\kappa^2+2\sigma^2}$.
\begin{proposition}
Define $\alpha:=(\lambda-\lambda_0)/\gamma,$
where $\lambda_0=a(\gamma-\kappa)/\sigma^2$ is the principal eigenvalue of the CIR pricing semigroup in $L^2((0,\infty),m)$ (see Appendix \ref{appendix_cir} in the e-companion).
(i) If $\alpha$ is not a non-positive integer, i.e. $\alpha\not=-n$, $n=0,1,\ldots$ (which means that $\lambda$ is not an $L^2((0,\infty),m)$-eigenvalue of the generator ${\mathcal A}$ of the pricing semigroup, i.e. $\lambda\not=\lambda_n=\gamma n +\lambda_0$), the two linearly independent solutions of Eq.\eqref{ODE_meanCIR} are:
\begin{equation}
\psi_{\lambda}(x)=e^{\frac{\kappa-\gamma}{\sigma^2}x}M(\alpha,\beta,\frac{2\gamma x}{\sigma^2}),\quad
\phi_{\lambda}(x)=e^{\frac{\kappa-\gamma}{\sigma^2}x}U(\alpha,\beta,\frac{2\gamma x}{\sigma^2}),
\end{equation}
where $M(a,b,z)$ and $U(a,b,z)$ are Kummer and Tricomi confluent hypergeometric functions.

(ii) If $\alpha$ is a non-positive integer, then the Kummer and Tricomi confluent hypergeometric functions $M$ and $U$ reduce to the generalized Laguerre polynomials and the two solutions $\psi_\lambda(x)$ and $\phi_\lambda(x)$ become linearly dependent and both reduce to the $L^2((0,\infty),m)$-eigenfunction $\varphi_n(x)$ of ${\mathcal A}$ given in Appendix \ref{appendix_cir} in the e-companion. Then one solution can be taken
to be $\varphi_n(x)$, while the other linearly independent solution differs in different cases (a complete study of the confluent hypergeometric equation can be found in \citet{slater_1960}, p.5-8 and is omitted here to save space).
\label{prop_1_meanCIR}
\end{proposition}

Using these linearly independent solutions, we can construct {\em local} martingales $\tilde{M}$ in the form \eqref{M}
with $\pi(x)=C_1 \psi_\lambda(x)+C_2\phi_\lambda(x)$
parameterized by $\lambda, C_1, C_2\in {\mathbb R}$. It is immediate that these processes are local martingales by the application of It\^{o}'s formula and the fact that $\pi$ is a solution of the Sturm-Liouville equation. The application of It\^{o}'s formula is justified since $\pi$ is $C^2((0,\infty))$, and $X$ stays strictly positive when Feller's condition is satisfied.
We now establish which of these local martingales are positive martingales. The proof of the following Theorem is in Appendix \ref{appendix_cir} in the e-companion.
\begin{theorem}
\label{positive_CIR}
$\tilde{M}^\pi$ is a positive martingale if and only if $\pi(x)=C_1\psi_\lambda(x)$ with $C_1>0$ and $\lambda\leq \lambda_0$ (correspondingly, $\alpha\geq 0$).
\end{theorem}

This result explicitly characterizes all positive eigenfunctions of the CIR pricing semigroup and, hence, all positive ${\mathbb Q}$-martingales in the form \eqref{M} in the CIR model. They are parameterized by a single parameter $\alpha\geq 0$ (equivalently, $\lambda\leq \lambda_0$). We will now look at the behavior of $X$ under the corresponding probability measures associated with these martingales.

First consider the solution
$\pi_0(x)=e^{-\frac{(\gamma-\kappa)}{\sigma^2}x}$
corresponding to $\alpha=0$ (hence, $\lambda=\lambda_0$). The state variable $X$ follows a mean-reverting CIR diffusion with the higher mean-reversion rate $\gamma=\sqrt{\kappa^2+2\sigma}>\kappa$ under $\mathbb{Q}^{\pi_0}$:
\be
dX_t=(a-\gamma X_t)dt+\sigma\sqrt{X_t}dB_t^{\mathbb{Q}^{\pi_0} },
\ee
where $B_t^{\mathbb{Q}^{\pi_0}}=B_t^{\mathbb{Q}}+\frac{\gamma-\kappa}{\sigma}\int_0^t \sqrt{X_s}ds$ is a standard Brownian motion under $\mathbb{Q}^{\pi_0}$.
Thus, $\pi_0$ is identified with the unique recurrent eigenfunction $\pi_R$. We also note that when $b=-\kappa<0$, the sufficient condition in part (v) of Theorem \ref{exist_theorem_1} is satisfied, and in this case the principal eigenvalue $\lambda_0$ gives the asymptotic yield of the zero-coupon bond in the CIR model $R_\infty=\lim_{t\rightarrow \infty}-t^{-1}\ln P(x,t)=\lambda_0=a(\gamma-\kappa)/\sigma^2.$
Moreover, it can be directly verified that this result also holds for the case $b=-\kappa>0$, even though the sufficient condition in (v) of   Theorem \ref{exist_theorem_1} is not satisfied in this case.

We now consider positive eigenfunctions of the pricing semigroup corresponding to $\alpha>0$ ($\lambda<\lambda_0$):
$$
\pi_\alpha(x)=\psi_{\lambda}(x)=e^{\frac{\kappa-\gamma}{\sigma^2}x}M(\alpha,\beta,\frac{2\gamma x}{\sigma^2}).
$$
It is easy to check that these solutions do {\em not} belong to $L^2((0,\infty),m)$ (it is easy to verify directly using the asymptotic properties of the Kummer function $M$ that they fail to be square-integrable with the CIR speed density $m$).
Under $\mathbb{Q}^{\pi_\alpha}$, $X_t$ solves the SDE with drift in \eqref{1dgpi0}:
\be
dX_t
=\left(a-\gamma X_t+\frac{2\alpha\gamma}{\beta}\frac{M(\alpha+1,\beta+1,2\gamma X_t/\sigma^2)}{M(\alpha,\beta,2\gamma X_t/\sigma^2)}X_t\right)dt+\sigma\sqrt{X_t}dB_t^{\mathbb{Q}^{\pi_\alpha} }.
\ee
This calculation uses the fact that $M'(\alpha,\beta,x)=(\alpha/\beta)M(\alpha+1,\beta+1,x)$. Using the asymptotic behavior of confluent hypergeometric functions, we obtain the following drift asymptotics:
$(2\alpha\gamma x/\beta)M(\alpha+1,\beta+1,2\gamma x/\sigma^2)/M(\alpha,\beta,2\gamma x/\sigma^2)\rightarrow 2\gamma x$ as $x \rightarrow +\infty$.
Thus, the drift asymptotically behaves as $a+\gamma x$ for large $x$ and the process is not mean-reverting under these probability measures.
Applying the test on page 234 of \citet{karlin_1981}, we verify that $+\infty$ is an attracting natural boundary in this case. Under the influence of the drift the process is asymptotically attracted to infinity and is transient.
This is an economically unstable behavior, resulting in asymptotically increasing interest rates.
The recurrence assumption rules out this behavior. To further illustrate this behavior, consider a special case with $\alpha=\beta$. In this case,
$\lambda=-\frac{a}{\sigma^2}(\kappa+\gamma)<0$  and
the confluent hypergeometric function reduces to the exponential function, $\pi(x)=\psi_\lambda(x)=e^{\frac{\kappa+\gamma}{\sigma^2}x}$.
Due to the fact that the CIR diffusion is an affine process, Eq.\eqref{martingale_meanCIR} can be verified directly by computing the expectation to verify that
$\tilde{M}^{\pi_\beta}_t = \exp\left(-\int_0^t X_u du +\frac{\gamma+\kappa}{\sigma^2}(X_t-X_0)-\frac{a}{\sigma^2}(\gamma+\kappa)t\right)$
is a martingale.
Applying Girsanov's theorem, we immediately see that under the corresponding measure change
$$
dX_t=(a+\gamma X_t)dt+\sigma\sqrt{X_t}dB_t^{\mathbb{Q}^{\pi_\beta} }.
$$
The coefficient in front of the linear term in the drift is now $+\gamma$, instead of $-\gamma$ in the mean-reverting case, and
the short rate is asymptotically attracted to infinity under this measure.

\subsubsection{Square-Root Model with Absorbing Boundary at Zero}

Consider the SDE $dX_t=b X_t dt+\sigma\sqrt{X_t}dB_t^{\mathbb Q}$
with $b\in {\mathbb R}$ and $\sigma>0$ and short rate $r_t=X_t$. It is a degenerate case of the CIR model with $a=0$. When started from $x=0$, $X_t=0$ for all $t\geq 0$ is a unique solution. When started from $x>0$, the solution hits zero by any positive time $t$ with positive probability, and $X_t=0$ for all $t\geq T_0$, where $T_0$ is the first hitting time of zero.
Thus, zero is an absorbing boundary.   
Clearly, there is no recurrent eigenfunction in this model, since a process with an absorbing boundary cannot be transformed into a recurrent process by a locally equivalent measure transformation.
To analyze all positive eigenfunctions in this case, consider the ODE:
\be
\frac{1}{2}\sigma^2 x \pi^{\prime\prime}+bx\pi^\prime - x\pi=-\lambda \pi.
\eel{ODE_absorbingCIR}
Considering it at $x=0$, we necessarily get that $\lambda=0$ for any positive solution with $\pi(0)>0$. Thus, zero eigenvalue is the only one consistent with an eigenfunction positive at $x=0$.
Substituting $\lambda=0$ back into Eq.\eqref{ODE_absorbingCIR}, the ODE reduced to $\frac{1}{2}\sigma^2 \pi_{xx}+b \pi_x - \pi=0$. It has two positive solutions
$\pi_{\pm}(x)=\exp\left\{(-b\pm\sqrt{b^2+2\sigma^2})x/\sigma^2\right\}.$
Using the affine property of the pricing semigroup, it is easy to directly verify that both of these solutions are invariant functions of the pricing semigroup, ${\mathscr P}_t \pi_{\pm}(x)=\pi_{\pm}(x)$. Thus, $e^{-\int_0^t X_sds}\pi_{\pm}(X_t)/\pi_{\pm}(x)$ are positive martingales.
Under $\mathbb{Q}^{\pi_\pm}$ the process $X_t$ solves the SDE:
\be
d X_t=\pm\sqrt{b^2+2\sigma^2}X_t dt+\sigma\sqrt{X_t}d B_t^{\mathbb{Q}^{\pi_\pm}}.
\ee
The process is still affine and has an absorbing boundary at zero under both of these measures.
When $X$ gets absorbed at zero, the interest rate is zero for all times after absorption.

We observe that the solutions $\pi_{\pm}(x)$ do not belong to $L^2([0,\infty),m)$ (the speed density is the same as CIR's with $a=0$). On the other hand, we observe that
$\pi(x)=xe^{\frac{-b-\gamma}{\sigma^2}x}$ is an eigenfunction with the eigenvalue $\lambda_0=\gamma$ and is square-integrable with $m$. However, it is not strictly positive, since it vanishes in the absorbing state $x=0$.
Thus, the corresponding martingale vanishes for all times $t\geq T_0$ and, hence, does not define an equivalent measure transformation. We thus conclude that there is no recurrent eigenfunction in this model. While we are able to construct two transition independent pricing kernels, $X$ gets absorbed at zero under both of them (an almost sure zero lower bound trap). \citet{linetsky_2014long} further show that the long bond exists and is identified with $\pi_-(X_t)/\pi_-(X_0)$ in this model.

\subsubsection{Vasicek Model}
\label{OUin5}

Consider an OU process under ${\mathbb Q}$ solving the SDE
$$dX_t=\kappa(\theta- X_t)dt+\sigma dB_t^{\mathbb Q}$$
with $\theta,\kappa\in {\mathbb R}$, $\kappa\neq0$, $\sigma>0$ and $r_t=X_t$.

First consider the case with $\kappa>0$. It is easy to check that $\pi_0(x)=e^{-x/\kappa}$ is the eigenfunction of the SL equation with the eigenvalue $\lambda_0=\theta-\sigma^2/(2\kappa^2)$. It is easy to check that it is square-integrable with the speed density $m(x)$ when $\kappa>0$ and is, thus, a positive $L^2({\mathbb R},m)$ eigenfunction of the pricing operator ${\mathscr P}_t$ with the eigenvalue $e^{-\lambda_0 t}$.
Girsanov's theorem immediately implies that $X$ solves
\be
dX_t=(\kappa\theta-\frac{\sigma^2}{\kappa}-\kappa X_t)dt+\sigma dB_t^{\mathbb{Q}^{\pi_0} }
\ee
under $\mathbb{Q}^{\pi_0}$.
Thus, $X$ is again a positively recurrent, mean-reverting OU process, but with the lower drift. Thus, $\pi_0$ is the unique recurrent eigenfunction.

Next consider the case with $\kappa<0$. It is easy to check that $\pi_0(x)=  e^{\frac{\kappa}{\sigma^2}x^2+(\frac{1}{\kappa}-\frac{2\kappa\theta}{\sigma^2})x}$ is the eigenfunction with the eigenvalue $\lambda=\theta-\kappa-\sigma^2/(2\kappa^2)$. It is easy to check that it is square-integrable with the speed density $m(x)$ when $\kappa<0$
and is, thus, a positive $L^2({\mathbb R},m)$ eigenfunction of the pricing operator ${\mathscr P}_t$. Girsanov's theorem immediately implies that $X$ solves
\be
dX_t=(\frac{\sigma^2}{\kappa}-\kappa\theta+\kappa X_t)dt+\sigma dB_t^{\mathbb{Q}^{\pi_0} }
\ee
under $\mathbb{Q}^{\pi_0}$. $X$ is a positive recurrent, mean-reverting OU process under ${\mathbb Q}^{\pi_0}$.
Thus, if the interest rate follows an OU process with mean-repelling drift under the risk-neutral measure, there still exists a unique recurrent eigenfunction. This is similar to what we have observed in the CIR model with $b>0$.

A complete analysis of all (non-recurrent) positive eigenfunctions in the OU model is given in Appendix \ref{appendix_ou} in the e-companion, where further examples of 1D diffusions are also given.

\subsection{Multi-dimensional Diffusion Models}
\subsubsection{Affine Models}
\label{exist_affine}



Affine diffusions are the most widely used class of term structure models in continuous-time finance due to their tractability (\citet{vasicek_1977equilibrium}, \citet{cox_1985_2}, \citet{duffie_1996}, \citet{duffie_2000}, \citet{dai_2000}, \citet{duffie_2003}). General multi-dimensional affine diffusion models (cf. \citet{filipovic_2009}) do not fall under our sufficient conditions in Section \ref{exist_3}. Nevertheless, we are able to give a detailed treatment of recurrent eigenfunctions in affine diffusion models due to their special properties. If the affine model is non-degenerate,  all eigenvalues of the slope matrix in the drift have strictly negative real parts, and an additional explicit sufficient condition on the parameters is verified, then there exists a unique recurrent eigenfunction and it has the exponential affine form $$\pi_R(x)=e^{u^\top x}.$$
Under the corresponding recurrent eigen-measure ${\mathbb Q}^{\pi_R}$, $X$ is a mean-reverting affine diffusion. Full details are given in Appendix \ref{affine_appendix} in the e-companion, where a sufficient condition for existence, an easy to implement numerical procedure to compute the vector $u$, an explicit expression for the eigenvalue $\lambda$, and an explicit expression for the affine diffusion $X$ under the recurrent eigen-measure ${\mathbb Q}^{\pi_R}$ are given.

\subsubsection{Quadratic Models}

Quadratic term structure models (\citet{beaglehole_1992}, \citet{constantinides_1992theory}, \citet{rogers_1997}, \citet{ahn_2002}, and \citet{chen_2004}) provide another important example of multi-dimensional diffusion models where the recurrent eigenfunction can be explicitly determined.
Suppose $X$ is a $d$-dimensional OU process and the short rate function is quadratic:
\be
r(x)=\gamma+\delta^\top x +x^\top \Phi x,
\ee
where the constant $\gamma$, vector $\delta$ and symmetric positive semi-definite matrix $\Phi$ are taken to be such that the short rate is non-negative for all $x\in {\mathbb R}^d$.
If $\Phi$ is strictly positive definite, then the QTSM satisfies the sufficient conditions in Theorem \ref{exist_3_main} (since $r(x)\rightarrow \infty$ as $\|x\|\rightarrow \infty$), and there is a unique recurrent eigenfunction. If $\Phi$ is merely positive semi-definite, this case is generally outside of the sufficient condition in Theorem \ref{exist_3_main}, but there may still be a unique recurrent eigenfunction.  A sufficient condition is given in Appendix \ref{appendix_quadratic} in the e-companion.
In either case, the recurrent eigenfunction takes the exponential quadratic form:
$$\pi_R(x)=e^{-u^\top x - x^\top V x}.$$
Appendix \ref{appendix_quadratic} in the e-companion provides a numerical procedure to determine
the vector $u$ and the symmetric positive semi-definite matrix $V$, and gives explicit expressions for the eigenvalue $\lambda_R$ and drift of the $d$-dimensional OU process $X$ under ${\mathbb Q}^{\pi_R}$ in terms of $u$ and $V$.

\subsection{CIR Model with Jumps}

Consider a CIR model with jumps under ${\mathbb Q}$
\be
dX_t=(a-\kappa X_t)dt+\sigma\sqrt{X_t}dB^{\mathbb Q}_t+dJ_t,
\ee
where $J_t$ is a compound Poisson process with L{\'e}vy measure $m(d\xi)=\frac{\varpi}{\mu}e^{-\xi/\mu}d\xi$ with the jump arrival rate $\varpi>0$  and positive exponential jumps with mean size $\mu>0$ (cf. \citet{duffie_2001risk}, \citet{filipovic_2001}). We consider this special case for simplicity as it leads to completely explicit results. Recurrent eigenfunctions in general affine jump-diffusion models will be investigated in a future publication.
The short rate is $r_t=X_t$.
The model is  affine in the sense that for any $z\leq 0$
\be
\mathbb{E}_x^{\mathbb Q}[e^{-\int_0^t X_sds + zX_t}]=e^{\phi(t,z)+\psi(t,z)x},
\ee
where the functions $\phi(t,z)$ and $\psi(t,z)$ satisfy
\be
\label{riccati}
\begin{split}
& \partial_t\phi(t,z)=F(\psi(t,z)),\quad \partial_t \psi(t,z)=R(\psi(t,z)),\quad \psi(0,z)=z,\\
&F(z)=\partial_t\phi(t,z)|_{t=0}=a z+\frac{\varpi\mu z}{1-\mu z},\quad
R(z)=\partial_t\psi(t,z)|_{t=0}=\frac{1}{2}\sigma^2 z^2-\kappa z-1.
\end{split}
\ee
Similar to the affine diffusion case, we look for the exponential affine eigenfunction $\pi(x)=e^{-ux}$ such that
$$
\mathbb{E}_x^{\mathbb Q}[e^{-\int_0^t X_sds - u X_t}]=e^{-\lambda t-ux}
$$
for some $\lambda$.
The constant $u$ has to satisfy:
$\frac{1}{2}\sigma^2u^2+\kappa u-1=0$. Take the larger root  $u=(-\kappa+\sqrt{\kappa^2+2\sigma^2})/\sigma^2$.
Then the principal eigenvalue is equal to
$$
\lambda_{JCIR}=\lambda_{CIR} +\frac{\varpi\mu u}{1+\mu u},
$$
where $\lambda_{CIR}=a (\gamma-\kappa)/\sigma^2$ is the principal eigenvalue of the CIR model without jumps in Section \ref{CIRin5}.
We find that under $\mathbb{Q}^\pi$ the process $X$ is again  CIR with jumps:
\be
dX_t=(a-\gamma X_t)dt+\sigma\sqrt{X_t}dB_t^{{\mathbb Q}^\pi}+dJ_t^{{\mathbb Q}^\pi}
\eel{JCIRRR}
with the mean reversion rate $\gamma=\sqrt{\kappa^2+2\sigma^2}$ and a  compound Poisson process $J^{{\mathbb Q}^\pi}$ having the L{\'e}vy measure $m^{{\mathbb Q}^\pi}(d\xi)=\frac{\varpi}{\mu}e^{-\xi(1/\mu+u)}d\xi$ under $\mathbb{Q}^\pi$.
Thus, under the measure change the arrival rate of jumps and the mean of the exponential jump size distribution change to:
$$
\hat{\varpi}=\frac{\varpi}{1+u\mu},\quad \hat{\mu}=\frac{\mu}{1+u\mu}\quad \text{where}\quad u=\frac{-\kappa+\sqrt{\kappa^2+2\sigma^2}}{\sigma^2}.
$$
To complete the proof that $\pi$ is the recurrent eigenfunction, we need to show that $X$ is recurrent in the sense of Definition \ref{recurrent_getoor} under $\mathbb{Q}^\pi$. This is done in Appendix \ref{appendix_cir_jump} in the e-companion.

\section{Conclusion}
\label{conclusion}

This paper has developed the spectral theory of Markovian asset pricing models where the underlying economic uncertainty follows a Borel right process and the stochastic discount factor is a positive semimartingale multiplicative functional of $X$.
A key result is the
uniqueness theorem for a positive eigenfunction of the Markovian pricing operator such that $X$ is recurrent under the eigen-measure associated with this eigenfunction (recurrent eigenfunction).
An application of this result yields uniqueness of the \citet{hansen_2009} eigen-factorization of the Markovian stochastic discount factor into the factor that discounts future cash flows at the stochastic rate of return earned from holding a security with the payoff defined by the recurrent eigenfunction (eigen-security) and an additional positive martingale that changes the probability measure to the eigen-measure. As a corollary, under the assumption of transition independence of the stochastic discount factor that effectively sets the martingale factor to unity, this factorization yields an extension of the Recovery Theorem of \citet{ross_2011} from discrete time, finite state irreducible Markov chains to
recurrent Borel right processes by identifying the physical probability measure with the recurrent eigen-measure.
Under the exponential ergodicity assumption we further obtain the long-term asymptotics of the pricing operator that identifies the asymptotic yield with the recurrent eigenvalue and expresses state dependence of the  finite holding period gross return on a security with the asymptotically long maturity in terms of the recurrent eigenfunction.

When an asset pricing model is specified by given risk-neutral probabilities together with a given short rate function of the Markovian driver, we gave
sufficient conditions for existence of a recurrent eigenfunction and provided
explicit examples in a number of models important in finance, including a variety of 1D diffusion models, affine and quadratic multi-dimensional diffusion models, and an affine model with jumps.

From the macro-finance perspective, these results deepen our understanding of long-term risk. Theorem 3.4 shows that under additional ergodicity assumptions the recurrent eigen-security can be identified with the pure discount bond of asymptotically long maturity and, thus, leads to the long-term factorization of the stochastic discount factor in Markovian models (\citet{alvarez_2005using}, \citet{hansen_2009}, \citet{martin_2013}, \citet{borovicka_2014mis}, \citet{linetsky_2014long}).
We refer to \citet{borovicka_2014mis} and  \citet{linetsky_2014long} for further developments in this direction.

From the empirical finance perspective, the results in this paper help set the stage for empirical recovery under assumptions more general than Ross'.
First, results in this paper allow us to replace discrete-time, finite-state Markov chains with continuous-time Markov processes. As seen in \citet{audrino_2014empirical} (and also \citet{tran_2013}), discrete specifications raise some non-trivial implementation issues requiring careful regularization. As is often the case, continuous specifications may lead to more stable, already regularized estimation procedures. In particular, results in this paper open avenues for working with affine and quadratic diffusion and jump-diffusion specifications in place of discrete Markov chains.
Furthermore, results in this paper help open avenues for testing more general recoveries relaxing the assumption of transition independence, thus allowing for non-trivial martingale components in the long-term factorization.
This will generally require combining historical time series data on underlying asset returns and/or macroeconomic variables  with the current market prices of options. In related literature, an alternative approach to recovery has recently been put forward \citet{martin2014expected}, who derived a lower bound for the expected excess return of an equity index in terms of equity index options, based on the negative covariance condition. \citet{bakshia2015inquiry} derived a similar lower bound for the expected excess return on the long bond. \citet{schneider2015almost} extended the bounds to other moments and also considered upper bounds.

Finally,
from the financial engineering perspective, the results in this paper yield explicit pricing of long-lived assets in Markovian models. These long-term pricing results can be viewed as extensions of pricing results in the eigenfunction expansion literature (in particular, \citet{davydov_2003} and other references cited in the introduction) to more general classes of Markovian asset pricing models. The price of the increased generality is that in these models we can identify only the principal eigenfunction and, hence, have the asymptotic pricing result suitable for long maturity assets, while in the eigenfunction expansion literature the entire spectral expansion is displayed, allowing one to price assets of all maturities, but under more restrictive modeling assumptions. Linking back with the empirical discussion above, the long-term pricing asymptotics open up a possibility to empirically identify the eigenvalue $\lambda_R$ and the eigenfunction $\pi_R$ from observing time series of market prices of long-lived assets, such as long-term bonds, as was also explained in \citet{martin_2013} in the discrete-time Markov chain framework.

%
\appendix

\section{Borel Right Processes}
\label{prelim_borel_right}

We refer the reader to
 \citet{blumenthal_1968}, \citet{sharpe_1988} and \citet{chen_2011} for more details. Here we follow the presentation in  Appendix A of \citet{chen_2011}.
Recall that a continuous-time Markov process on a measurable space $(E,{\mathscr E})$ is a quadruplet\\ $(\Omega,{\mathscr F},(X_t)_{t\geq 0},({\mathbb P}_x)_{x\in E})$, where $(\Omega,{\mathscr F})$ is a measurable space, for each starting point $x\in E$  $(\Omega,{\mathscr F},(X_t)_{t\geq 0},{\mathbb P}_x)$ is a stochastic process with state space $(E,{\mathscr E})$ and continuous time parameter such that, for each $t\geq 0$ and $B\in {\mathscr E}$, ${\mathbb P}_x(X_t \in B)$ is ${\mathscr E}$-measurable as a function of $x\in E$, there exists an {\em admissible filtration} $({\mathscr F}_t)_{t\geq 0}$ such that the {\em Markov property} holds with respect to it, i.e.
${\mathbb P}_x(X_{s+t} \in B|{\mathscr F}_t)={\mathbb P}_{X_t}(X_{s} \in B),$ ${\mathbb P}_x$-a.s.,
and ${\mathbb P}_x(X_0=x)=1$ (the  {\em normality} of the Markov process $X$ indicating that the probability measure ${\mathbb P}_x$ governs the behavior of the process started from $x$ at time $0$). A Markov process is said to be {\em conservative} if ${\mathbb P}_x(X_t \in E)=1$ for all $x\in E$ and $t\geq 0$ (the process stays in $E$).
Since our stochastic driver $X$ is conservative, we do not deal with killing and do not adjoin the cemetery state to our state space. For a Markov process $X$ the {\em transition function} is defined by $P_t(x,B):={\mathbb P}_x(X_t\in B)$  for $t\geq 0$, $x\in E$, $B\in {\mathscr E}$. For a Markov process $X$, we define the {\em minimum admissible filtration}
generated by $X$  by ${\mathscr F}_t^0:=\sigma\{X_s,s\leq t\}$ and ${\mathscr F}_\infty^0:=\sigma\{X_s,s\geq 0\}$. $X$ has the Markov property with respect to $({\mathscr F}_t^0)_{t\geq 0}$, and, for any $\Lambda\in {\mathscr F}_\infty^0$, ${\mathbb Q}_x(\Lambda)$ is an  ${\mathscr E}$-measurable function of $x$.
For every probability measure $\mu$ on $(E,{\mathscr E})$, the integral ${\mathbb P}_\mu(\Lambda)=\int_E {\mathbb P}_x(\Lambda)\mu(dx)$, $\Lambda\in {\mathscr F}_\infty^0$, defines a probability measure on $(\Omega,{\mathscr F}_\infty^0)$, which is called the {\em probability law of the Markov process $X$ with the initial distribution} $\mu$ because ${\mathbb P}_\mu(X_0\in B)=\mu(B)$ for $B\in {\mathscr E}$.

In the generic definition of a Markov process, the state space $(E,{\mathscr E})$ is only assumed to be a measurable space. In this paper we assume that $E$ is a Lusin topological space equipped with the Borel sigma-field ${\mathscr E}$. Namely, $E$ is homeomorphic to a Borel subset of some compact metric space.
The (conservative) Markov process $X$ on a Lusin space $(E,{\mathscr E})$ is called a (conservative) {\em Borel right process} if it satisfies the following  conditions: (i) For each $t\geq 0$, there exists a {\em shift operator} $\theta_t: \Omega\rightarrow \Omega$ such that $X_s\circ \theta_t = X_{s+t}$ for every $s\geq 0$. (ii) For each $\omega\in \Omega$, the sample path $t\rightarrow X_t(\omega)\in E$ is right continuous on $[0,\infty)$. (iii) $X$ is a strong Markov process (recall that a Markov process is called {\em strong Markov} if there exists a right-continuous admissible filtration $({\mathscr M}_t)$ for which the strong Markov property holds, i.e. for any $({\mathscr M}_t)$-stopping time
$\sigma$, an  initial distribution $\mu$, $s\geq 0$, and $B\in {\mathscr E}$,
${\mathbb P}_\mu(X_{\sigma+s} \in B|{\mathscr M}_\sigma)={\mathbb P}_{X_\sigma}(X_{s} \in B),$ ${\mathbb P}_\mu$-a.s. on $\{\sigma<\infty\}$).
Since the stochastic driver $X$ is conservative in this paper, we do not deal with killing and the cemetery state.

{\em Borel} in Borel right process indicates that the state space $E$ of $X$ is homeomorphic to a Borel subset of a compact metric space and is equipped with the Borel sigma-field ${\mathscr E}$  so that the transition function $P_t f$ is Borel measurable for every $f\in {\cal B}_b(E)$ (the space of bounded Borel measurable functions). {\em Right} in the name refers to {\em right processes}, strong Markov processes with right-continuous paths as defined in \citet{sharpe_1988} or \citet{chen_2011} Definition A.1.35 and Theorem A.1.37, where the state space $E$ is taken to be a more general Radon topological space, i.e. $E$ is homeomorphic to a universally measurable subset of some compact metric space (a set is universally measurable if it is measurable with respect to all finite measures on $E$).

The above definition of a BRP apparently depends on an arbitrary choice of a right continuous admissible filtration $({\mathscr M}_t)_{t\geq 0}$ for $X$ describing the strong Markov property. However, it actually depends only on the minimum admissible filtration $({\mathscr F}^0_t)$ for $X$ due to the fact that the BRP is strong Markov with respect to $({\mathscr F}_{t+}^0)_{t\geq 0}$ defined by
${\mathscr F}_{t+}^0:=\cap_{t^\prime>t}{\mathscr F}_{t^\prime}^0,$ $t\geq 0$.
The minimum admissible filtration $({\mathscr F}^0_t)$ can be completed as follows.
Denote by ${\mathscr F}_\infty^\mu$ the ${\mathbb P}_\mu$-completion of ${\mathscr F}_\infty^0$ and by ${\cal N}$ the family of all null sets in ${\mathscr F}_\infty^\mu$ (recall that ${\mathbb P}_\mu(\Lambda)=\int_E {\mathbb P}_x(\Lambda)\mu(dx)$). We then let ${\mathscr F}_t^\mu=\sigma({\mathscr F}_t^0,{\cal N})$ for each $t\geq 0$. We further let ${\mathscr F}_t=\cap_{\mu}{\mathscr F}_t^\mu$, where $\mu$ run through all
 probability measures on ${\mathscr E}$. The resulting  filtration is called the {\em minimum augmented admissible filtration} of the BRP $X$. By Theorem A.1.18 on p.443 of \citet{chen_2011}, the minimum augmented admissible filtration of the BRP $X$ is already right continuous, and $X$ is strong Markov with respect to it. It thus satisfies the usual hypothesis of stochastic calculus.

Stochastic calculus of semimartingales defined over a right process has been developed in \citet{cinlar_1980} (see also Chapter VI of \citet{sharpe_1988}). As shown in these references, stochastic calculus for semimartingales over a right process can be set up so that all key properties hold simultaneously for all ${\mathbb P}_x$, $x\in E$. Specifically, let $Y$ be a process which is a semimartingale over $(\Omega,{\mathscr F},({\mathscr F}_t)_{t\geq 0},{\mathbb P}_x)$ for every $x\in E$.
Theorem 3.12 of  \citet{cinlar_1980} shows that its decomposition as a sum of a local martingale and a process of finite variation, its quadratic variation process, its continuous local martingale part, and stochastic integrals with respect to it are all the same for all  ${\mathbb P}_x$, $x\in E$. Moreover, $Y$ is then a semimartingale over $(\Omega,{\mathscr F},({\mathscr F}_t)_{t\geq 0},{\mathbb Q}_\mu)$ for
 all initial distributions $\mu$, and the above mentioned decompositions and processes are also fitted to ${\mathbb P}_\mu$.

\section{On Recurrence of Markov Process}
\label{relation_recur}
In this section, we give several recurrence definitions which are used in this paper and discuss their relation. We call the recurrence in the sense of Definition \ref{recurrent_getoor} $\textbf{(R0)}$ and we will label each recurrence definition as we proceed.

We first consider an alternative definition of recurrence for a BRP in \citet{tweedie_1994}.
We start with the definition of $\varphi$-irreducibility of \citet{tweedie_1994} and \citet{tweedie_1993}.
\begin{definition}{\bf (Irreducibility)}
$X$ is called $\varphi$-{\em irreducible} if there exists a non-trivial sigma-finite measure $\varphi$ on $(E,\mathscr{E})$ such that the mean occupation time of any set $B\in \mathscr{E}$ with $\varphi(B)>0$ does not vanish, i.e.
$\varphi(B)>0 \Rightarrow R(x,B)>0$ for all $x\in E$.
\end{definition}
That is, starting from any point $x\in E$ the Markov process $X$ on average spends a positive amount of time in each Borel set of positive measure $\varphi(B)>0$ (it can be infinite).
Irreducibility measures are not unique, nor are they equivalent. Some measures charge more sets than others. However,
if the process $X$ is $\varphi$-irreducible, then there exists a {\em maximal irreducibility measure} $\psi$ such that for any measure $\varphi^\prime$, the process is $\varphi^\prime$-irreducible if and only if $\varphi^\prime$ is absolutely continuous with respect to $\psi$, and
\be
\psi(B)=0\Rightarrow\psi\{x\in E:R(x,B)>0\}=0.
\eel{psi_dominate}
(Theorem 2.1 in \citet{tweedie_1994}).
For a given process $X$, the maximal irreducibility measure is unique up to measure equivalence. From now on, $\psi$ always refers to the maximal irreducibility measure.

\begin{definition}
\label{recurrent_set_def}
(\citet{tweedie_1994}, p.179)
A set $B$ is called {\em recurrent} if $R(x,B)=\infty$, $x\in B$, and {\em uniformly transient} if there exists  a constant $M<\infty$ such that $R(x,B)\leq M, x\in E$; and transient if it can be covered by countably many uniformly transient sets.
\end{definition}
A nice formulation of a dichotomy between recurrence and transience for a Markov process can be made as follows using the notion of $\psi$-irreducibility (\citet{tweedie_1994}, Theorem 2.3).
\begin{definition}
\label{recurrent_tweedie}{\bf (R1)}
Suppose $X$ is $\psi$-irreducible. Then $X$ is recurrent in the sense that every set $B\in {\mathscr E}$ with $\psi(B)>0$ is recurrent, or $X$ is transient in the sense that $E$ is a transient set.
\end{definition}

Next we introduce the  definition of recurrence for one-dimensional diffusions in \citet{borodin_2002}, p.20. Here we consider the setting for 1D diffusions  under assumptions made in Section \ref{exist_2}.
\begin{definition}
\label{borodin_recu}{\bf (R2)}
A 1D diffusion $X$ is said to be recurrent if $\mathbb{P}_x(H_y<\infty)=1$ for all $x,y\in I$,
where $H_y=\inf\{t\geq0 : X_t=y\}$. Otherwise, it is called transient.
\end{definition}
The intuition is that starting from any point in $I$ a recurrent 1D diffusion hits any other point $y$ in $I$ in finite time with probability one.

Next we introduce the definition of recurrence for diffusions in ${\mathbb R}^d$ in \citet{pinsky_1995} Chapter 2.7 which is similar to the previous one, but is suitable for multi-dimensional diffusions.
Here $X$ is a diffusion in ${\mathbb R}^d$ in the sense of Section \ref{exist_3} (with the assumptions in Section \ref{exist_3} assumed to hold).
\begin{definition}
\label{pinsky_recu}{\bf (R3)}
$X$ is said to be recurrent if $\mathbb{P}_x(\sigma_{B_{\epsilon}(y)}<\infty)=1$ for all $x,y\in E$ and $\epsilon>0$, where $\sigma_{B_{\epsilon}(y)}=\inf\{t\geq0: |X_t-y|\leq\epsilon\}$. Otherwise $X$ is said to be transient.
\end{definition}
The intuition is that starting from any point in ${\mathbb R}^d$ a recurrent diffusion hits any open ball centered at any point in finite time with probability one.

The different definitions of recurrence are cast in different contexts under different sets of assumptions and are not generally equivalent. Before we discuss the relationships between these definitions, we introduce an important absolute continuity assumption (also known as Meyer's hypothesis (L)) for the resolvent of a Markov process (see Definition A.2.16 (AC)' on p.422 in \citet{chen_2011} or 10.25 on p.56 of \citet{sharpe_1988}).

For $\alpha\geq0$, define $R_\alpha$ as the {\em resolvent} or $\alpha$-{\em potential} operator of the Markov process:
\be
R_\alpha g(x):={\mathbb E}_x\left[\int_0^\infty e^{-\alpha t}g(X_t)dt\right].
\eel{resolvent}
The resolvent measures $(R_\alpha(x,B):=(R_\alpha {\bf 1}_B)(x))_{\alpha>0}$ are all finite measures and are equivalent to each other. Furthermore, if for some $x\in E$, some set $B$ and some $\alpha_0\geq0$ we have that $R_{\alpha_0}(x,B)=0$, then $R_\alpha(x,B)=0$ for all $\alpha\geq0$ (recall that $R_0(x,B)=R(x,B)$ is the Green's measure). Thus, by \eqref{psi_dominate} we have
\be
\psi(B)=0\Rightarrow\psi\{x\in E:R_\alpha(x,B)>0\}=0.
\eel{psi_dominate2}

\begin{assumption}
\label{ac_condition}{\bf (Absolute Continuity Assumption)}
For some $\alpha>0$ (and thus for all $\alpha>0$) and all $x\in E$, the resolvent measure is absolutely continuous with respect to the maximal irreducibility measure, i.e.
$R_\alpha(x,\cdot)\prec \psi(\cdot)$.
\end{assumption}

\begin{proposition}
\label{relation_recu}
(1) Suppose $X$ is a $\psi$-irreducible BRP. Then $\bf{(R0)}\Rightarrow\bf{(R1)}$. Conversely, if $X$ is $\psi$-irreducible and Assumption \ref{ac_condition} holds, then $\bf{(R1)}\Rightarrow\bf{(R0)}$.\\
(2) Suppose $X$ is a 1D diffusion under the setting of Section \ref{exist_2}. Then $\bf{(R0)}\Leftrightarrow\bf{(R1)}\Leftrightarrow\bf{(R2)}$.\\
(3) Suppose $X$ is a diffusion on $\mathbb{R}^d$ under the setting of Section \ref{exist_3}. Then $\bf{(R0)}\Leftrightarrow\bf{(R1)}\Leftrightarrow\bf{(R3)}$.\\
\end{proposition}

{\bf Proof.}
(1) Suppose $X$ is $\psi$-irreducible and $\bf{(R0)}$. For any set $B$ with $\psi(B)>0$, by $\psi$-irreducibility we have $R(x,B)>0$. Hence, by $\bf{(R0)}$, $R(x,B)=\infty$, which implies $\bf{(R1)}$.

Conversely, suppose $X$ is $\psi$-irreducible, Assumption \ref{ac_condition} holds and $\bf{(R1)}$. For any set with $\psi(B)>0$, $R(x,B)=\infty$ for all $x\in E$ by recurrence in the sense of Definition \ref{recurrent_tweedie}. For any set $B$ with $\psi(B)=0$, we have $R_\alpha(x,B)=0$ for all $x\in E$ and $\alpha>0$ by Assumption \ref{ac_condition}. As discussed above Eq.\eqref{psi_dominate2}, it also holds that $R(x,B)=0$ for all $x\in E$.
Thus, for any $\psi$-measurable set $B$, either $R(x,B)=0$ or $R_0(x,B)=\infty$. Hence, since the measure $\psi$ is sigma-finite, the same holds for any universally measurable set $B\in {\mathscr E}^*$, which implies $\bf{(R0)}$.

(2) As we discussed in Section \ref{exist_2}, $X$ satisfies Assumption \ref{ac_condition}, where $\psi$ is the speed measure $m$. Thus, $\bf{(R0)}\Leftrightarrow\bf{(R1)}$. We only need to prove $\bf{(R1)}\Leftrightarrow\bf{(R2)}$. Let $p(t,x,y)$ denote the density of $X$ and $G(x,y):=\int_0^\infty p(t,x,y)dt$. From p.20 of \citet{borodin_2002}, we can see if $G(x_0,y_0)<\infty$ for any interior point $(x_0,y_0)\in I\times I$, then $G$ is continuous at that point. We have the relation $R(x,B)=\int_0^\infty \int_B p(t,x,y)dydt=\int_B G(x,y)dy$.

Suppose $\bf{(R1)}$. Choose a finite interval $B$ in the interior of $I$ with $m(B)>0$. Then $R(x,B)=\int_B G(x,y)dy=\infty$. Thus, there exists $y$ such that $G(x,y)=\infty$. By p.20 of \citet{borodin_2002}, $X$ is recurrent in the sense of $\bf{(R2)}$.

Conversely, suppose $\bf{(R2)}$. Again by p.20 of \citet{borodin_2002}, $G(x,y)=\infty$ for all $x,y\in I$. Thus $R(x,B)=\int_B G(x,y)dy=\infty$ for all $B$ with $m(B)>0$, which implies $\bf{(R1)}$.

(3) We first observe that under our assumptions $X$ has a positive density with respect to Lebesgue measure. Thus, it is irreducible and satisfies Assumption \ref{ac_condition}. Thus, by (1) we have $\bf{(R0)}\Leftrightarrow\bf{(R1)}$ and we only need to showe $\bf{(R1)}\Leftrightarrow\bf{(R2)}$.

Suppose $\bf{(R2)}$. Then by proof of \citet{pinsky_1995} Theorem 2.1 on p.130, $R(x,B)=\infty$ for all $x\in E$ and every ball $B$. Thus every point is topologically recurrent (cf. \citet{tweedie_1994} Section 4). Since $X$ is a Lebesgue-irreducible Feller process taking values in $\mathbb{R}^d$, by Theorem 7.1 of \citet{tweedie_1994} it is a $T$ model. Then $\bf{(R1)}$ holds by Theorem 4.1 of \citet{tweedie_1994}.

Conversely, suppose $\bf{(R2)}$ does not hold. Then again by \citet{pinsky_1995} Theorem 2.1 on p.130, $R(x,B)<\infty$ for all $x\in E$ and every ball $B$. It is then clear that $\bf{(R1)}$ does not hold. $\Box$\\

\section{Proof of Theorem \ref{recover_short}}
\label{proof_short}
Let
$$
\Lambda_t:=\exp\left(\int_0^t r(X_s)ds-\lambda t\right)\frac{h(X_t)}{h(X_0)}=\left.\frac{d\mathbb{Q}}{d\mathbb{Q}^\pi}\right|_{{\mathscr F}_t}.
$$
Applying It\^{o}'s integration by parts, we can write:
$$
h(X_0)(\Lambda_t-\Lambda_0) = e^{\int_0^t r(X_s)ds-\lambda t}h(X_t)-h(X_0)
$$
$$
=\int_0^t (r(X_s)-\lambda)e^{\int_0^s r(X_u)du -\lambda s}h(X_s) ds   +
\int_0^t e^{\int_0^s r(X_u)du -\lambda s}dA^h_s
+\int_0^t e^{\int_0^s r(X_u)du -\lambda s}dM^h_s,
$$
where by Theorem 3.18 of  \citet{cinlar_1980} we decompose the additive semimartingale functional $h(X_t)-h(X_0) = A^h_t + M_t^h$  into an additive functional of finite variation $A^h$ and an additive $\mathbb{Q}^\pi$-local martingale functional $M^h$ ($M_t^h$ should not be confused with $M_t^\pi$). Since $\Lambda_t$ is a $\mathbb{Q}^\pi$-martingale, it must hold that
$$\int_0^t (r(X_s)-\lambda) e^{\int_0^s r(X_u)du -\lambda s}h(X_s)ds   +
\int_0^t e^{\int_0^s r(X_u)du -\lambda s}dA^h_s=0.$$
Thus, the additive functional of finite variation has to be absolutely continuous, $A_t^h=\int_0^t f(X_s)ds$ with some $f(x)$, and
it must hold that
$[r(x)-\lambda]h(x)+f(x)=0$, so that $r(x)=\lambda-f(x)/h(x)$. Furthermore, since $M_t^h=h(X_t)-h(X_0) -\int_0^t f(X_s)ds$ is a local martingale, the pair $(h,f)$ belongs to the domain of the extended generator ${\cal G}^\pi$ of $X$ under ${\mathbb Q}^\pi$ by definition, $f(x)={\cal G}^\pi h(x)$, and Eq.\eqref{shortrate} is verified.

The expression for $Q_t$ is obtained from writing the ${\mathbb P}$-martingale
$$M_t=e^{\int_0^t r(X_s)ds}S_t = e^{\int_0^t r(X_s)ds-\lambda t}\frac{\pi(x)}{\pi(X_t)}M_t^\pi$$ and
$$
Q_tf(x)={\mathbb E}^{{\mathbb Q}}_x[f(X_t)]={\mathbb E}^{{\mathbb P}}_x[M_t f(X_t)]={\mathbb E}^{{\mathbb Q}^\pi}_x\left[ e^{\int_0^t r(X_s)ds-\lambda t}\frac{\pi(x)}{\pi(X_t)}f(X_t)\right].
$$
The result for the generator is then well known (see Proposition 3.4 on page 351 in \citet{revuz_1999} for diffusions, or \citet{palmowski_2002} and references therein for the general setting). $\Box$

\section{Proofs of Sufficient Conditions for Existence}
\label{appendix_existence}
\subsection{Proof of Theorem \ref{exist_theorem_1}}
\label{proof_zhang}

(i) For each Borel set $B$ with $m(B)>0$,
\be
\mathbb{E}_x^{\mathbb Q}[\int_0^\infty 1_B(X_t)dt]\geq\int_0^\infty \mathbb{E}_x^{\mathbb Q}[e^{-A_t} 1_B(X_t)] dt=\int_0^\infty dt\int_B \hat{p}(t,x,y)m(dy)>0.
\ee
Thus $X$ is $m$-irreducible. Assumption \ref{ac_condition} is obviously satisfied.\\
\\
Part (ii) is proved by the application and slight extension of the arguments in \citet{zhang_2013}, Section 2, pages 2-4 as follows.
Identifying the pricing semigroup with the transition semigroup of $X^r$ and arguing as on page 2 of \citet{zhang_2013}, the pricing operators form a strongly continuous contraction semigroup on $L^2(E,m)$. The assumption \eqref{exist_2_assumption1} implies that, for any $t>0$, the pricing operator ${\mathscr P}_t$ and its dual $\hat{\mathscr P}_t$ are Hilbert-Schmidt operators. Let ${\mathscr G}$ and $\hat{\mathscr G}$ be the infinitesimal generators of the semigroup $({\mathscr P}_t)_{t\geq 0}$ and its dual $(\hat{\mathscr P}_t)_{t\geq 0}$ on $L^2(E,m)$, respectively. Under our assumptions, it follows from Jentzsch's theorem (Theorem V.6.6 on page 337 of \citet{schaefer_1974}) that the common value $\lambda:=\inf {\rm Re}(\sigma(-{\mathscr G}))=\inf {\rm Re}(\sigma(-\hat{\mathscr G}))$ is non-negative and an eigenvalue of multiplicity one for both $-{\mathscr G}$ and $-\hat{\mathscr G}$, and that an eigenfunction $\pi(x)$ of $-{\mathscr G}$ and an eigenfunction $\hat{\pi}(x)$ of $-\hat{\mathscr G}$ can be chosen to be strictly positive $m$-a.e. on $E$ with $\|\pi\|_2=\|\hat{\pi}\|_2=1$ (here $\sigma({\mathscr A})$ denotes the spectrum of ${\mathscr A}$).
Following \citet{zhang_2013}, the application of Assumption \eqref{exist_2_assumption2} yields that
 the eigenfunctions are, in fact, bounded, continuous and, thus, strictly positive everywhere on $E$. Therefore \eqref{exist_2_eigenfunction} are valid for every $x\in E$ and $t>0$.

We note that \citet{zhang_2013} impose \eqref{exist_2_assumption2} for all $t>0$. However, their proof remains unchanged if it is relaxed to require that there exists a constant $T>0$ such that \eqref{exist_2_assumption2} hold for all $t\geq T$. This relaxation of the assumption is useful for us, as it accommodates the CIR model, as will be shown in Section \ref{CIRin5}. \\
\\
Part (iii) directly follows from Lemma 2.1 in \citet{zhang_2013}.\\
\\
(iv) The transition function $Q_t^\pi(x,dy)$ of $X$ under $\mathbb{Q}^\pi$ has a positive and continuous density with respect to $m$:
\be
p^\pi(t,x,y)=e^{\lambda t}\frac{\pi(y)}{\pi(x)}p(t,x,y).
\ee
By (iii) we have for the Green's measure of $X$ under $\mathbb{Q}^\pi$:
$$
R^\pi(x,B)=\int_0^\infty \int_B p^\pi(t,x,y) m(dy)dt
$$
$$
\geq
\int_T^\infty \int_B C^{-1}
\pi(y)\hat{\pi}(y) m(dy)dt-\int_T^\infty\int_B \frac{c}{C}e^{-\alpha t}\frac{\pi(y)}{\pi(x)}m(dy)dt=\infty
$$
for any $B\in {\mathscr E}$ with $0<m(B)<\infty$.
Thus, $X$ is recurrent under $\mathbb{P}_x$ in the sense of Definition \ref{recurrent_getoor} {\bf(R0)}. Due to Assumption \ref{ac_condition}, it is then also recurrent in the sense of Definition \ref{recurrent_tweedie} {\bf(R1)}.
Finally, (iii) and (iv) imply that the transition density $p^\pi(t,x,y)$ of $X$ under ${\mathbb P}^\pi$ converges to the stationary density $C^{-1}\pi(y) \hat{\pi}(y)$ as $t\rightarrow \infty$. Thus, under the assumptions in this section, $X$ is positive recurrent under ${\mathbb Q}^\pi$  with the stationary distribution $C^{-1}\pi(y) \hat{\pi}(y)m(dy)$.\\
\\
(v) If $m(E)<\infty$, then constants are in $L^2(E,m)$, and from the estimate for the density in part (iii) we obtain the large time estimate for $\mathscr{P}_tf(x)$.
$\Box$


\subsection{Proof of Theorem \ref{1d_sufficient} and Sufficient Conditions for 1D Diffusions}
\label{proof_1D}

{\bf Proof of Theorem \ref{1d_sufficient}.} (i) The $L^2(I,m)$-principal eigenfunction of a regular or singular SL problem does not have zeros in $(l,r)$. Inaccessible boundaries are not included in the state space $I$ and, hence, we do not need to check whether the principal eigenfunction vanishes at inaccessible boundaries. Regular instantaneously reflecting boundaries are included in the state space.
If $l$ is an instantaneously reflecting boundary, the $L^2(I,m)$ principal eigenfunction satisfies  $\lim_{x\downarrow l}\frac{\pi_0^\prime(x)}{s(x)}=0$ and $\pi_0(l)>0$.  Similarly, if $r$ is an instantaneously reflecting boundary, the $L^2(I,m)$ principal eigenfunction satisfies  $\lim_{x\uparrow r}\frac{\pi_0^\prime(x)}{s(x)}=0$ and $\pi_0(r)>0$.
Due to the functional calculus form of the spectral theorem, if $\pi$ is an eigenfunction of the non-negative self-adjoint operator $-{\cal A}$ with the eigenvalue $\lambda\geq 0$, it is also an eigenfunction of the symmetric semigroup ${\mathscr P}_t=e^{t{\cal A}}$ on $L^2(I,m)$ generated by  ${\cal A}$ with the eigenvalue $e^{-\lambda t}\leq 1$.

(ii) Let $Q^{\pi_0}_t$ denote the transition semigroup of $X$ under $\mathbb{Q}^{\pi_0}$, i.e.
$Q_t^{\pi_0} f=\mathbb{E}^{\mathbb{Q}^{\pi_0}}_x[f(X_t)].$
Since ${\mathbb Q}^{\pi_0}|_{{\mathscr F}_t}=\tilde{M}_t^\pi {\mathbb Q}|_{{\mathscr F}_t}$, we can write
\be
Q_t^{\pi_0} f(x)=\frac{1}{\pi_0(x)}\mathbb{E}^\mathbb{Q}_x[e^{-\int_0^t r(X_s)ds+\lambda t}f(X_t)\pi_0(X_t)]=\frac{1}{\pi_0(x)}{\mathscr P}^\lambda_t (\pi_0 f)(x),
\ee
where ${\mathscr P}_t^\lambda=e^{\lambda t}{\mathscr P}_t$.
The generator ${\cal G}^{\pi_0}$ of $(P_t^{\pi_0})_{t\geq 0}$ is then:
$$
{\cal G}^{\pi_0} f=\frac{1}{\pi_0}{\cal A}_\lambda (\pi_0 f).
$$
The expression in (ii) then follows by using
 the fact that $\pi$ satisfies the SL equation ${\cal A}_\lambda \pi(x)=({\cal A}-\lambda)\pi(x)=0$. The speed density $m^{\pi_0}(x)$ and scale density $s^{\pi_0}(x)$ can be easily calculated as $m^{\pi_0}(x)=\pi_0^2(x)m(x)$ and $s^{\pi_0}(x)=s(x)/\pi_0^2(x)$. Since $\pi_0(x)\in L^2(I,m)$, the speed measure of $X$ under $\mathbb{Q}^{\pi_0}$ is finite, i.e. $\int_l^r m^{\pi_0}(y)dy<\infty$. Thus it is also the stationary distribution for the diffusion. The only thing left is to prove $X$ is recurrent under $\mathbb{Q}^{\pi_0}$.  We can split it into three cases.\\
 (1) Both ends of $I$ are reflecting boundary. Then it is covered in Theorem \ref{exist_theorem_1}.\\
 (2) Only one end of $I$ is reflecting boundary (wlog we assume it is the left end). Since $X$ is conservative under $\mathbb{Q}^{\pi_0}$, by \citet{pinsky_1995} Theorem 5.1.5, for any $x_0\in (l,r)$,
 \be
 \int_{x_0}^r dx\  s^{\pi_0}(x)\int_{x_0}^x m^{\pi_0}(y)dy=\infty.
 \ee
 Since $\int_l^r m^{\pi_0}(y)dy<\infty$, we have $\int_{x_0}^r s^{\pi_0}(x)dx=\infty$. By p.20 of \citet{borodin_2002}, $X$ is recurrent $\bf{(R2)}$ under $\mathbb{Q}^{\pi_0}$. By Proposition \ref{relation_recu}, $\bf{(R0)}$ also holds.\\
 (3)There is no reflecting boundary. Similarly, by \citet{pinsky_1995} Theorem 5.1.5, for any $x_0\in (l,r)$,
 \be
 \int_{l}^{x_0} dx s^{\pi_0}(x)\int_x^{x_0}m^{\pi_0}(y)dy=\infty \quad\text{and}\quad \int_{x_0}^r dx s^{\pi_0}(x)\int_{x_0}^x m^{\pi_0}(y)dy=\infty.
 \ee
 Using the fact that $\int_l^r m^{\pi_0}(y)dy<\infty$, we have $\int_l^{x_0}s^{\pi_0}(x)dx=\infty$ and $\int_{x_0}^r s^{\pi_0}(x)dx=\infty$. By p.20 of \citet{borodin_2002}, $X$ is recurrent $\bf{(R2)}$, and thus $\bf{(R0)}$ under $\mathbb{Q}^{\pi_0}$

 (iii) By   \citet{mckean_1956} (see also \citet{linetsky_2004} Eq.(12)), under our assumptions the pricing operator can be represented by
 \be
 \mathscr{P}_t f(x)=\int_I p(t,x,y)f(y)m(dy)
 \ee
 where $m(dy)$ is the speed measure and the density $p(t,x,y)$ has the spectral representation
 \be
 p(t,x,y)=e^{-\lambda_0 t}\pi_0(x)\pi_0(y)+\int_{\lambda_0+\alpha}^\infty e^{-\lambda t}\sum_{i,j=1}^2 u_i(x,\lambda) u_j(y,\lambda)\rho_{ij}(d\lambda).
 \ee
 Thus for all $t\geq 2T>0$
 \be
 \begin{array}{ll}
\displaystyle{|p(t,x,y)-e^{-\lambda_0 t}\pi_0(x)\pi_0(y)|} & \displaystyle{=\left|\int_{\lambda_0+\alpha}^\infty e^{-\lambda t}\sum_{i,j=1}^2 u_i(x,\lambda) u_j(y,\lambda)\rho_{ij}(d\lambda)\right|}\\
& \displaystyle{\leq\sum_{i,j=1}^2  \left(\int_{\lambda_0+\alpha}^\infty e^{-\lambda t} u_i^2(x,\lambda)\rho_{ij}(d\lambda)\right)^{1/2}\left(\int_{\lambda_0+\alpha} e^{-\lambda t} u_j^2(y,\lambda)\rho_{ij}(d\lambda)\right)^{1/2}}\\
&\displaystyle{\leq\sum_{i,j=1}^2 \left(e^{-(\lambda_0+\alpha)(t-2T)}\int_{\lambda_0+\alpha}^\infty e^{-\lambda 2T}u_i^2(x,\lambda)\rho_{ij}(d\lambda)\right)^{1/2}\left(e^{-(\lambda_0+\alpha)(t-2T)}\right.}\\
&\displaystyle{\enskip\left. \int_{\lambda_0+\alpha}^\infty e^{-\lambda 2T}u_j^2(y,\lambda)\rho_{ij}(d\lambda)\right)^{1/2}}\\
&\displaystyle{\leq \sum_{i,j=1}^2 e^{-(\lambda_0+\alpha)(t-2T)}p^{1/2}(2T,x,x)p^{1/2}(2T,y,y)}\\
&:=c e^{-(\lambda_0+\alpha) t}p^{1/2}(2T,x,x)p^{1/2}(2T,y,y).\\
 \end{array}
 \ee
By the symmetry of the density $p(t,x,y)=p(t,y,x)$, Eq.\eqref{exist_2_assumption1} for $t=T$ implies $\int_I p(2T,y,y)m(dy)<\infty$. Thus for any $L^2(I,m)$ payoff $f$, we have for $t\geq 2T>0$:
\be
\begin{array}{ll}
\mathscr{P}_t f(x) &\displaystyle{ =\int_I p(t,x,y)f(y)m(dy)}\\
& \displaystyle{\leq\int_I |p(t,x,y)-e^{-\lambda_0 t}\pi_0(x)\pi_0(y)|f(y)m(dy)+ \int_I e^{-\lambda_0 t}\pi_0(x)\pi_0(y)f(y)m(dy)}\\
& \displaystyle{\leq\int_I ce^{-(\lambda_0+\alpha)t}p^{1/2}(2T,x,x)p^{1/2}(2T,y,y)f(y)m(dy)+c_fe^{-\lambda_0 t}\pi_0(x)}\\
&\displaystyle{\leq c_fe^{-\lambda_0 t}\pi_0(x)+ce^{-(\lambda_0+\alpha)t}p^{1/2}(2T,x,x)\int_I p^{1/2}(2T,y,y)f(y)m(dy)}\\
&\displaystyle{\leq c_fe^{-\lambda_0 t}\pi_0(x)+ce^{-(\lambda+\alpha)t}p^{1/2}(2T,x,x)\left(\int_I p(2T,y,y)m(dy)\right)^{1/2}\left(\int_I f^2(y)m(dy)\right)^{1/2}}\\
& \displaystyle{\leq c_fe^{-\lambda_0 t}\pi_0(x)+Kp^{1/2}(2T,x,x)\|f\|_{L^2(I,m)}e^{-(\lambda+\alpha)t}},\\
\end{array}
\ee
where $K$ is a constant independent of $f$, $x$ and $t$.
$\Box$\\
\\
Next we give some sufficient conditions for the existence of an $L^2(I,m)$-principal eigenvalue.
We first need to recall some results from Sections 3.4-3.6 of \citet{linetsky_2008} (see \citet{linetsky_2004} for references and proofs).
\begin{definition}
For a given real $\lambda$, equation \eqref{SLeq} is said to be {\em oscillatory} at an endpoint $e\in \{l,r\}$ if and only if every solution has infinitely many zeros clustering at $e$. Otherwise it is called {\em non-oscillatory} at $e$.
\end{definition}
This classification is mutually exclusive for a fixed $\lambda$, but can vary with $\lambda$. For equation \eqref{SLeq}, there are two distinct possibilities at each endpoint.
\begin{proposition}{\bf (Oscillatory/Non-oscillatory Classification of Boundaries)}
Let $e\in \{l,r\}$ be an endpoint of equation \eqref{SLeq}. Then $e$ belongs to one and only one of the following two cases:\\
(i) Equation \eqref{SLeq} is non-oscillatory at $e$ for {\em all} real $\lambda$. Correspondingly, the endpoint $e$ is said to be non-oscillatory. \\
(ii) There exists a real number $\Lambda\geq 0$ such that equation \eqref{SLeq} is oscillatory at $e$ for all $\lambda > \Lambda$ and non-oscillatory at $e$ for all $\lambda < \Lambda$. Correspondingly, $e$ is said to be {\em oscillatory with cutoff} $\Lambda$.
Equation \eqref{SLeq} can be either oscillatory or non-oscillatory at $e$ for $\lambda=\Lambda>0$. It is always non-oscillatory for $\lambda=0$.
\end{proposition}
Based on the oscillatory/non-oscillatory classification of boundaries, the spectrum of the non-negative operator $-{\cal A}$ is classified as follows.
\begin{proposition} ({\bf Spectral Classification})\\
\label{spectral_class}
(i)
{\rm {\bf Spectral Category I.}}
If both endpoints are non-oscillatory, then the spectrum of $-{\cal A}$ is simple, non-negative and purely discrete.\\
(ii)
{\rm {\bf Spectral Category II.}}
If one of the endpoints is non-oscillatory and the other endpoint is oscillatory
with cutoff $\Lambda\geq 0$,
then the spectrum is simple and non-negative, the
essential spectrum is nonempty, $\sigma_e(-{\cal A})\subset [\Lambda,\infty)$, and $\Lambda$ is the
lowest point of the essential spectrum.
If the SL equation is non-oscillatory at the oscillatory endpoint for $\lambda=\Lambda\geq 0$, then there is a {\em finite} set of simple eigenvalues in $[0,\Lambda]$ (it may be empty).
If the SL equation is oscillatory at the oscillatory endpoint for $\lambda=\Lambda>0$, then there is an {\em infinite} sequence of simple eigenvalues in $[0,\Lambda)$ clustering at $\Lambda$.\\
(iii)
{\rm {\bf Spectral Category III.}} If $l$ is oscillatory with cutoff $\Lambda_l\geq 0$ and $r$ is oscillatory with cutoff $\Lambda_r\geq 0$, then the essential spectrum is nonempty, $\sigma_e(-{\cal A})\subset [\underline{\Lambda},\infty)$,
$\underline{\Lambda}:=\min\{\Lambda_l, \Lambda_r\}$, and $\underline{\Lambda}$ is the lowest point of the essential spectrum. The spectrum is simple (has multiplicity one) below $\overline{\Lambda}:=\max\{\Lambda_l, \Lambda_r\}$ and is not simple above $\overline{\Lambda}$. If the SL equation is non-oscillatory for $\lambda= \underline{\Lambda}\geq 0$, then there is a {\em finite} set of simple eigenvalues in $[0,\underline{\Lambda}]$ (it may be empty).
If the SL equation is oscillatory for $\lambda=\underline{\Lambda}>0$, then there is an {\em infinite} sequence of simple eigenvalues in $[0,\underline{\Lambda})$ clustering at $\underline{\Lambda}$.
\end{proposition}

Based on this spectral classification, we can establish the following result.
\begin{theorem}
\label{1d_sufficient2}
Under the assumptions on $X$ and $r$ in this section, the operator $-{\cal A}$ has an $L^2(I,m)$-principal eigenfunction $\pi_0(x)$ with a spectral gap above the corresponding principal eigenvalue $\lambda_0$ if {\em  one} of the following sufficient conditions holds:
\begin{itemize}
  \item{(1)} Both boundaries $l$ and $r$ are non-oscillatory.
  \item{(2)} One of the boundaries is non-oscillatory, and the other boundary is oscillatory with cutoff $\Lambda>0$, and the SL equation is oscillatory at the oscillatory endpoint for $\lambda=\Lambda$.
  \item{(3)} Both boundaries $l$ and $r$ are oscillatory with cutoffs $\Lambda_l$ and $\Lambda_r$ with $\underline{\Lambda}=\min\{\Lambda_l,\Lambda_r\}>0$, and the SL equation is oscillatory  for $\lambda=\underline{\Lambda}$ at the end-point corresponding to $\underline{\Lambda}$.
\end{itemize}
\end{theorem}
{\bf Proof.}
(1) According to Proposition \ref{spectral_class}, when both boundaries are non-oscillatory, the spectrum of the SL problem is purely discrete. Thus, there is an infinite sequence of simple eigenvalues. The eigenfunction corresponding to the lowest eigenvalue is the principal eigenfunction we are interested in (the {\em ground state}).

(2) When one of the boundaries is non-oscillatory and the other boundary is oscillatory with a positive cutoff $\Lambda$ {\em and} the SL equation is oscillatory at $\lambda=\Lambda$, by Proposition  \ref{spectral_class} there is an infinite  sequence of simple eigenvalues in $[0,\Lambda)$ clustering at $\Lambda$. Again, we are interested in the principal eigenfunction corresponding to the lowest eigenvalue.

(3) When both boundaries are non-oscillatory with cutoffs $\Lambda_l>0$ and $\Lambda_r>0$ and the SL equation is oscillatory at $\underline{\Lambda}=\min\{\Lambda_l,\Lambda_r\}>0$, according to Proposition  \ref{spectral_class} there is an infinite sequence of simple eigenvalues in $[0,\underline{\Lambda})$ clustering at $\underline{\Lambda}$. This shows the three sufficient conditions (1)-(3) in Theorem \ref{1d_sufficient2}. Finally, we note that in the cases where the cutoffs are non-oscillatory, the spectral classification in Proposition \ref{spectral_class} states that there is a finite sequence of discrete eigenvalues in $[0,\Lambda]$, but it may be empty. Thus, we cannot conclude whether or not there is a principal eigenfunction in $L^2(I,m)$ in those cases.

Under each of the three sufficient conditions, there is a spectral gap between the $L^2(I,m)$-principal eigenvalue $\lambda$ and the infimum of the spectrum of the non-negative self-adjoint SL operator $-{\cal A}$ that lies above $\lambda$.
$\Box$\\
\\
Next we give easy to verify explicit sufficient conditions for the boundaries to be non-oscillatory or to be oscillatory with positive cutoff $\Lambda>0$ such that the SL equation is oscillatory at $\lambda=\Lambda$, as those cases give sufficient conditions in Theorem \ref{1d_sufficient} for the existence of a positive principal eigenfunction. We start with the following result (cf. \citet{linetsky_2008}, p.236).
\begin{proposition}
Entrance or instantaneously reflecting boundaries are non-oscillatory.
\end{proposition}
In contrast, natural boundaries can be either non-oscillatory or oscillatory with cutoff $\Lambda\geq 0$.
To determine when a natural boundary is non-oscillatory or oscillatory with cutoff $\Lambda$, it is convenient to transform the SL equation to the {\em Liouville normal form} (cf. \citet{everitt_2005}, p.280 or \citet{linetsky_2008}). In order to do this transform, we further assume that $\mu$ is once continuously differentiable and $\sigma$ is twice continuously differentiable on $(l,r)$.
Fix some $x_0\in (l,r)$ and consider a mapping $g: (l,r)\rightarrow (g(l),g(r))$: $g(x):=\int_{x_0}^x dz/\sigma(z).$
Since $\sigma(x)>0$ on $(l,r)$ (this follows from $m(x)>0$ on $(l,r)$), $g(x)$ is strictly increasing on $(l,r)$. Let $g^{-1}$ denote its inverse.
Now we
transform the independent and dependent variables in the SL equation as follows:
\begin{equation}
y=g(x)=\int_{x_0}^x \frac{dz}{\sigma(z)},\,\,\,
v(y)=\left\{\left.\frac{u(x)}{\sqrt{\sigma(x)s(x)}}\right\}\right|_{x=g^{-1}(y)},
\end{equation}
where $s(x)$ is the scale density.
Then the function $v(y)$ satisfies the SL equation in the {\em Liouville normal form}:
\be
-\frac{1}{2}v''(y)+Q(y)v(y)=\lambda v(y),\,\,\,y\in(g(l),g(r)),
\eel{Shrodinger}
where the {\em potential function} $Q(y)$ is given by
\be
Q(y)=U(g^{-1}(y)),\quad
U(x):= \frac{1}{8}(\sigma'(x))^2-\frac{1}{4}\sigma(x)\sigma''(x)+\frac{\mu^2(x)}{2\sigma^2(x)}+\frac{1}{2}\mu'(x)-\frac{\mu(x)\sigma'(x)}{\sigma(x)}+r(x).
\eel{Potential}
This transformation of the dependent and independent variables is called the {\em Liouville transformation} in the Sturm-Liouville theory. It reduces the SL  equation \eqref{SLeq} to the Liouville normal form \eqref{Shrodinger}. The SL equation in the Liouville normal form has the form of the celebrated (stationary) {\em one-dimensional Schr\"{o}dinger equation}.

The oscillatory/non-oscillatory classification of boundaries of the SL equation remains invariant under the Liouville transformation, i.e., the SL equation \eqref{SLeq} is non-oscillatory at an end-point $e\in \{l,r\}$ for a particular $\lambda$ if and only if the Schr\"{o}dinger equation \eqref{Shrodinger} is non-oscillatory at $g(e)$ for that $\lambda$. The oscillatory/non-oscillatory classification of the Schr\"{o}dinger equation depends on the behavior of the potential function $Q$ near the endpoints. We have the following classification result (\citet{linetsky_2004}).

\begin{proposition}
\label{natural_class}
{\bf (Oscillatory/Non-Oscillatory Classification of Natural Boundaries)}\\
Suppose $e\in \{l,r\}$ is a natural boundary, $U(x)$ is defined in Eq.\eqref{Potential}, and the limit $\lim_{x\rightarrow e}U(x)$
exists (it is allowed to be infinite).\\
(i) If $e$ is transformed into a finite endpoint by the Liouville transformation, i.e., $g(e)=\int_{x_0}^e \frac{dz}{\sigma(z)}$ is finite, then $e$ is non-oscillatory.\\
(ii) Suppose $e$ is transformed into $-\infty$ or $+\infty$ by the Liouville transformation.
If $\lim_{x\rightarrow e}U(x)=+\infty$, then $e$ is non-oscillatory. If
$\lim_{x\rightarrow e}U(x)=\Lambda$ for some finite $\Lambda$,
then $e$ is oscillatory with cutoff $\Lambda$. Since the operator $-{\cal A}$ is non-negative, it follows that $\Lambda\geq 0$. If $\Lambda>0$ and $\lim_{x\rightarrow e}g^2(x)(U(x)-\Lambda)>-1/4$, then $e$ is non-oscillatory for $\lambda=\Lambda >0$. If $\Lambda>0$ and $\lim_{x\rightarrow e}g^2(x)(U(x)-\Lambda)<-1/4$, then $e$ is oscillatory for $\lambda=\Lambda >0$. If $\Lambda=0$, $e$ is always non-oscillatory for $\lambda=\Lambda=0$.
\end{proposition}

Proposition \ref{natural_class} gives explicit oscillatory/non-oscillatory classification of natural boundaries in terms of the asymptotic behavior of $\sigma$, $\mu$ and $r$ near the boundary. Combined with Theorem \ref{1d_sufficient}, Proposition \ref{spectral_class} and Theorem \ref{1d_sufficient2}, it gives explicit sufficient conditions for the existence of a positive $L^2(I,m)$-principal eigenfunction corresponding to a principal eigenvalue with a spectral gap above it for the pricing operator in a risk-neutral asset pricing model where $X$ is a one-dimensional diffusion and $r(x)$ is a short rate under assumptions in this section. We stress that these conditions are merely sufficient. First, in the oscillatory with cutoff $\Lambda$ natural boundary case with non-oscillatory $\Lambda$, a principal eigenvalue may exist in $[0,\Lambda]$. Unfortunately we do not have an explicit sufficient condition for the existence of an eigenvalue in this case. Such cases have to be checked case by case.
Moreover, if the principal eigenvalue $\lambda$ {\em does} exist in $[0,\Lambda]$ and it is not equal to $\Lambda$, then there is a spectral gap between $\lambda$ and the portion of the spectrum above $\lambda$. The eigen-measure ${\mathbb Q}^\pi$ corresponding to this eigenvalue is recurrent (the proof is similar to the proof of  Theorem \ref{1d_sufficient}).
Finally, even if the SL equation possesses {\em no}  $L^2(I,m)$-eigenfunctions, it is possible that the pricing operator ${\mathscr P}_t$ still possesses a positive eigenfunction {\em outside} of $L^2(I,m)$.

\subsection{Proof of Theorem \ref{exist_3_main}}
\label{proof_mD}


We first observe that under our assumption, the diffusion has a positive density with respect to Lebesgue measure. Thus, it is irreducible and satisfies Assumption  \ref{ac_condition} with with respect to Lebesgue measure. Also under our assumptions the {\em generalized principal eigenvalue} (cf.  \citet{pinsky_1995}, p.147 for the definition) $\lambda_c({\mathbb R}^d)$ of the operator ${\cal G}-r(x)$ on ${\mathbb R}^d$ is finite and satisfies $\lambda_c({\mathbb R}^d)\leq -r_0$, where $r_0:=\inf\{r(x):x\in {\mathbb R}^d\}$ ($r_0>-\infty$ by our assumptions on $r(x)$).
This follows from Theorems 3.2 and 3.3 of  \citet{pinsky_1995}, pp.146-8.

We next prove that under our assumptions on $r(x)$ the operator ${\cal G}-r(x)-\lambda_c({\mathbb R}^d)$ is {\em critical} (cf. \citet{pinsky_1995}, p.145 for the definition).
Let $(D_n)_{n\geq 1}$ be the exhausting domain sequence such that $r_n\rightarrow \infty$ and consider the operator ${\cal G}-r(x)$ defined on domains $\mathbb{R}^d-D_n$ with the vanishing Dirichlet boundary conditions on the boundary $\partial D_n$. By Theorems 3.2 and 3.3 of  \citet{pinsky_1995}, pp.146-8, the corresponding generalized principal eigenvalues are finite and satisfy $\lambda_c({\mathbb R}^d-D_n)\leq -r_n$. Since $r_n\rightarrow \infty$, $\lambda_{c,\infty}({\mathbb R}^d):=\inf\{\lambda_c(D^\prime): D^\prime\quad \text{a domain satisfying} \quad {\mathbb R}^d - D^\prime \subset \subset {\mathbb R}^d\}=-\infty$ (cf. \citet{pinsky_1995}, p.176 for the definition of this quantity). Thus, $-\infty= \lambda_{c,\infty}({\mathbb R}^d)<\lambda_{c}({\mathbb R}^d)\leq -r_0<\infty$. Therefore, the operator ${\cal G}-r(x)-\lambda_c({\mathbb R}^d)$ is critical by Theorem 7.2, p.176 of \citet{pinsky_1995}.

It then follows that the operator ${\cal G}-r(x)-\lambda_c({\mathbb R}^d)$ on ${\mathbb R}^d$ possesses a unique {\em positive harmonic function} $\pi(x)\in C^{2,\alpha}({\mathbb R}^d)$, i.e. $({\cal G}-r(x)-\lambda_c({\mathbb R}^d))\pi(x)=0$ and $\pi(x)>0$ for all $x\in {\mathbb R}^d$ (cf. \citet{pinsky_1995}, p.148 Theorem 3.4).  Here $C^{2,\alpha}({\mathbb R}^d)$ is the space of functions having continuous second derivatives with all their partial derivatives up to the second order H\"{o}lder continuous  with exponent $\alpha$ on ${\mathbb R}^d$.

We can associate a positive ${\mathbb Q}$-local martingale $e^{-\int_0^t (\lambda_c({\mathbb R}^d)+r(X_s))ds} \frac{\pi(X_t)}{\pi(x)}$ with this positive harmonic function. We need to ensure that this process is, in fact, a martingale. To this end, it is sufficient to show that
$$
{\mathbb E}^\mathbb{Q}_x[e^{-\int_0^t(\lambda_c({\mathbb R}^d)+r(X_s))ds} \pi(X_t)]=\pi(x)
$$
for all $x\in {\mathbb R}^d$ and all $t>0$, i.e., $\pi(x)$ is a positive {\em invariant function} of the semigroup generated by ${\cal G}-r(x)-\lambda_c({\mathbb R}^d)$. Under our assumptions, this follows from Theorem 8.6 of \citet{pinsky_1995}, p.182.
The corresponding eigen-measure
$({\mathbb Q}_x^\pi)_{x\in \mathbb{R}^d}$ solves the martingale problem for the $h$-{\em transform} ${\cal G}^\pi$ of the operator ${\cal G}-r(x)-\lambda_c({\mathbb R}^d)$ with $h=\pi$ (cf. \citet{pinsky_1995}, p.126).

Finally, we need to show that $X$ is recurrent under ${\mathbb Q}^\pi$. By Proposition \ref{relation_recu}, we only need to prove $\bf{(R3)}$. By Theorem 2.1 of \citet{pinsky_1995} on p.130, $X$ is recurrent $\bf{(R3)}$ under ${\mathbb Q}^\pi$ if and only if the operator ${\cal G}^\pi$ does not possess Green's measure on ${\mathbb R}^d$, i.e. ${\mathbb E}_x^{{\mathbb Q}^\pi}[\int_0^\infty {\bf 1}_D(X_s)ds]=\infty$ for all $x\in {\mathbb R}^d$ and all open set $D\subset {\mathbb R}^d$ .  Since ${\cal G}^\pi$ is the $h$-transform of ${\cal G}-r(x)-\lambda_c({\mathbb R}^d)$, by Proposition 2.2 of \citet{pinsky_1995}, p.133 ${\cal G}^\pi$ does not possess the Green's measure if and only if ${\cal G}-r(x)-\lambda_c({\mathbb R}^d)$ does not possess the Green's measure. However, we have already proved that ${\cal G}-r(x)-\lambda_c({\mathbb R}^d)$ is critical. Hence, by definition of criticality (cf. \citet{pinsky_1995}, pp.145-6) it does not possess the Green's measure. This completes the proof.
$\Box$

\section{Complements on 1D Diffusion Examples}
\subsection{Complements on the CIR Model}
\label{appendix_cir}
Consider the CIR SDE \eqref{SDE1_meanCIR}. The drift is Lipschitz and the volatility $\sigma(x)=\sigma \sqrt{x}$ satisfies the Yamada-Watanabe condition, so the SDE has a unique strong solution for any $x\geq 0$.
Since in the degenerate case with $a=0$ and $x=0$ the solution is $X_t=0$ for all $t\geq 0$, by the comparison theorem for one-dimensional SDEs the solution for $a>0$ and $x\geq 0$ stays non-negative, $X_t\geq 0$ for all $t\geq 0$.
When the Feller condition is satisfied,
$2a \geq \sigma^2$,
the solution stays strictly positive when started from any $x>0$, i.e. $T_0=\infty$ a.s., where $T_0$ is the first hitting time of zero. It can also be started from $x=0$, in which case it instantaneously enters the interval $(0,\infty)$ and stays strictly positive for all $t>0$. 
We take the state space to be $I=(0,\infty)$ in this case and do not include zero in $I$ (we consider only positive starting values in this case, $x>0$).
When the Feller condition is not satisfied, $0< 2a < \sigma^2$,
the solution can reach zero in finite time when started from $x>0$, and zero is an instantaneously reflecting boundary.
In this case the state space  is the interval $I=[0,\infty)$ since zero can be reached from the interior and is included in the state space. The scale and speed densities on $I$ are:
\begin{equation}
\label{CIR_speed}
s(x)=x^{-\beta}e^{-\frac{2b x}{\sigma^2}}, \quad\quad m(x)=\frac{2}{\sigma^2}x^{\beta-1}e^{\frac{2b x}{\sigma^2}},\quad\text{where}\quad \beta:=\frac{2a}{\sigma^2}.
\end{equation}
When $b<0$, the process is mean-reverting with mean-reversion rate $\kappa:=-b$. It is positive recurrent with the stationary measure  with the Gamma density $C^{-1}m(x)$ on $I$ with $C=\frac{1}{\kappa}(\frac{\sigma^2}{2\kappa})^{\beta-1} \Gamma(\beta)$ normalizing the speed density to be a probability density.
When $b>0$, the drift is linearly increasing in the state variable, the process is non-mean-reverting and is transient.
When $b=0$, the process reduces to a squared Bessel process (namely, $\frac{4}{\sigma^2}X$ is  BESQ$^{(\nu)}$ with index $\nu=\beta-1$,  cf. \citet{jeanblanc_2009}). It is transient if $\beta>1$ and recurrent if $\beta\leq 1$
(this can be established by examining convergence of the integral $\int_0^\infty p(t,x,y)dt$ with the transition density of the squared Bessel process).

The symmetric, continuous positive density of the CIR pricing semigroup $\mathscr{P}_tf(x)=\mathbb{E}_x^{\mathbb Q}[e^{-\int_0^t X_s ds}f(X_t)]$ with respect to the speed measure $m(x)dx$ is known in closed form (for $b<0$ this density appeared in \citet{cox_1985}; the same expression holds for $b\geq 0$):
\be
p(t,x,y)=\frac{\gamma e^{-\lambda_0 t-\frac{b}{\sigma^2}(x+y)}}{(1-e^{-\gamma t})}\Big(\frac{1}{xy e^{-\gamma t}}\Big)^{\frac{\beta-1}{2}}\exp\Big(-\frac{\gamma(1+e^{-\gamma t})}{\sigma^2(1-e^{-\gamma t})}(x+y)\Big)I_{\beta-1}\Big(\frac{4\gamma\sqrt{xy e^{-\gamma t}}}{\sigma^2(1-e^{-\gamma t})}\Big),
\ee
where
\be
\gamma:=\sqrt{b^2+2\sigma^2},\quad \lambda_0:=\frac{a}{\sigma^2}(\gamma+b),
\ee
and
$I_\alpha(x)$ is the modified Bessel function.
Applying the Hille-Hardy formula to expand the Bessel function into the bilinear series of generalized Laguerre polynomials (cf. Section 6 of \citet{rafael_2013}) yields the bilinear eigenfunction expansion of the density
$$
p(t,x,y)=\sum_{n=0}^\infty e^{-\lambda_n t}\varphi_n(x)\varphi_n(y)
$$
with the eigenvalues and eigenfunctions
 (cf. \citet{davydov_2003}, \citet{gorovoi_2004}, or \citet{rafael_2013}; here we label the eigenvalues starting from zero):
\be
\lambda_n=\gamma n+\frac{a}{\sigma^2}(\gamma+b),\quad
\varphi_n(x)=\Big[\Big(\frac{2}{\sigma^2}\Big)^{\beta-1}\frac{\gamma^\beta n!}{\Gamma(\beta+n)}\Big]^{\frac{1}{2}}e^{-\frac{(b+\gamma)x}{\sigma^2}}L_n^{(\beta-1)}\left(\frac{2\gamma}{\sigma^2}x\right),\quad n=0,1,\ldots.
\ee
The eigenfunctions are continuous, bounded, and form an orthonormal system in $L^2(I,m)$.
The eigenvalues satisfy the trace class condition in Section \ref{exist_1_1}, and Eq.\eqref{exist_2_assumption1} is automatically satisfied in this case. We now verify Eq.\eqref{exist_2_assumption2}. Since $p(t,x,x)$ is continuous, by \eqref{l2assumption_equ} we only need to show that it remains bounded at the boundaries at zero and infinity. Using the asymptotics of the Bessel function $I_\alpha(x)=O(x^\alpha)$ as  $x\rightarrow 0$ and $I_\alpha(x)=O(x^{-\frac{1}{2}}e^x)$ as $x\rightarrow\infty$, we have $p(t,x,x)=O(1)$ as $x\rightarrow0$ and
$p(t,x,x)=O\Big(x^{-\beta+\frac{1}{2}}e^{-\frac{2f(t)}{\sigma^2}x}\Big)$ as  $x\rightarrow\infty$, where
$f(t)=b+\gamma\frac{1-e^{-\frac{\gamma t}{2}}}{1+e^{-\frac{\gamma t}{2}}}$. Thus, for $b\geq 0$ (non-mean-reverting case) Eq.\eqref{exist_2_assumption2} is verified for all $t>0$.
For $b=-\kappa<0$ (mean-reverting case), Eq.\eqref{exist_2_assumption2} is verified for all $t\geq T$ with $T=\frac{2}{\gamma} \ln\left(\frac{\gamma + \kappa}{ \gamma - \kappa}\right)$.
Therefore the CIR model satisfies Assumption \ref{L2assumption} for all $b\in {\mathbb R}$, Theorem \ref{exist_theorem_1} applies, and the unique recurrent eigenfunction is the $L^2(I,m)$-principal eigenfunction
\be
\pi_R(x)=\varphi_0(x)=
e^{-\frac{b+\gamma}{\sigma^2}x}.
\ee
The state variable $X$ follows a mean-reverting CIR diffusion under the measure $\mathbb{Q}^{\pi_R}$:
\be
dX_t=(a-\gamma X_t)dt+\sigma\sqrt{X_t}dB_t^{\mathbb{Q}^{\pi_R}},
\ee
where $\gamma=\sqrt{b^2+2\sigma^2}$ and $B_t^{\mathbb{Q}^{\pi_R}}=B_t^\mathbb{Q}+\frac{\gamma+b}{\sigma}\int_0^t \sqrt{X_s}ds$ is a standard Brownian motion under $\mathbb{Q}^{\pi_R}$. Clearly $X$ is recurrent under $\mathbb{Q}^{\pi_R}$.\\
\\
{\bf Proof of Theorem \ref{positive_CIR}.} To verify martingality, we need to verify that
$\pi(x)$ is an invariant function of the semigroup $({\mathscr P}_{t}^\lambda:=e^{\lambda t}{\mathscr P}_{t})_{t\geq 0}$:
\be
{\mathscr P}_{t}^\lambda \pi(x)=e^{\lambda t}\int_0^{+\infty}\pi(y)p(t,x,y)m(y)dy=\pi(x),
\eel{martingale_meanCIR}
where $p(t,x,y)$ is the density of the CIR pricing semigroup with respect to the speed measure.
To verify this, we use  Theorem 5.1.8 of \citet{pinsky_1995}
 that  gives the necessary and sufficient conditions for a positive harmonic function of the second-order differential operator on an open interval to also be a positive invariant function of the semigroup generated by this operator. We first state Pinsky's result.
\begin{theorem} (Theorem 5.1.8 of \citet{pinsky_1995})
Let the second-order differential operator
\be
A=\frac{a(x)}{2}\frac{d^2}{dx^2}+b(x)\frac{d}{dx}+V(x)
\ee
satisfy $a,b,V\in C_\text{loc}^{0,\alpha}$ and $a>0$ on $I=(\alpha, \beta),$ where $-\infty\leq\alpha<\beta\leq+\infty$. Let $x_0\in(\alpha, \beta)$. Then a positive harmonic function $\phi$, $A\phi=0$, is an invariant function of the semigroup generated by $A$ if and only if the following two conditions hold:
\be
\int_{\alpha}^{x_0}\frac{dx}{\phi^2(x)}\exp\Big(-\int_{x_0}^{x}\frac{2b}{a}(z)dz\Big)\int_x^{x_0}dy\frac{\phi^2(y)}{a(y)}\exp\Big(\int_{x_0}^y\frac{2b}{a}(z)dz\Big)=\infty,
\eel{cond1_meanCIR}
\be
\int_{x_0}^{\beta}\frac{dx}{\phi^2(x)}\exp\Big(-\int_{x_0}^{x}\frac{2b}{a}(z)dz\Big)\int_{x_0}^x dy\frac{\phi^2(y)}{a(y)}\exp\Big(\int_{x_0}^y\frac{2b}{a}(z)dz\Big)=\infty.
\eel{cond2_meanCIR}
\label{key_theorem_meanCIR}
\end{theorem}

In our case, $a(x)=\sigma^2 x$, $b(x)=a-\kappa x$, $V=-x+\lambda$. Explicitly analyzing the asymptotic behavior of solutions $\psi_\lambda(x)$ and $\phi_\lambda(x)$ establishes that only the solution $\psi_\lambda(x)$ leads to a martingale, so we must have $C_2=0$. The solution $\phi_\lambda(x)$ leads to strict local martingales. We omit the relevant calculations to save space.
Next we check positivity.  The positive zeros of Kummer and Tricomi confluent hypergeometric functions are known and the result is as follows (it can be found in \citet{erdelyi_1953}, p.289):
$\psi_\lambda(x)>0$ for all $x>0$ if and only if $\alpha\geq 0$ (hence, $\lambda\leq \lambda_0$).
Thus, we arrive at the result that $\pi(x)=C_1\psi_\lambda(x)+C_2\phi_\lambda(x)$ is a positive invariant function of the semigroup $({\mathscr P}_t^\lambda)_{t\geq 0}$ if and only if $C_1>0$, $C_2=0$ and $\lambda\leq \lambda_0$. $\Box$\\


\subsection{The 3/2 Model}

Consider the 3/2 model where the short rate solves the SDE under ${\mathbb Q}$:
\be
dX_t=\kappa(\theta-X_t)X_tdt+\sigma X_t^{3/2} dB_t^{\mathbb Q},
\ee
with $\kappa,\theta,\sigma>0$. The solution stays strictly positive for all positive parameter values, and is recurrent with a stationary density equal to the normalized speed density $m(x)=x^{-2\alpha-1}e^{-\frac{\beta}{x}}$, where $\alpha:=\frac{k}{\sigma^2}+1$ and $\beta:=\frac{2\kappa\theta}{\sigma^2}$. This interest rate model is studied in \citet{ahn_1999}. Applying It\^{o} formula it can be shown that this model is the reciprocal of the CIR model in the sense that it can be written in the form $X_t=1/Y_t$, where $Y_t$ follows a CIR process.
Similar to the CIR model, the pricing semigroup has purely discrete spectrum and is trace class (cf. \citet{linetsky_2004}, Section 6.3.3 for details). However, $p(t,x,x)$ is unbounded as $x\downarrow 0$, so the condition (iii) in Assumption \ref{L2assumption} is not satisfied for any $t>0$. Nevertheless, the condition (ii) in Theorem \ref{1d_sufficient} is satisfied (see also Theorem \ref{1d_sufficient2}), and there exists a recurrent eigenfunction.
To determine it explicitly, consider the SL ODE:
\be
\frac{1}{2}\sigma^2 x^3\pi^{\prime\prime}+\kappa(\theta-x)x\pi^{\prime} - x\pi=-\lambda \pi.
\ee
It can be easily checked that $\pi_0(x)= x^{\alpha-\mu-1/2}$ is the positive eigenfunction with the eigenvalue $\lambda_0=\kappa \theta(\mu-\alpha+1/2)$, where $\mu:=\sqrt{\left(\frac{\kappa}{\sigma^2}+\frac{1}{2}\right)^2+\frac{2}{\sigma^2}}$.
By Theorem \ref{key_theorem_meanCIR}, it is easy to verify that $\pi_0$ is also an eigenfunction of the pricing semigroup.
This can also be seen from the fact that $\pi_0(x)$ is square-integrable with the speed density and is, in fact, the principal eigenfunction of the generator of the pricing semigroup in $L^2((0,\infty),m)$ (cf. \citet{linetsky_2004}, Section 6.3.3 for details) and, hence, the eigenfunction of the pricing operator with eigenvalue $e^{-\lambda_0 t}$.  Therefore, it defines a measure $\mathbb{Q}^{\pi_0}$. By Girsanov's theorem, it is easy to directly verify that  under $\mathbb{Q}^{\pi_0}$:
\be
\begin{array}{rl}
d X_t=\tilde{\kappa}(\tilde{\theta} -X_t)X_t dt+\sigma X_t^{3/2}dB_t^{\mathbb{Q}^{\pi_0}}
\end{array}
\ee
with $\tilde{\kappa}=(\mu-1/2)\sigma^2$ and $\tilde{\theta}=\kappa \theta/\tilde{\kappa}$.
Since $\mu-1/2>0$, this is again a mean-reverting 3/2 model and is positive recurrent. Thus $\pi_0$ is the unique recurrent eigenfunction.

\subsection{Vasicek Model}
\label{appendix_ou}

Consider an OU process under ${\mathbb Q}$ solving the SDE
$dX_t=\kappa(\theta- X_t)dt+\sigma dB_t^{\mathbb Q}$
with $\theta,\kappa\in {\mathbb R}$, $\kappa\neq0$, $\sigma>0$ and $r_t=X_t$.
Solutions of the OU Sturm-Liouville ODE
\be
\frac{1}{2}\sigma^2 \pi^{\prime\prime}+\kappa(\theta- x)\pi^\prime-x\pi=-\lambda\pi,
\eel{ODE_OU}
can be expressed in terms of Weber parabolic cylinder functions. We treat both cases $\kappa>0$ and $\kappa<0$ together.
\begin{proposition}
Define:
$\alpha:=\sigma\sqrt{\frac{2}{|\kappa|^3}},$ $z:=\frac{\sqrt{2|\kappa|}}{\sigma}(\theta-x),$ $\varepsilon:=\rm{sign}(\kappa)$, and $\mu:=\frac{1}{\kappa}(\lambda-\theta+\frac{\sigma^2}{2\kappa^2})$ if $\kappa>0$ and $\mu:=\frac{1}{|\kappa|}(\lambda-\theta+\frac{\sigma^2}{2\kappa^2}+\kappa)$ if $\kappa<0$.
If $\mu$ is not a non-negative integer, which means $\lambda$ is not an $L^2({\mathbb R},m)$-eigenvalue of ${\mathcal A}$, then two linearly independent solution of Eq.\eqref{ODE_OU} are :
\begin{equation}
\psi_{\lambda}(x)=e^{\varepsilon \frac{z^2}{4}}D_\mu(z-\alpha),\quad
\phi_{\lambda}(x)=e^{\varepsilon \frac{z^2}{4}}D_\mu(\alpha-z),
\end{equation}
where $D_\mu(z)$ is the Weber parabolic cylinder function (these solutions are chosen to satisfy the square-integrablity with the speed measure $m$ on $(-\infty,0]$ and on $[0,\infty)$, respectively).

If $\mu$ is a non-negative integer, then the Weber functions reduce to Hermite polynomials and the two solutions given above become linearly dependent and reduce to $L^2({\mathbb R},m)$ eigenfunctions of the pricing semigroup. One solution can still be taken in the form
$\psi_{\lambda}(x)=e^{\varepsilon \frac{z^2}{4}}D_\mu(z-\alpha)$.
The other linearly independent solution differs in different cases (we omit explicit expressions to save space).
\label{prop_2_OU}
\end{proposition}
Similar to our analysis of the CIR model, we now apply Theorem \ref{key_theorem_meanCIR} to establish which of the local martingales $\tilde{M}^\pi$ \eqref{M} are positive martingales.
\begin{theorem}
When $\kappa>0$ or $\kappa<0$, the local martingale $\tilde{M}^\pi$ \eqref{M} is a positive martingale if and only if $\mu<0$ and $C_1,C_2\geq0$ with $C_1C_2\neq0$,  or $\mu=0$ and $C_1>0$, $C_2=0$.
\end{theorem}
The proof strategy is similar to the  CIR in Appendix F.1. It is based on the application of Theorem \ref{key_theorem_meanCIR} with $a(x)=\sigma^2$, $b(x)=\kappa(\theta-x)$, $V(x)=-x+\lambda$ on ${\mathbb R}$, where the integral conditions are verified by considering the asymptotics of solutions at $\pm\infty$, together with the verification of the positiveness of solutions in turn based on the asymptotics and zeros of the Weber parabolic functions (cf.  \citet{erdelyi_1953_2}, p.122-123, \citet{erdelyi_1953}, p.262,  \citet{erdelyi_1953_2}, p.126). We omit details to save space.

When $\kappa>0$, the solution with $\mu=0$ reduces to $\pi_0(x)=e^{-x/\kappa}$ due to the reduction of the Weber function in this case. It is easy to check that it is square-integrable with the speed density $m(x)$.
Girsanov's theorem immediately implies that $X$ solves
\be
dX_t=(\kappa\theta-\frac{\sigma^2}{\kappa}-\kappa X_t)dt+\sigma dB_t^{\mathbb{Q}^{\pi_0} }
\ee
under $\mathbb{Q}^{\pi_0}$.
Thus, $X$ is again a positively recurrent, mean-reverting OU process, but with lower drift. Thus, $\pi_0$ is the unique recurrent eigenfunction.

When $\kappa<0$, the solution with $\mu=0$ reduces to $\displaystyle{\psi_\lambda(x)=  Ce^{\frac{\kappa}{\sigma^2}x^2+(\frac{1}{\kappa}-\frac{2\kappa\theta}{\sigma^2})x}}$ ($D_0(x)=e^{-x^2/4}$). It is easy to check that it is square-integrable with the speed density $m(x)$ when $\kappa<0$
and is, thus, a positive $L^2({\mathbb R},m)$ eigenfunction of the pricing operator ${\mathscr P}_t$ with the eigenvalue $e^{-\lambda_0 t}$ with $\lambda_0=\theta-\kappa-\frac{\sigma^2}{2\kappa^2}$. Girsanov's theorem immediately implies that $X$ solves
\be
dX_t=(\frac{\sigma^2}{\kappa}-\kappa\theta+\kappa X_t)dt+\sigma dB_t^{\mathbb{Q}^{\pi_0} }
\ee
under $\mathbb{Q}^{\pi_0}$. $X$ is a positive recurrent, mean-reverting OU process under ${\mathbb Q}^{\pi_0}$.
Thus, we see that if the interest rate follows an OU process with mean-repelling drift under the risk-neutral measure, there still exists a unique recurrent eigenfunction. This is similar to what we have observed in the CIR model with $b>0$.

Next we consider positive eigenfunctions corresponding to the solutions $\pi(x)=C_1 \psi_\lambda(x)+C_2\phi_\lambda(x)$ with $\mu<0$, $C_1, C_2\geq0$ with $C_1C_2\neq 0$. Consider the case with $\kappa>0$. By \eqref{1dgpi0}, under $\mathbb{Q}^\pi$ $X$ solves:
$$
dX_t=\left(\kappa\theta-\frac{\sigma^2}{\kappa}-\kappa X_t+ \sqrt{2\kappa}\mu\sigma\frac{C_2D_{\mu-1}(\alpha-z)-C_1D_{\mu-1}(z-\alpha)}{C_2D_\mu(\alpha-z)+C_1D_\mu(z-\alpha)}\right)dt+\sigma dB_t^{\mathbb{Q}^{\pi} }
$$
(the expression in the drift follows from  the fact that
$\frac{d}{dz}(e^{\frac{1}{4}z^2}D_\mu(z))=\mu e^{\frac{1}{4}z^2}D_{\mu-1}(z)$).
Using the asymptotic behavior of the Weber parabolic cylinder function, we obtain that for $C_1C_2\neq 0$
\begin{center}
$\displaystyle{\sqrt{2\kappa}\mu\sigma\frac{C_2D_{\mu-1}(\alpha-z)-C_1D_{\mu-1}(z-\alpha)}{C_2D_\mu(\alpha-z)+C_1D_\mu(z-\alpha)}}\rightarrow 2\kappa x(1+o(1))$ \quad as $|x| \rightarrow \infty$.
\end{center}
Thus, we observe that $X$ is no longer mean-reverting, but mean-repelling.
In particular, consider a special case with
$\lambda=-\frac{\sigma^2}{2\kappa}+\theta-\kappa,$ $\pi(x)=e^{\frac{\kappa}{\sigma^2}x^2+(\frac{1}{\kappa}-\frac{2\kappa\theta}{\sigma^2})x}$.
By Girsanov theorem, we immediately see that  $X$ follows under ${\mathbb{Q}^{\pi} }$
$$
dX_t=(-\kappa\theta+\frac{\sigma^2}{\kappa}+\kappa X_t)dt+\sigma dB_t^{\mathbb{Q}^{\pi} }.
$$
The coefficient in front of $X_t$ in the drift is now $\kappa>0$. Thus, $X$ is mean-repelling. Analysis of solutions with $\mu<0$ in the case when $\kappa<0$ is similar. $X$ is mean-repelling under ${\mathbb{Q}^{\pi}}$.

\subsection{Merton's Model with Brownian Short Rate}
\label{BMin5}

Let $X_t=x+at+\sigma B_t^{\mathbb Q}$ be a Brownian motion with drift
$a\in {\mathbb R}$ and volatility $\sigma>0$ and consider the Brownian short rate $r_t=X_t$.
This is the historically earliest continuous-time stochastic model of the term structure of interest rates first considered by \citet{merton_1973}. This model can be viewed as a degenerate case of the OU model with $b=0$ ($\kappa=0$).
The SL ODE reads
\be
\frac{1}{2}\sigma^2 \pi^{\prime\prime}+a\pi^{\prime} - x\pi=-\lambda \pi.
\eel{ODE_degOU}
In this case the Weber parabolic cylinder functions reduce to the Airy functions.
Define $\alpha:=\Big(\displaystyle{\frac{2}{\sigma^2}}\Big)^{1/3}(\lambda-\displaystyle{\frac{a^2}{2\sigma^2}}),\enskip z=\Big(\frac{2}{\sigma^2}\Big)^{1/3}x$.
Two linearly independent solution of Eq.\eqref{ODE_degOU} are:
\begin{equation}
\label{BM_solution}
\psi_{\lambda}(x)=e^{-a(\frac{1}{2\sigma^4})^{1/3}z}Ai(z-\alpha),\quad
\phi_{\lambda}(x)=e^{-a(\frac{1}{2\sigma^4})^{1/3}z}Bi(z-\alpha).
\end{equation}
Since the Airy functions $Ai(z)$ and $Bi(z)$ both have infinitely many zero on the negative half-line, by Sturm's separation theorem, any linear combination of $Ai(z)$ and $Bi(z)$ has infinitely many zero for $z<0$. Thus, for any $\lambda$ there is no positive eigenfunction.


\subsection{Merton's Short Rate Model with Quadratic Drift}

Consider the SDE
\be
dX_t=\kappa(\theta-X_t)X_t dt+\sigma X_t d B^{\mathbb{Q}}_t
\ee
with $\kappa,\theta,\sigma>0$. This process has been deduced by \citet{merton_1975} as a model for the short rate from his economic growth model. Define $\beta:=\frac{2\kappa\theta}{\sigma^2}$.
Applying Feller's tests, both the origin and infinity are inaccessible (natural) boundaries.
When $\beta>1$, the speed density is
$m(x)=x^{\beta-2}e^{-2\kappa x/\sigma^2}$ and is integrable on $(0,\infty)$. Thus, the process is positive recurrent with a stationary gamma density. When $\beta=1$, the speed measure is not integrable. However, applying Theorem 5.1.1 of \citet{pinsky_1995}, p.208, we establish that the process is recurrent. Thus, it is necessarily null recurrent. When $\beta<1$, applying Theorem 5.1.1 of \citet{pinsky_1995}, we establish that the process is transient, and the origin is an attracting boundary (the process is asymptotically attracted to the origin with probability one, ${\mathbb Q}_x(\lim_{t\rightarrow \infty}X_t=0)=1$).  \citet{lewis_1998} has obtained closed-form solutions for zero-coupon bonds in the short rate model with $r_t=X_t$.
As shown in \citet{lewis_1998}, the pricing semigroup has some non-empty continuous spectrum and, hence, is not Hilbert-Schmidt and does not satisfy sufficient conditions in Section \ref{exist_1_1}. It also does not satisfy any of the sufficient conditions (1)-(3) in Theorem \ref{1d_sufficient2}. Nevertheless,
as shown in \citet{lewis_1998}, when $\beta>1+\frac{2}{\kappa}$ the SL equation
$$
\frac{1}{2}\sigma^2 x^2 \pi^{\prime\prime}+\kappa(\theta-x)x\pi^{\prime} - x\pi=-\lambda \pi
$$
has a positive $L^2((0,\infty),m)$-eigenfunction
$\pi_0(x)=x^{-1/\kappa}$ with the corresponding eigenvalue $\lambda_0=\theta-\frac{(1+\kappa)\sigma^2}{2\kappa^2}$ and the spectral gap above it. It is easy to check that the eigenfunction is square-integrable with the speed density.
Thus, when $\beta>1+\frac{2}{\kappa}$, by Theorem \ref{1d_sufficient} this is a recurrent eigenfunction. By Girsanov theorem, $X$ follows the process under ${\mathbb Q}^{\pi_0}$:
\be
dX_t=\kappa(\tilde{\theta}- X_t)X_tdt+\sigma X_t dB_t^{{\mathbb Q}^{\pi_0}}
\eel{mertonq_p_sde}
with $\tilde{\theta}=\theta-\frac{\sigma^2}{\kappa^2}$.
Since $\tilde{\beta}:=\frac{2\kappa\tilde{\theta}}{\sigma^2}>1$, $X$ is positive recurrent under $\mathbb{Q}^{\pi_0}$, and we verify that $\pi_0$ is indeed the recurrent eigenfunction.

When $\beta\leq 1+\frac{2}{\kappa}$, $\pi_0(x)=x^{-1/\kappa}$ fails to be square-integrable with the speed density. However, by Theorem \ref{key_theorem_meanCIR}, $\pi_0(x)$ is nevertheless an eigenfunction of the pricing semigroup and, thus, defines a positive martingale and a corresponding eigen-measure. By Girsanov theorem, $X$ follows Eq.\eqref{mertonq_p_sde} under $\mathbb{Q}^{\pi_0}$. It is null recurrent when $\tilde{\beta}=1$ ($\beta= 1+\frac{2}{\kappa}$). Thus, $\pi_0$ is the recurrent eigenfunction.
When $\beta<1+\frac{2}{\kappa}$, $\tilde{\beta}<1$ and $X$ is transient under $\mathbb{Q}^{\pi_0}$ and zero is an attracting boundary. Thus, in this case $\pi_0$ fails to be recurrent (there is no recurrent eigenfunction in this case).



\subsection{One-Dimensional Diffusions on Bounded Intervals with Reflection}

Our last example is a model where the state variable is a
diffusion with drift $\mu(x)$ and volatility $\sigma(x)$ on a finite interval $[l,r]$, and the short rate $r_t=r(X_t)$ is a function of $X$. We assume that $\mu$ and  $\sigma$ are continuous on the closed interval $[l,r]$ and $\sigma(x)>0$ on $[l,r]$. The boundaries at $l$ and $r$ are regular for the diffusion process, and we specify them to be
instantaneously reflecting. $r(x)$ is assumed non-negative and continuous on $[l,r]$.
The generator of the pricing semigroup is
${\cal A}\pi=\frac{1}{2}\sigma^{\prime\prime}(x) \pi^{\prime\prime}+\mu(x)\pi^{\prime} - r(x)\pi$
with Neumann boundary conditions at $l$ and $r$ corresponding to instantaneous reflection,
$\pi^\prime(l)=0$ and $\pi^\prime(r)=0$.
By the regular Sturm-Liouville theory, the spectrum in $L^2([l,r])$ is purely discrete, the pricing semigroup in $L^2([l,r])$ is trace class with the eigenvalues $e^{-\lambda_n t}$ with $\lambda_n$ increasing at the rate proportional to $n^2$, the eigenfunctions $\varphi_n(x)$ are continuous on $[l,r]$,  and the density $p(t,x,y)$ of the pricing semigroup is continuous on $[l,r]\times [l,r]$. In this case Assumptions  5.1 are satisfied, Theorem 5.1 holds,
and there exists a unique recurrent eigenfunction which is the principal $L^2([l,r])$-eigenfunction $\pi(x)=\varphi_0(x)$. All higher eigenfunctions $\varphi_n(x)$ with $n\geq 1$ are not strictly positive on $[l,r]$. Under $\mathbb{Q}^\pi$, $X$ follows a diffusion on $[l,r]$ with drift
$\mu(x)+\sigma^2(x)\pi^\prime(x)/\pi(x)$ and with instantaneous reflection at both boundaries. Thus $X$ is recurrent under $\mathbb{Q}^\pi$, and $\pi$ is the recurrent eigenfunction. Note that the risk premium term in the drift $\sigma^2 \pi^\prime(x)/\pi(x)$ vanishes at the boundaries $l$ and $r$ due to the Neumann boundary conditions for $\pi$. Thus, near the boundaries the process under $\mathbb{Q}^\pi$ behaves like the original process under the risk-neutral measure. However, inside the interval we have recovered a non-trivial risk premium term in the drift.
This is in agreement with the result of \citet{carr_2012} on Ross recovery for 1D diffusions on bounded intervals with regular boundaries (see also \citet{dubynskiy_2013} for more on Ross recovery with reflecting boundaries).

In contrast to the previous diffusion examples in this section leading to singular SL problems, in this case the regular SL equation on $[l,r]$ does not have any non-$L^2([l,r])$ solutions since continuous functions on $[l,r]$ are square-integrable on $[l,r]$. Thus, there are no additional non-$L^2([l,r])$ positive continuous eigenfunctions in this case.

\section{Recurrent Eigenfunction in Affine Diffusion Term Structure Models}
\label{affine_appendix}

The process we work with in this section takes values in the state space $E=\mathbb{R}_+^m\times\mathbb{R}^n$ for some $m,n\geq 0$ with $m+n=d$, where
$\mathbb{R}_+^m=\big\{x\in \mathbb{R}^m : x_i\geq 0$ for $i=1,...,m\big\}$, where it
 solves the following SDE:
\be
d X_t=b(X_t)dt+\rho(X_t)d B^\mathbb{Q}_t,\quad X_0=x,
\eel{affinesde}
where the diffusion matrix $\alpha(x)=\rho(x)\rho(x)^\top$ and the drift $b(x)$ are affine in $x$:
\be
\alpha(x)=a+\displaystyle{\sum_{i=1}^d}x_i\alpha_i,\quad
b(x)=b+\displaystyle{\sum_{i=1}^d}x_i\beta_i=b+Bx
\ee
for some $d\times d$-matrices $a$ and $\alpha_i$ and $d$-dimensional vectors $b$ and $\beta_i$, where we denote by $B=(\beta_1,...,\beta_d)$ the $d\times d$-matrix with $i$-th column vector $\beta_i$, $1\leq i\leq d$. The first $m$ coordinates of $X$ are CIR-type and are non-negative, while the last $n$ coordinates are OU-type.  Define the index sets
$\emph{I}=\{1,...,m\}$  and $\emph{J}=\{m+1,...,m+n\}$.
For any vector $\mu$ and matrix $\nu$, and index sets $\emph{M},\emph{N}\in \{I,J\}$, we denote by
$\mu_\emph{M}=(\mu_i)_{i\in \emph{M}},$ $\nu_{\emph{M}\emph{N}}=(\nu_{ij})_{i\in \emph{M},j\in \emph{N}}$
the respective sub-vector and sub-matrix. Without loss of generality (cf. Theorem 7.2 of \citet{filipovic_2009}), we assume the affine diffusion is in the canonical form in the following sense:
\be
dX_I(t)=(b_I+B_{II}X_I(t)+B_{IJ}X_J(t))dt+\rho_{II}(X(t))dB^\mathbb{Q}_I(t),\quad X_I(0)=x_I,
\eel{canonsde1}
\be
dX_J(t)=(b_J+B_{JI}X_I(t)+B_{JJ}X_J(t))dt+\rho_{JJ}(X(t))dB^\mathbb{Q}_J(t),\quad X_J(0)=x_J,
\eel{canonsde2}
where the matrix $\rho(x)$ is block-diagonal with $\rho_{IJ}(x)\equiv 0$, $\rho_{JI}(x)\equiv 0$, and
\be
\rho_{II}(x)={\rm diag}(\sqrt{x_1},\ldots,\sqrt{x_q},0,\ldots,0),\quad {\rho}_{JJ}(x) {\rho}_{JJ}(x)^\top = a_{JJ}+\sum_{i=1}^mx_i\alpha_{i,JJ}
\eel{canonrho1}
for some integer $0\leq q\leq m$.
To ensure the process stays in the domain $E={\mathbb R}_+^m\times {\mathbb R}^n$ and rule out degeneracy, we make the following assumptions on the coefficients.
\begin{assumption}{\bf (Admissibility and Non-Degeneracy)}\\
(1) $a_{JJ}$ and $\alpha_{i,JJ}$ are symmetric positive semi-definite for all $i=1,2,...,m$,\\
(2) $q=m$ and $a_{JJ}+\sum_{i=1}^m\alpha_{i,JJ}$ is non-singular,\\
(3) $b_I>0$, $B_{IJ}=0$, and $B_{II}$ has non-negative off-diagonal elements.
\label{admi_and_nonde}
\end{assumption}
The non-degeneracy assumption (2) ensures that none of the risk factors are redundant. The positivity of the constant vector $b_I>0$ in the drift of the CIR-type components in assumption (3) ensures that when the process reaches the boundary of the state space $\partial E$, it will return to the interior of the state $\mathring{E}={\mathbb R}_{++}^m\times {\mathbb R}^n$ ($\mathbb{R}^m_{++}:=\big\{x\in \mathbb{R}^m : x_i>0$ for $i=1,...,m\big\}$) and will not reduce to the diffusion on the boundary. This condition is commonly imposed in the literature on affine models (cf.  \citet{glasserman_2008} and \citet{dai_2000}).

For any parameters satisfying  Assumption \ref{admi_and_nonde}, there exists a unique solution to SDE \eqref{canonsde1}-\eqref{canonsde2} taking values in $E$ (cf. Theorem 8.1 of \citet{filipovic_2009}). Denote by ${\mathbb Q}_x$ the law of the solution $X^x$ of the affine SDE  for $x\in E$, ${\mathbb Q}_x(X_t\in A):={\mathbb Q}(X^x_t\in A)$.
Then $Q_t(x,A)={\mathbb Q}_x(X_t\in A)$ defined for all $t\geq 0$, Borel subsets $A$ of $E$, and $x\in E$ defines a Markov transition semigroup $(Q_t)_{t\geq 0}$  on the Banach space of Borel measurable bounded functions on $E$ by $Q_tf(x):=\int_E f(y)Q_t(x,dy)$. As shown in  \citet{duffie_2003}, this semigroup is {\em Feller}, i.e. it leaves the space of continuous functions vanishing at infinity invariant. Thus, the Markov process $(\Omega,{\mathscr F},(X_t)_{t\geq 0},({\mathbb Q}_x)_{x\in E})$ is a {\em Feller process} on $E$. It has continuous paths in $E$ and has the strong Markov property (cf. \citet{yamada_1971}, Corollary 2, p.162). Thus, it is a Borel right process (in fact, a Hunt process).

In an affine diffusion term structure model, the short rate process $r_t$ is specified to be affine
\be
r_t=r(X_t)=\gamma+\delta^\top X_t
\eel{affinerate}
for some constant $\gamma$ and a $d$-dimensional vector $\delta$. The model is called affine due to the following result (cf. Theorem 4.1 of \citet{filipovic_2009}).
\begin{proposition} {\bf (Zero-Coupon Bonds)}
Let $\tau>0$. The following statements are equivalent: \\
(i) ${\mathbb E}^\mathbb{Q}[e^{-\int_0^\tau r(X^x_s)ds}]<\infty$ for all $x\in {\mathbb R}_+^m\times {\mathbb R}^n$. \\
(ii) There exists a unique solution  $(\Phi(\cdot,u),\Psi(\cdot,u)):[0,\tau]\rightarrow {\mathbb C}\times {\mathbb C}^d$ of the following Riccati system of equations up to time $\tau$
\be
\begin{split}
&\partial_t\Phi(t,u)=\frac{1}{2}\Psi_J(t,u)^\top a_{JJ}\Psi_J(t,u)+b^\top\Psi(t,u)-\gamma, \quad \Phi(0,u)=0,\\
&\partial_t\Psi_i(t,u)=\frac{1}{2}\Psi(t,u)^\top \alpha_i\Psi(t,u)+\beta_i^\top\Psi(t,u)-\delta_i,\quad i\in\emph{I},\\
&\partial_t\Psi_J(t,u)=B_{JJ}^\top\Psi_J(t,u)-\delta_J,\quad \Psi(0,u)=u\\
\end{split}
\eel{riccati_d}
for $u=0$. In either case, there exists an open convex neighborhood $U$ of $0$ in ${\mathbb R}^d$ such that the system of Riccati equations \eqref{riccati_d} admits a unique solution $(\Phi(\cdot,u),\Psi(\cdot,u)):[0,\emph{}\tau]\rightarrow {\mathbb C}\times {\mathbb C}^d$ for all $u\in {\cal S}(U):=\{z\in\mathbb{C}^d|\enskip \text{real part of }z\in U\}$, and the following {\em affine representation} holds:
\be
\mathbb{E}^\mathbb{Q}\Big[e^{-\int_t^{T} r(X^x_s)ds+u^\top X^x_{T}}\left| {\mathscr F}_t\right. \Big]=e^{\Phi(T-t,u)+\Psi(T-t,u)^\top X^x_t}
\eel{representation}
for all $u\in {\cal S}(U)$, $t\leq T\leq t+\tau$ and $x\in {\mathbb R}_+^m\times {\mathbb R}^n$.
\end{proposition}

We next consider a sub-class of mean-reverting affine diffusions.
\begin{definition}
\label{def_meanrevert}{\bf (Mean-Reverting Affine Diffusions)}
Under Assumption \ref{admi_and_nonde}, an affine diffusion $X$ is called mean-reverting if its drift vector has the following form:
\be
b(x)=B(x-\theta),
\label{mean_reverting}
\ee
where all eigenvalues of the matrix $B$ have strictly negative real parts and $\theta\in \mathbb{R}_{++}^m\times \mathbb{R}^n$.
\end{definition}

In the mean-reverting case the unique solution of the ODE system
$dY(t)/dt=B(Y(t)-\theta)$
starting from any initial condition $Y_0=y\in E$ converges to $\theta$ as $t\rightarrow+\infty$.
Since $B_{\emph{I}\emph{J}}=0$, the condition that all eigenvalues of $B$ have strictly negative real parts implies that $B_{\emph{I}\emph{I}}$ and $B_{\emph{J}\emph{J}}$ also have eigenvalues with strictly negative real parts.
Together with the condition that $B_{II}$ has non-negative off-diagonal elements it implies that $-B_{II}$ is a non-singular {\em M-matrix}. For the properties of M-matrices we refer to \citet{berman_1994} or Appendix \ref{vector_equation}. By Proposition \ref{nonsingular_M}, under the assumption that $b_I>0$, if $-B_{II}$ is a non-singular M-matrix and all eigenvalues of $B_{JJ}$ have strictly negative real parts, we have that $-B^{-1}b\in \mathbb{R}_{++}^m\times \mathbb{R}^n$. Thus $b(x)$ can be written in the form Eq.\eqref{mean_reverting} with $\theta=-B^{-1}b$. Therefore, we have the following.
\begin{proposition}
\label{mean_revert_pro}{\bf (Necessary and Sufficient Conditions for Mean-reversion)}
Under Assumptions  \ref{admi_and_nonde}, $X$ is a mean-reverting affine diffusion if and only if $-B_{II}$ is a non-singular M-matrix and all eigenvalues of $B_{JJ}$ have strictly negative real part.
\end{proposition}
We can prove the following result for mean-reverting affine diffusions that will be of key importance in our study of recurrent eigenfunctions.
\begin{proposition}
\label{recurrent_equ_mean}{\bf (Recurrence of Mean-Reverting Affine Diffusions)}
Under Assumptions  \ref{admi_and_nonde}, if the solution $X=(X^x)_{x\in E}$ of the affine SDE is mean-reverting and $({\mathbb Q}_x)_{x\in E}$ denote the probability laws of $X^x$, then the Borel right process $(X,({\mathbb Q}_x)_{x\in E})$ is recurrent in the sense of Definition \ref{recurrent_getoor}.
\end{proposition}
While this result is not surprising in light of the result of \citet{glasserman_2008} on the existence of a stationary distribution for a mean-reverting affine diffusion,
due to some measurability issues the proof of recurrence in the sense of Definition \ref{recurrent_getoor} is quite technical and is omitted here.
The difficulty stems from the fact that affine diffusions are degenerate from the point of view of diffusion theory, with $\rho_{ii}(x)=\sqrt{x_i}$  vanishing on the boundary $x_i=0$.

We will now look for positive eigenfunctions of the pricing operator of affine term structure model in the exponential affine form. 
We start with the ansatz\footnote{A general study of locally equivalent measure transformations and risk premia in affine modes that preserve the affine property is given in \citet{cheridito_2004}. }
$$\pi(x)=e^{u^\top x}$$
for a positive eigenfunction  of the pricing operator ${\mathscr P}_t$. Then the vector $u$ has to satisfy the equation for each $x\in E$ and $t>0$:
\be
\mathbb{E}^\mathbb{Q}\Big[e^{-\int_0^t r(X^x_s)ds+u^\top X^x_t}\Big]=e^{u^\top x-\lambda t}.
\eel{long}
Using Eqs.\eqref{representation} and \eqref{riccati_d}, Eq.\eqref{long} yields the following (in this case $\Phi(t,u)=-\lambda t$, $\Psi(t,u)=u$):
\be
\lambda=\gamma-\frac{1}{2}u_J^\top a_{JJ}u_J-b^\top u,\quad
\frac{1}{2}u^\top \alpha_i u+\beta_i^\top u-\delta_i=0\quad \forall i\in\emph{I},\quad
B_{JJ}^\top u_J-\delta_J=0.
\eel{riccati_a}
Suppose all eigenvalues of $B_{JJ}$ have strictly negative real parts. Then $B_{JJ}$ is non-singular and the OU component ($n$-dimensional vector)  of the $(m+n)$-dimensional vector $u$ is immediately found from the last equation
$$u_J=(B_{JJ}^\top)^{-1}\delta_J.$$
Substituting this result into the second equation, we obtain a {\em quadratic vector equation} for the $m$-dimensional vector $u_I$ (the CIR component of $u$):
\be
Mu_I=c+\frac{1}{2}u_I^2,
\eel{affine_vector}
where $u_I^2$ denotes the $m$-dimensional vector with components $(u_I^2)_i=u_i^2$, $i=1,\ldots,m$, and the $m\times m$-matrix $M$ and  the $m$-dimensional vector $c$ are given by (recall that, without loss of generality, we assume that $\rho(x)$ is in the canonical form \eqref{canonrho1}):
\be
M=-B_{\emph{I}\emph{I}}^\top, \quad
c_i=-\delta_i+\sum_{j=m+1}^{m+n} B_{ji} u_j + \frac{1}{2}\sum_{j,k=m+1}^{m+n} \alpha_{i,jk}u_j u_k.
\eel{M_c_def}
In general, the quadratic vector equation \eqref{affine_vector} can have no, one, or multiple solutions.
If $u_I$ is a solution to the quadratic vector equation \eqref{affine_vector}, then $\pi(x)$ is a positive eigenfunction of the pricing operator with the eigenvalue $e^{-\lambda t}$ with $\lambda$ given by the first equation in \eqref{riccati_a}.
The process
\be
M_t^x=e^{\lambda t-\int_0^t r(X_s^x)ds}\frac{\pi(X_t)}{\pi(x)}=e^{u^\top (X_t^x-x)+\lambda t-\int_0^t (\gamma+\delta^\top X_s^x) ds}
\eel{expmart}
is then a unit-mean positive ${\mathbb Q}_x$-martingale for each $x\in E$ and can be used to define a new probability measure
 ${\mathbb Q}_x^\pi$ for each $x\in E$. It remains to verify whether $X$ is recurrent under this measure.
Appendix \ref{vector_equation} gives some key results about quadratic vector equations of the form \eqref{affine_vector}. Based on these results, we are able
to give sufficient conditions to ensure that there exists a solution $u$ such that  $X$ is a mean-reverting affine diffusion under ${\mathbb Q}^\pi$ associated with the positive eigenfunction $\pi(x)=e^{u^\top x}$. By Proposition \ref{recurrent_equ_mean}, a mean-reverting affine diffusion is recurrent in the sense of Definition \ref{recurrent_getoor}. Thus, $X$ is recurrent under the corresponding ${\mathbb Q}^\pi$. Finally,
by Theorem \ref{unique_theorem_1} such a solution is unique, and $\pi$ is the unique recurrent eigenfunction.
\begin{theorem}
\label{exist_theorem_affine} {\bf (Existence of a Recurrent Eigenfunction in Affine Diffusion Models)}
Let $X$ be an affine diffusion \eqref{canonsde1}-\eqref{canonsde2} satisfying Assumption \ref{admi_and_nonde}  under the risk-neutral probability measure ${\mathbb Q}$ and $r$ is the short rate \eqref{affinerate}.
Suppose the coefficients of $X$ satisfy the following additional assumptions:
(i) If $n>0$, then all eigenvalues of the matrix $B_{JJ}$  have strictly negative real parts;
(ii) If $m>0$, then there exists a vector $y\in {\mathbb R}^m$, such that
the following inequality holds (here the $m$-vector $y^2$  has components $(y^2)_i=y_i^2$ and $M$ and $c$ are defined in \eqref{M_c_def}):
$$My-c-\frac{1}{2}y^2>0.$$
Then there exists a unique positive eigenfunction of the affine pricing operator ${\mathscr P}_t$ and it has the exponential affine form
\be
\pi(x)=e^{u^\top x}=e^{{u}_I^\top x_I+u_J^\top x_J}
\eel{eigen_affine}
with the eigenvalue $e^{-\lambda t}$ with
$$\lambda=\gamma-\frac{1}{2}{u}_J^\top a_{JJ}{u}_J-b^\top {u},$$
where
$u_I=u^*_I$
is the {\em minimal solution} of the quadratic vector equation Eq.\eqref{affine_vector} guaranteed to exist under the assumptions (i) and (ii), and
$$u_J= (B_{JJ}^\top)^{-1}\delta_J,$$ such that under the corresponding probability measure ${\mathbb Q}^{\pi}$ the process $X$
follows a mean-reverting affine diffusion  \eqref{canonsde1}--\eqref{canonsde2}
with the following drift parameters:
\be
\tilde{b}_I=b_I,\quad \tilde{B}_{II}=B_{II}+{\rm diag}(u_1^*,...,u_m^*),\quad \tilde{b}_J=b_J+a_{JJ}u_J,
\eel{pparameters1}
\be
\tilde{B}_{JJ}=B_{JJ},\quad (\tilde{B}_{JI})_{ki}=(B_{JI})_{ki}+\sum_{l=m+1}^{m+n} \alpha_{i,kl}u_l,\quad i=1,...,m,\,k=m+1,...,m+n,
\eel{pparameters2}
and driven by an $(m+n)$-dimensional ${\mathbb Q}^\pi$-standard Brownian motion
$B^{{\mathbb Q}^\pi}_t=B^\mathbb{Q}_t-\int_0^t \Lambda_s ds$
with the market price of risk process given by
\be
\Lambda_i(t)=(\Lambda_I)_i(t) = u_i^*\sqrt{X_i(t)},\quad i=1,\ldots,m,\quad \Lambda_J(t)=\rho_{JJ}^\top u_J.
\eel{market_price2}
\end{theorem}
{\bf Proof.}
First suppose that $u_I^*$ is a solution to the quadratic vector equation (not necessarily minimal), $M_t^x$ is the corresponding martingale \eqref{expmart}, and ${\mathbb Q}^\pi_x$ is the corresponding probability measure. By It\^{o}'s formula,
\be
d M^x_t=M^x_t\Lambda_t^\top d B^\mathbb{Q}_t,
\ee
where $\Lambda_t=(\Lambda_I(t),\Lambda_J(t))$ is given by Eq.\eqref{market_price2}. By Girsanov's theorem, the process $B^{{\mathbb Q}^\pi}$ is a standard Brownian motion under $\mathbb{Q}^\pi$. It is then immediate that $X$ solves the affine SDE \eqref{canonsde1}-\eqref{canonsde2} with drift parameters \eqref{pparameters1}-\eqref{pparameters2} and the unchanged $\rho(x)$ under $\mathbb{Q}^\pi$. It is straightforward to verify directly that the drift parameters under $\mathbb{Q}^\pi$ satisfy  Assumption \ref{admi_and_nonde}.

We next verify that the conditions (i) and (ii) are indeed sufficient both to ensure that the minimal solution $u_I^*$ exists, and that the affine process $X$ is mean-reverting under $\mathbb{Q}^\pi$.
First we observe that $M$ is a {\em Z-matrix}.
By Theorem \ref{my_solution},  Eq.\eqref{affine_vector} has a solution if and only if there exists a vector $y$, such that $My-c-\frac{1}{2}y^2\geq 0$. If a solution exists, then there exists a minimal solution $u_I^*$ such that
$M-{\rm diag}(u_1^*,...,u_m^*)$ is an {\em M-matrix} (cf. Appendix \ref{vector_equation}).
If furthermore there exists a vector $y$ satisfying a strict inequality $My-c-\frac{1}{2}y^2>0$, then $M-{\rm diag}(u_1^*,...,u_m^*)$ is  a {\em non-singular M-matrix}. Thus, the condition (ii) guarantees that both the minimal solution $u_I^*$ exists and that the matrix $M-{\rm diag}(u_1^*,...,u_m^*)$ is a non-singular $M$-matrix. By Proposition \ref{mean_revert_pro} $X$ is a mean-reverting affine diffusion under ${\mathbb P}^\pi$ if and only if $-\tilde{B}_{II}=-B_{II}-{\rm diag}(u_1^*,...,u_m^*)$ is a non-singular $M$-matrix (ensured by the condition (ii)) and all eigenvalues of $B_{JJ}$ have strictly negative real parts (ensured by the condition (i)).
Thus, we have verified that
the condition (ii) is sufficient for the minimal solution $u_I^*$ to exists and that
 the conditions (i) and (ii) together are sufficient for  $X$ to be mean-reverting under the measure change $\mathbb{Q}^\pi$ corresponding to this solution. By
Proposition \ref{recurrent_equ_mean} $X$ is recurrent under $\mathbb{Q}^\pi$. Finally, by Theorem \ref{unique_theorem_1} a recurrent eigenfunction is unique. $\Box$\\
\\
There are two special cases where the condition (ii) in Theorem \ref{exist_theorem_affine} automatically holds without any restrictions on the coefficients. The first case is $\delta_\emph{I}>0$, $\delta_\emph{J}=0$. In this case the short rate depends only on the CIR factors $X^I$ and is independent of the OU factors $X^J$. The second case is $m=0$, so that $X$ is an $n$-dimensional OU process with no CIR factors.

The condition (ii) is merely a non-degeneracy condition.
Its weaker version, requiring that there exists $y$, such that $My-c-\frac{1}{2}y^2\geq 0$ (denote it as (ii)*),
still ensures that Eq.\eqref{affine_vector} has a solution. But under the change of measure corresponding to this solution $X$ is not necessarily mean-reverting.
The strict inequality in (ii), together with the assumption (i), ensures the mean-reverting property of $X$ under the measure change corresponding to the solution, thus ensuring recurrence.


As explained in the proof of Theorem \ref{my_solution} in Appendix \ref{vector_equation}, Eq.\eqref{affine_vector} can be easily solved numerically by Newton's iterations. Start from some negative initial vector $u^0<0$. Given $u^k$, the vector $u^{k+1}$ is found by solving the linear system
$$
J(u^k)(u^{k+1}-u^k)=-F(u^k)
$$
with $F(u)=Mu-c-\frac{1}{2}u^2$ and Jacobian $J(u)=M-{\rm diag}(u_1,\ldots,u_m)$.
If Newton's iterations converge to a solution $u$, we then check if the matrix $J(u)=M-{\rm diag}(u_1,\ldots,u_m)$ is a non-singular $M$-matrix. If it is, then this is a minimal solution, and $X$ is recurrent under the corresponding recovery.
If Newton's iterations converge, but the solution is such that $J(u)=M-{\rm diag}(u_1,\ldots,u_m)$ is not a non-singular $M$-matrix, then this solution is not minimal, i.e. $u^* < u$, and we continue our search for a minimal solution by selecting a new starting point $<u_0$ and repeat the algorithm. If Newton's iterations do not converge, then either our starting point $u_0>u^*$, or there is no solution (condition (ii)* is not satisfied). We then select a new starting point $<u_0$ and repeat the algorithm. Details are given in Appendix \ref{vector_equation}.

\section{Quadratic Vector Equations}
\label{vector_equation}

Here we study quadratic vector equations of the form \eqref{affine_vector}.
We start with the definition of an M-matrix (see \citet{berman_1994}).
\begin{definition}
\label{definition_M}
If a matrix $A$ can be expressed as $A=sI-B$ with $s>0$ and $B\geq0$ with $s\geq\rho(B)$, the spectral radius of $B$, then $A$ is called an M-matrix. If $s>\rho(B)$, then $A$ is called a non-singular M-matrix.
\end{definition}

\begin{proposition}
If $A$ is a Z-matrix (i.e. a matrix with non-positive off-diagonal entries), then the following condition is equivalent to $A$ being an M-matrix.
If $x\neq0$ and $y=Ax$, then for some subscript $i$, $x_i\neq0$ and $x_iy_i\geq0$.
\label{singular_M}
\end{proposition}
\begin{proposition}
If $A$ is a Z-matrix, then the following conditions are equivalent to $A$ being a non-singular M-matrix:\\
(i) The real part of each eigenvalue of A is positive .\\
(ii) $A^{-1}$ exists and $A^{-1}\geq0$.\\
(iii) If $A\geq \epsilon I+M$ for an M-matrix $M$ and $\epsilon>0$.\\
(iv) There exists $x\geq0$ such that $Ax>0$.\\
\label{nonsingular_M}
\end{proposition}
Let $F(x)=Mx-c-\frac{1}{2}x^2$ and $J(x)=M-{\rm diag}(x_1,\ldots,x_m)$ (the Jacobian). We consider the equation $F(x)=0$.
A solution $x^*$ is called minimal if for any other solution $x$ we necessarily have that $x^*\leq x$. We have the following results.

\begin{theorem}
\label{my_solution}
If $M$ is a Z-matrix, then:\\
\emph{(i)} $F(x)=0$ has a solution if and only if there exists a vector $y$ such that $F(y)\geq 0$.\\
\emph{(ii)} If $F(x)=0$ has at least one solution, there is a minimal solution $x^*$. Furthermore, $J(x^*)$ is an M-matrix.\\
\emph{(iii)} If $F(x)=0$ has a minimal solution $x^*$, then starting from any $x^0<x^*$ Newton's iterations $(x^k)_{k\geq 0}$ solving
\be
J(x^k)(x^{k+1}-x^k)=-F(x^k)
\eel{NI}
converge monotonically to $x^*$, $x^1<x^2<\ldots<x^*$.\\
\emph{(iv)} If there exists a vector $y$ such that $F(y)\geq0$, then the minimal solution $x^*\leq y$.\\
\emph{(v)} If there exists a vector $y$ such that $F(y)>0$, then $F(x)=0$ has a minimal solution $x^*<y$. Furthermore,  $J(x^*)$ is a non-singular M-matrix.
\end{theorem}
{\bf Proof.}
(i) Let $\xi+\theta=x$, where $\theta$ is a non-positive constant vector. Substituting it into $F(x)=0$, we get:
\be
[M-\text{diag}(\theta_1,\ldots,\theta_m)]\xi=[c+\frac{1}{2}\theta^2-M\theta]+\frac{1}{2}\xi^2.
\eel{transform}
We can choose $\theta_k=-s$, so that $M-\text{diag}(\theta_1,\ldots,\theta_m)=sI+M$. By Definition \ref{definition_M} and Proposition \ref{nonsingular_M} (iii), we can choose $s$ large enough so that $M-\text{diag}(\theta_1,\ldots,\theta_m)$ is a non-singular M-matrix. Furthermore, since $c+\frac{1}{2}\theta^2-M\theta$ is quadratic and strictly convex, we can choose $s$ large enough so that $c+\frac{1}{2}\theta^2-M\theta>0$. Finally, there exists a finite number of solutions of $F(x)=0$, so if $s$ is large enough, every solution $x$ of $F(x)=0$ satisfies $x>\theta$. We choose $s$ that satisfies all the properties above.
Now let us define
\be
\tilde{M}=M-\text{diag}(\theta_1,\ldots,\theta_m),\enskip \tilde{c}=c+\frac{1}{2}\theta^2-M\theta,
\enskip \tilde{F}(\xi)=\tilde{M}\xi-\tilde{c}-\frac{1}{2}\xi^2.\enskip
\tilde{J}(\xi)=\tilde{M}-\text{diag}(\xi_1,\ldots,\xi_m).
\ee
Then $F(x)=0$ is equivalent to
\be
\tilde{M}\xi=\tilde{c}+\frac{1}{2}\xi^2,
\eel{quadratic_2}
where $\tilde{M}$ is a non-singular M-matrix and $\tilde{c}\geq0$ so that it can only have strictly positive solution. By Lemma 3.4 of \citet{poloni_2010}, Eq.\eqref{quadratic_2} has a positive solution if and only if there exists a vector $y\geq0$ such that $\tilde{F}(y)\geq0$. Suppose there exists a vector $y$ such that $F(y)\geq0$. Since $\tilde{F}(\xi)=F(x)$ when $\xi+\theta=x$, we can choose $\theta<y$ and $\eta=y-\theta>0$, such that $\tilde{F}(\eta)\geq0$. This means that Eq.\eqref{quadratic_2} (hence $F(x)=0$) has a solution. Conversely, if $F(x)=0$ has a solution, there exists an $y$ such that $F(y)\geq0$. This proves (i).\\
(ii) Since $\tilde{J}(\xi)=J(x)$ when $\xi+\theta=x$, (ii) follows from Theorem 3.1 and 3.2 of \citet{poloni_2010}.\\
(iii) Choose $x^0$ such that $x^0<x^*$. We prove by induction that $x^k< x^*$. For $k=0$, $x^0<x^*$, so the base step holds.
Suppose $x^k< x^*$. Since $J(x^*) $ is an M-matrix and $x^k< x^*$, by Proposition \ref{nonsingular_M} (iii) and (ii), $J(x^k)=M-\text{diag}(x^k_1,\ldots,x^k_m)$ is a non-singular M-matrix and $(J(x^k))^{-1}\geq0$. Thus, Newton's iterations are well-defined. To show monotonicity, we first note that for each $k$
\be
J(x^k)(x^*-x^k)+F(x^k)=\frac{1}{2}(x^*-x^k)^2>0.
\ee
Since $(J(x^k))^{-1}\geq0$, this implies
$-(J(x^k))^{-1}F(x^k)< x^*-x^k,$
which implies $x^{k+1}< x^*$ for each $k$. This completes the induction.

By Newton's iteration \eqref{NI}, $F(x^{k})=-\frac{1}{2}(x^{k}-x^{k-1})^2\leq 0$ for all $k\geq1$. We have showed that $(J(x^k))^{-1}\geq0$, thus by \eqref{NI}, we get $x^{k+1}\geq x^k$ for all $k\geq1$. This means $x^k$ is a monotonic sequence bounded above by $x^*$, hence it must converge. Denote its limit as $x^\infty\leq x^*$. Then by taking the limit in Eq.\eqref{NI} we get $F(x^\infty)=0$, which means $x^\infty$ is a solution to Equation $F(x)=0$. By the definition of minimal solution, $x^\infty=x^*$.\\
(iv) Choose $x^0$ such that $x^0<x^*$, $x^0<y$ and use it as the starting point of Newton's iterations. As in the proof of last statement, we can prove by induction that $x^k<y$. Then taking the limit gives (iv).\\
(v) By (i) and (ii), we know $F(x)=0$ has a minimal solution $x^*$. Consider another quadratic vector equation $\bar{F}(x)=Mx-\bar{c}-\frac{1}{2}x^2$ where $\bar{c}=c+F(y)$ (note that $y$ is given in the assumption and is fixed). Since $\bar{F}(y)=0$, again by (i) and (ii), we know $\bar{F}(y)=0$ has a minimal solution $\bar{x}^*\leq y$. Since $F(\bar{x}^*)=\bar{F}(\bar{x}^*)+F(y)=F(y)>0$, by (iv) the minimal solution $x^*$ of $F(x)=0$ satisfies $x^*\leq\bar{x}^*$. Let $\xi=\bar{x}^*-x^*\geq0$. We can verify that
\be
J(x^*)\xi=F(y)+\frac{1}{2}\xi^2>0.
\ee
Since $J(x^*)$ is a Z-matrix, by Proposition \ref{nonsingular_M} (iv), $J(x^*)$ is non-singular M-matrix. Since $J(x^*)$ is Z-matrix, $\xi\geq0$ and $J(x^*)\xi>0$, we see that $\xi>0$. Thus $x^*<\bar{x}^*\leq y$.
$\Box$

\section{Recurrent Eigenfunction in Quadratic Term Structure Models}
\label{appendix_quadratic}

In this section we study recurrent eigenfunctions in quadratic term structure models (\citet{beaglehole_1992}, \citet{constantinides_1992theory}, \citet{rogers_1997}, \citet{ahn_2002}, and \citet{chen_2004}).
Suppose $X$ is a $d$-dimensional OU process solving the SDE under ${\mathbb Q}$:
\be
dX_t=(b+BX_t)dt+\rho dB_t^{\mathbb Q},
\ee
where $b$ is a $d$-dimensional vector, $B$ is a $d\times d$ matrix, and $\rho$ is a non-singular $d\times d$ matrix, so that the diffusion matrix $a=\rho\rho^\top$ is strictly positive definite.
The short rate function is taken to be
\be
r(x)=\gamma+\delta^\top x +x^\top \Phi x,
\ee
where the constant $\gamma$, vector $\delta$ and symmetric positive semi-definite matrix $\Phi$ are taken to be such that the short rate is non-negative for all $x\in {\mathbb R}^d$.

If $\Phi$ is strictly positive definite, then the QTSM satisfies the sufficient conditions in Theorem \ref{exist_3_main} (since $r(x)\rightarrow \infty$ as $\|x\|\rightarrow \infty$), and there is a unique recurrent eigenfunction. If $\Phi$ is merely positive semi-definite, this case is generally outside the sufficient condition in Theorem \ref{exist_3_main}, but there may still be a unique recurrent eigenfunction. Below we establish a sufficient condition.
We assume that $\Phi$ is not zero (if it is, then the model reduces to the affine model with only OU factors and no CIR factors and is covered in the last section).
Consider an exponential quadratic function
$f_{u,V}(x):=e^{-u^\top x -x^\top V x},$
where the vector $u$ and symmetric positive semi-definite matrix $V$ are such that $u^\top x +x^\top V x\geq c$ for all $x\in {\mathbb R}^d$ and some real constant $c$.
Then the following holds (cf. Theorem 3.6 of \citet{chen_2004}):
\be
\mathbb{E}_x^{\mathbb Q}\Big[e^{-\int_0^T r(X_s)ds}f_{u,V}(X_T)\Big]=e^{-l(T,u,V)- m(T,u,V)^\top x- x^\top N(T,u,V)x},
\eel{quadratic_term}
where the scalar $l$, vector $m$ and symmetric matrix $N$ satisfy the Riccati equations ($tr(\cdot)$ denotes the  matrix trace):
\be
\begin{split}
&\partial_t l(t,u,V)=F(m(t,u,V),N(t,u,V)),\enskip l(0,u,V)=0,\\
&\partial_t  m(t,u,V)=R(m(t,u,V),N(t,u,V)),\enskip m(0,u,V)=u,\\
&\partial_t  N(t,u,V)=T(m(t,u,V),N(t,u,V)),\enskip N(0,u,V)=V,\\
\end{split}
\eel{quadratic_riccati}
with $F(m,N)=-\frac{1}{2}m^\top a m+tr(a N)+m^\top b +\gamma,$
$R(m,N)=-2N a m+B^\top m+2N b+\delta$ and
$T(m,N)=-2N a N +B^\top N+NB+\Phi.$

We look for a positive eigenfunction $\pi(x)$ of the pricing operator in the exponential quadratic form $f_{u,V}(x)$:
\be
\mathbb{E}\Big[e^{-\int_t^T r(X_s)ds+\lambda T}f_{u,V}(X_t)\Big]=f_{u,V}(x).
\ee
Using Eqs.\eqref{quadratic_term} and \eqref{quadratic_riccati}, we see that the matrix $V$ satisfies the so-called
{\em continuous-time algebraic Riccati equation (CTARE)} (such equations are  well studied in the stochastic control literature, cf. \citet{lancaster_1995}) $$2VaV-B^\top V-VB-\Phi=0.$$
Given a CTARE solution $V$, the vector $u$ satisfies the linear equation
$$2V a u-B^\top u-2V b-\delta=0.$$
Given the solutions $u$ and $V$, the eigenvalue is
$$\lambda=\gamma-\frac{1}{2}u^\top a u+tr(a V)+u^\top b.$$

Suppose first that $\Phi$ is positive definite (this is the case considered in Example 3.2 in \citet{rogers_1997}). Then by the CTARE theory there exists a unique solution $V$ such that all of the eigenvalues of the matrix $B-2a V$ have negative real parts. Standard numerical algorithms are available to determine the solution numerically, including in Matlab.
Since the matrix $B-2aV$ is non-singular, there is also a unique solution for $u$. Then we have a positive ${\mathbb Q}$-martingale
$\tilde{M}^\pi_t=e^{-\int_0^t r(X_s)ds+\lambda t}f_{u,V}(X_t)/f_{u,V}(X_0)$ and can define a new measure ${\mathbb Q}^\pi$. By It\^{o}'s formula we obtain $d \tilde{M}^\pi_t=-\tilde{M}^\pi_t \Lambda(X_t)^\top dB_t^{\mathbb Q},$
where $\Lambda(X_t)=\rho^\top(-u-2VX_t)$. By Girsanov's theorem, $B_t^{{\mathbb Q}^\pi}=B_t^{{\mathbb Q}}-\int_0^t \Lambda(X_s)ds$ is a standard Brownian motion under $\mathbb{Q}^\pi$. Thus, under $\mathbb{Q}^\pi$
\be
d X_t
=(b-a u+ (B-2a V)X_t)dt+\rho d B_t^{{\mathbb Q}^\pi}.
\ee
Since the eigenvalues of the matrix $B-2a V$ have non-negative real parts, $X$ is mean-reverting and, hence, recurrent under $\mathbb{Q}^\pi$ (cf. Appendix \ref{affine_appendix}). Thus $\pi=f_{u,V}$ is the unique recurrent eigenfunction.

Now consider the case where $\Phi$ is merely positive semi-definite.
By the CTARE theory (cf. \citet{lancaster_1995} p.234), if the pair of matrices $(B^\top,\Phi)$ is {\em stabilizable}\footnote{The pair of matrices $(B^\top,\Phi)$ is stabilizable if and only if there exists a matrix $K$ such that all of the eigenvalues of the matrix $B^\top + \Phi K$ have negative real parts.}, where $B$ is the matrix in the drift of the OU process under the risk-neutral measure ${\mathbb Q}$, then there exists a unique positive semi-definite solution $V$  of the CTARE such that all of the eigenvalues of the matrix $B-2aV$ have strictly negative real parts (this solution can be found numerically by standard numerical algorithms).
Thus, in this case $X$ is mean-reverting under ${\mathbb Q}^\pi$ corresponding to this solution $(u,V)$, and $\pi=f_{u,V}$ is the recurrent eigenfunction.

\section{Recurrence of a CIR Process with Jumps}
\label{appendix_cir_jump}
\citet{lingfe_2014density} obtain an analytical representation for the transition density of the JCIR process. From their results it follows that the JCIR process has a positive density.
Thus, it is irreducible with respect to the Lebesgue measure.

On the other hand, we know that affine jump-diffusion processes have the Feller property (cf. \citet{duffie_2003} Theorem 2.7). By Theorem 3.2 (ii)$\rightarrow$(iii) of \citet{schilling_1998}, its transition semigroup maps bounded continuous functions to bounded continuous functions. By Theorem 7.1 of \citet{tweedie_1994}, $X$ is a $T$-model. Take an arbitrary point $x_0\in\mathbb{R}_+$. By Theorem 4.1 of \citet{tweedie_1994}, $X$ is recurrent in the sense of Definition \ref{recurrent_tweedie} if and only if $x_0$ is topologically recurrent (a point $x$ is topologically recurrent if $R(x,O)=\infty$ for all neighborhoods $O$ of $x$).
By Theorem 2.6 of \citet{martin_2011}, $X$ has a limiting distribution under $\mathbb{Q}^\pi$, and this limiting distribution is also a stationary distribution. Since JCIR process has a positive transition density, it is then clear that its stationary measure must charge every open neighborhood of $x_0$ (and, in fact, every set with positive Lebesgue measure). Thus it is easy to see that $R(x_0,O)=\infty$ for all neighborhoods $O$ of $x_0$, i.e. $x_0$ is topologically recurrent. Thus, $X$ is recurrent in the sense of Definition \ref{recurrent_tweedie} {\bf(R1)}. Since $X$ has a positive transition density with respect to Lebesgue measure, it satisfies Assumption \ref{ac_condition}. Thus by Proposition \ref{relation_recu} $X$ is also recurrent in the sense of Definition \ref{recurrent_getoor} {\bf(R0)}.

\bibliography{mybib7}

\end{document}